\documentclass[a4paper,11pt,fleqn]{article}
\usepackage[DIV13]{typearea}
\usepackage{amsmath}
\usepackage{amssymb} 
\usepackage{amsfonts}
\usepackage{cite}
\usepackage{bbold}
\usepackage[footnotesize]{caption2}
\usepackage{graphicx}
\usepackage[center,footnotesize,hang]{subfigure}
\usepackage{url}

\newcommand{\GeV}{\ensuremath{\,\mathrm{GeV}}}
\newcommand{\TeV}{\ensuremath{\,\mathrm{TeV}}}

\newcommand{\Cl}[1]{\mathcal{C} _{#1}}

\newcommand{\Ord}[2]{\; ^{\circ} \mathrm{#1}_{#2}  \;}
\newcommand{\OrdCl}[1]{\; ^{\circ} \mathcal{C} _{#1} \;}
\newcommand{\Rep}[1]{\underline{\mbox{\textbf{#1}}}}
\newcommand{\MoreRep}[2]{\underline{\mbox{\textbf{#1}}} _{\mbox{\textbf{#2}}}}
\newcommand{\Groupname}[2]{$ {#1} _{#2} $}
\newcommand{\Doub}[2]{$ {#1} _{#2} ^{\prime} $}

\DeclareMathOperator{\diag}{diag}

\newcommand{\Eqref}[1]{Eq.\eqref{#1}}

\newcommand{\Tabref}[1]{Table \ref{#1}}

\newcommand{\Secref}[1]{Section \ref{#1}}
\newcommand{\Appref}[1]{Appendix \ref{#1}}

\newcommand{\VEV}[1]{\langle #1 \rangle}

\begin{document}
\begin{titlepage}
\vspace*{10mm}
\begin{center}
{\Large\sffamily\bfseries
\mathversion{bold} $\theta_{C}$
 from the Dihedral Flavor Symmetries 
\Groupname{D}{7} and \Groupname{D}{14}
\mathversion{normal}}
\\[13mm]
{\large
A.~Blum\footnote{E-mail: \texttt{alexander.blum@mpi-hd.mpg.de}}, 
C.~Hagedorn\footnote{E-mail: \texttt{claudia.hagedorn@mpi-hd.mpg.de}} and
A.~Hohenegger\footnote{E-mail: \texttt{andreas.hohenegger@mpi-hd.mpg.de}}} 
\\[5mm]
{\small \textit{
Max-Planck-Institut f\"{u}r Kernphysik\\ 
Postfach 10 39 80, 69029 Heidelberg, Germany
}}
\vspace*{1.0cm}
\end{center}
\normalsize
\begin{abstract}
\noindent In \cite{dntheorypaper} it has been shown that the Cabibbo angle
$\theta_{C}$ might arise from a dihedral flavor symmetry which is broken to different
(directions of) subgroups in the up and the down quark sector. This leads to
a prediction of $\theta_{C}$ in terms of group theoretical quantities only, i.e.
the index $n$ of the dihedral group \Groupname{D}{n}, the index $\rm j$ of the
fermion representation $\MoreRep{2}{j}$ and the preserved subgroups indicated by
$m_{u}$ and $m_{d}$. Here we construct a low energy model which incorporates this idea. The 
gauge group is the one
of the Standard Model and \Groupname{D}{7}
$\times$ $Z_{2} ^{(aux)}$ serves as flavor symmetry. The additional $Z_{2} ^{(aux)}$ is 
necessary in order to maintain two sets of Higgs fields, one which couples only
to up quarks and another one coupling only to down quarks.
We assume that \Groupname{D}{7} is broken spontaneously at the electroweak
scale by vacuum expectation values of $SU(2)_{L}$ doublet Higgs fields. The quark masses and
mixing parameters can be accommodated well. Furthermore, the potential of the Higgs fields
is studied numerically in order to show that the required configuration of the
vacuum expectation values can be achieved. We also comment on more minimalist models which 
explain the Cabibbo angle in terms of group theoretical quantities, while
$\theta_{13}^{q}$ and $\theta_{23} ^{q}$ vanish at leading order. 
Finally, we perform a detailed numerical study of the lepton mixing matrix
$V_{MNS}$ in which one of its elements is entirely determined by the
group theory of a dihedral symmetry. Thereby, we show that nearly tri-bi-maximal
mixing can also be produced by a dihedral flavor group with preserved subgroups. 
\end{abstract}

\end{titlepage}

\setcounter{footnote}{0}

%%%%%%%%%%%%%%%%%%%%%%%%%%%%%%%%%%
\section{Introduction}
%%%%%%%%%%%%%%%%%%%%%%%%%%%%%%%%%%
\label{sec:intro}

\noindent Discrete groups have been widely used as flavor symmetry.
However, only in some special cases there is a direct connection
between the flavor group $G_{F}$ and the resulting mixing pattern for
the fermions, i.e. a correlation which does not rely on further parameter
equalities not induced by $G_{F}$. Such cases occur 
in the \Groupname{A}{4} (\Doub{T}{}) models 
\cite{a4,tprime} as well as in our systematic study of the dihedral
groups \cite{dntheorypaper} where the key feature
is the fact that $G_{F}$ is broken
in a non-trivial way, i.e. one has to demand that certain of its subgroups
 are preserved in different sectors of the theory. Especially,
the fact that sizable mixing results from the mismatch of
two different (directions of) subgroups is used in the \Groupname{A}{4} (\Doub{T}{})
models as well as in an application of the group \Groupname{D}{7}
shown in \cite{dntheorypaper}. In the group
 \Groupname{A}{4} (\Doub{T}{}) which has been studied in great detail
 tri-bi-maximal mixing (TBM) in the lepton sector is predicted, if one assumes
that the left-handed leptons transform as a triplet under \Groupname{A}{4} (\Doub{T}{}),
and the left-handed conjugate leptons, $e^{c}$, $\mu^{c}$ and $\tau^{c}$, as the three non-equivalent
one-dimensional representations of the group. There exist two sets of gauge singlets which transform
non-trivially under \Groupname{A}{4} (\Doub{T}{}): one set only couples to 
neutrinos at the leading order (LO), while the other one only to charged
leptons (fermions). The first one breaks \Groupname{A}{4} (\Doub{T}{})
spontaneously down to \Groupname{Z}{2} (\Groupname{Z}{4}) and the latter one down to
\Groupname{Z}{3}. The lepton mixing then stems from
two sectors in which different subgroups of \Groupname{A}{4} (\Doub{T}{}) are 
conserved.
In contrast to this, the up quark and down quark mass matrix preserve the
same subgroup at LO. Similarly, it has recently been shown that such
a mechanism can also be implemented with other discrete groups, for
example the dihedral groups \Groupname{D}{n} and \Doub{D}{n}.
In a first application we observed in \cite{dntheorypaper} that
the Cabibbo angle $\theta_{C}$ or equivalently the CKM matrix element
$|V_{us}|$ can be predicted in terms of group theoretical
indices only, such as the index $n$ of the group \Groupname{D}{n}, the index $\rm j$
of the representation under which the (left-handed) quarks transform and the
misalignment of the two different (directions of) subgroups 
$Z_{2} = <\mathrm{B} \, \mathrm{A} ^{m_{u}}>$ and $Z_{2} = <\mathrm{B} \, \mathrm{A} ^{m_{d}}>$:
\begin{equation}
|V_{us}|= \left| \cos \left( \frac{\pi \, (m_{u} - m_{d}) \, \mathrm{j}}{n} \right) \right|
\end{equation}
\noindent There is a  crucial difference between the models using a dihedral symmetry 
and \Groupname{A}{4} (\Doub{T}{}) as flavor symmetry, namely the issue whether the
representations under which the Higgs (flavon) fields transform are chosen or not. 
In our study on dihedral
symmetries \cite{dntheorypaper} we always assumed that for each representation $\mu$
which (has a component which) transforms trivially under the relevant subgroup
 there exist(s) (a) Higgs field(s) transforming as $\mu$ and acquiring a non-vanishing 
vacuum expectation value (VEV).
Due to this the resulting mass matrices are only determined by the choice of the
fermion representations, the dihedral group and the preserved subgroups,
but not by the choice of the scalar fields. This makes our results less
arbitrary. However, in the case of the \Groupname{A}{4} (\Doub{T}{}) it is 
necessary to choose the transformation properties of the scalar fields
properly, i.e. one has to exclude scalars which transform
as non-trivial singlets under \Groupname{A}{4} (\Doub{T}{}) and couple to neutrinos at LO, 
in order to arrive at the TBM scenario \cite{AF,tprime}.\\
\noindent In this paper we investigate
the idea of \cite{dntheorypaper} by constructing a viable (low energy) model for the quark
sector. The gauge group is chosen to be the one of the Standard Model (SM). 
We study the mass matrices numerically in order to demonstrate that all quark
masses and mixing parameters can be accommodated. We
discuss the Higgs potential under the assumption that all involved
fields are copies of the SM Higgs doublet. Furthermore, instead of accommodating all
quark mixing angles at LO it is also worth studying setups
in which the Cabibbo angle is predicted in terms of group theoretical 
quantities, while the two other mixing angles are zero. This can be done in at 
least two ways: $a.)$ one can choose the
representations under which the scalar fields should transform or
$b.)$ one can look at cases in which the preserved subgroup in each sector is not
only a  \Groupname{Z}{2}, but rather is itself a dihedral group
\Groupname{D}{q}, $q>1$. Finally, we motivate
possible extensions of the model to the lepton sector by
performing a detailed numerical study. Additionally, we show that 
nearly TBM can be also accommodated by using a dihedral flavor symmetry.\\
\noindent The paper is organized as follows: in \Secref{sec:basics} we review
the findings of \cite{dntheorypaper} which we explore in more
detail; \Secref{sec:vckm} treats the mixing matrix $V_{CKM}$ only - in an
analytic way as well as numerically; in \Secref{sec:quarksector} we
study a model for the quark sector which incorporates the idea
presented in \cite{dntheorypaper} and show that it accommodates both
quark mixings and masses; in \Secref{sec:Higgs} the Higgs potential, belonging to one of 
the models of \Secref{sec:quarksector}, is discussed and a numerical analysis proves
that the advocated VEV structure can be achieved. 
\Secref{sec:alternatives} is devoted to ans\"{a}tze in which only the Cabibbo
angle is generated at LO. In \Secref{sec:vmns}
we perform the same analysis, as for the quark mixing matrix $V_{CKM}$ in
\Secref{sec:vckm}, for the lepton mixing matrix $V_{MNS}$ in order to see
whether the fact that one element of the mixing matrix is described only in terms
of group theoretical quantities is also applicable here. Thereby, we assume that the
neutrinos are Dirac particles as all the other fermions and possess the same mass ordering, i.e. that they are normally ordered. Finally, we summarize our results in
\Secref{sec:summary}. \Appref{app:vmix} contains the possible forms
of the mixing matrices $V_{CKM}$ and $V_{MNS}$, in \Appref{app:grouptheory} 
 the group theory of \Groupname{D}{7}, i.e. the flavor group used in \Secref{sec:quarksector} and \Secref{sec:Higgs},
is presented. Further details of the study of the Higgs sector are delegated to \Appref{app:Higgs}.

%%%%%%%%%%%%%%%%%%%%%%%%%%%%%%%%%%%%%
\section{Basics}
%%%%%%%%%%%%%%%%%%%%%%%%%%%%%%%%%%%%%
\label{sec:basics}

\noindent In this section we repeat the findings of \cite{dntheorypaper}
concerning the possible structure of (Dirac) mass matrices with a non-vanishing determinant. 
They are of the form:
\begin{eqnarray}
\label{eq:generalmatrices}
&M_{1} = \left( \begin{array}{ccc} 
		A & 0 & 0 \\
		0 & B & 0 \\
		0 & 0 & C 
	\end{array} \right) \; , \;\; 
M_{2} = \left( \begin{array}{ccc}
		A & 0 & 0\\				
		0 & 0 & B\\
		0 & C & 0
		\end{array} \right)&\\ 
\label{eq:blockmatrix}
& M_{3} = \left( \begin{array}{ccc}
		A & 0 & 0\\				
		0 & B & C\\
		0 & D & E
		\end{array} \right) \; ,&\\ 
\label{eq:M4andM5fordowns}
& M_{4} = \left( \begin{array}{ccc}
		0 & A & B\\				
		C & D & E\\
		-C \, \mathrm{e}^{- i \, \phi \, \mathrm{j}} & D \, \mathrm{e}^{-i \, \phi \, \mathrm{j}} 
		& E \, \mathrm{e}^{-i \, \phi \, \mathrm{j}}
		\end{array} \right)
\;\;\; \mbox{and} \;\;\; 
M_{5} = \left( \begin{array}{ccc}
		A & C & C \, \mathrm{e}^{-i \, \phi \, \mathrm{k}}\\				
		B & D & E\\
		B \, \mathrm{e} ^{-i \, \phi \, \mathrm{j}}& 
		E \, \mathrm{e}^{-i \, \phi \, (\mathrm{j}- \mathrm{k})} &
		D \, \mathrm{e}^{-i \, \phi \, (\mathrm{j} + \mathrm{k})}
		\end{array} \right) \;\;
&
\end{eqnarray}
\noindent where $A,B,C,D,E$ are complex numbers which are products of Yukawa couplings
and VEVs, $\phi= \frac{2 \, \pi}{n} \, m$ ($n$: index of the dihedral group,
$m$: index of the breaking direction) 
and $\rm j,k$ are indices of representations. 
Regarding $M_{4}$ notice that we
presented in \cite{dntheorypaper} the transpose of this matrix. However,
a transposition in general only corresponds to the exchange of the transformation 
properties of the left-handed and left-handed conjugate fields under the 
flavor symmetry and therefore does not change the group theoretical part of
the discussion about the preserved subgroups. Furthermore, these matrices are determined up
to permutations of columns and rows which also only correspond to permutations
among the three generations of the fields. As in \cite{dntheorypaper} we work
in the SM and with the assumption that all Higgs fields $H$ in the model
are copies of the SM one. Therefore the displayed mass matrices are those
for down-type fermions, i.e. down quarks and charged leptons. The corresponding
ones for up-type fermions, i.e. up quarks and (Dirac) neutrinos, require some
changes due to the fact that only the conjugates of the Higgs fields, 
$\epsilon \, H^{\star}$, couple to up-type fermions and we use complex
matrices for the two-dimensional representations of \Groupname{D}{n}. According to the rules of
\cite{dntheorypaper} on how to deduce the up-type fermion mass matrices from the shown ones, 
$M_{4}$ and $M_{5}$ are of the form
\begin{equation}
\label{eq:M4andM5forups}
M_{4} = \left( \begin{array}{ccc}
		0 & A & B\\				
		C \, \mathrm{e}^{i \, \phi \, \mathrm{j}} & D \, \mathrm{e}^{i \, \phi \, \mathrm{j}} 
		& E \, \mathrm{e}^{i \, \phi \, \mathrm{j}}\\
		-C & D & E 
		\end{array} \right)
\;\;\; \mbox{and} \;\;\; 
M_{5} = \left( \begin{array}{ccc}
		A & C \, \mathrm{e}^{i \, \phi \, \mathrm{k}} & C\\				
		B \, \mathrm{e} ^{i \, \phi \, \mathrm{j}} 
		& D \, \mathrm{e}^{i \, \phi \, (\mathrm{j} + \mathrm{k})}
		& E \, \mathrm{e}^{i \, \phi \, (\mathrm{j}- \mathrm{k})}\\
		B & E & D
		\end{array} \right)
\end{equation}
\noindent An explicit example is given in 
\Secref{sec:quarksector}, where a model for quark masses and mixings is presented.\\
\noindent We concentrate on the last two forms, $M_{4}$ and $M_{5}$.
This we do for two reasons: first we want to accommodate all masses
and mixing parameters at tree level in the first part of the work, 
i.e. we do not want to rely on the
fact that one mixing angle is only generated by
higher order effects; second we would like to have the same mass matrix
structure for up quarks (Dirac neutrinos) and down quarks
(charged leptons).\\
\noindent Let us briefly mention the origin of the matrix structures 
$M_{4}$ and $M_{5}$. The flavor symmetry is a single-valued dihedral
group \Groupname{D}{n} with arbitrary index $n$. The preserved subgroup
is in both cases $Z_{2} = <\mathrm{B} \mathrm{A}^{m}>$ where $m=0,1,...,n-1$. 
This subgroup
allows non-vanishing VEVs for the following one-dimensional representations:
$\MoreRep{1}{1}$ (is always allowed to have a VEV), $\MoreRep{1}{3}$ for
$m$ even and $\MoreRep{1}{4}$ for $m$ odd. All two-dimensional representations
acquire a so-called structured VEV, i.e. for two fields $\psi_{1,2}$ 
transforming as an irreducible two-dimensional representation $\MoreRep{2}{p}$
their VEVs have to have the correlation: $<\psi_{1}> = <\psi_{2}> \, 
\mathrm{e}^{-\frac{2 \, \pi \, i \, \mathrm{p} \, m}{n}}$.
The notation of the representations used here is according to the one given in \cite{dntheorypaper}.
\\
\noindent In case of $M_{4}$ we take the left-handed fields $L$ to transform as $\MoreRep{1}{k} + \MoreRep{2}{j}$
under the dihedral group, and the left-handed conjugate fields $L^{c}$ 
transform as the three singlets $\MoreRep{1}{$\mathrm{i}_{1}$}+ \MoreRep{1}{$\mathrm{i}_{2}$} + \MoreRep{1}{$\mathrm{i}_{3}$}$. A complete study of 
all possible assignments shows that one of the entries in the first row
needs to be zero in order to prevent the determinant of the matrix from being
zero. 
The matrix structure $M_{5}$ arises, if both left-handed and left-handed
conjugate fermions transform as $\Rep{1}+\Rep{2}$, $L \sim (\MoreRep{1}{i},
\MoreRep{2}{j})$ and $L^{c} \sim (\MoreRep{1}{l},\MoreRep{2}{k})$. Here the
constraint $\mathrm{det} (M) \neq 0$ enforces the $(11)$ element of the
mass matrix to be non-zero, i.e. $\MoreRep{1}{i} \times \MoreRep{1}{l}$
has to have a non-vanishing VEV. Note that the indices of the representations
do not need to coincide, i.e. $\mathrm{i} \neq \mathrm{l}$ and 
$\mathrm{j} \neq \mathrm{k}$  is possible, although it might not be
favorable from the viewpoint of a (partially) unified model.\\
\noindent To study the mixing matrices arising from $M_{4}$ and $M_{5}$ 
for down-type as well as up-type fermions we observe 
that the products $M_{i} \, M_{i} ^{\dagger}$, $i=4,5$, can be written in
the general form
\[
\left( \begin{array}{ccc}
	a & b \, \mathrm{e} ^{i \, \beta} & b \, \mathrm{e} ^{i \, (\beta + \phi \, \mathrm{j})}\\
	b \, \mathrm{e} ^{- i \, \beta} & c & d \, \mathrm{e} ^{i \, \phi \, \mathrm{j}}\\
	b \, \mathrm{e} ^{-i \, (\beta + \phi \, \mathrm{j})} & d \, \mathrm{e}^{-i \, \phi \, \mathrm{j}} & c 
\end{array}
\right)
\]
where $a$, $b$, $c$, $d$ and $\beta$ are real functions of $A$, $B$, $C$, $D$ and $E$. The phase
$\beta$ lies in the interval $[ 0, 2 \, \pi )$. Since we work in the basis in which the left-handed fields are
on the left-hand side and the left-handed conjugate fields on the right-hand side, the unitary matrix
which diagonalizes $M_{i} \, M_{i}^{\dagger}$ acts on the left-handed fields and therefore determines the physical mixing matrices, i.e. the CKM
matrix and the MNS matrix.
The three eigenvalues are given as $(c-d)$, $\frac{1}{2} \, (a+c+d -
\sqrt{(a-c-d)^2 + 8 \, b^2})$ and  $\frac{1}{2} \, (a+c+d +\sqrt{(a-c-d)^2 + 8 \, b^2})$. Assuming this
ordering of the eigenvalues the mixing matrix $U$ which fulfills 
$U^{\dagger} \, M_{i} \, M_{i} ^{\dagger} \, U= \diag$ is of the form:
\[
U= \left( \begin{array}{ccc}
	0 & \cos (\theta) \, \mathrm{e} ^{i \, \beta} 
	& \sin (\theta) \, \mathrm{e} ^{i \, \beta}\\
	-\frac{1}{\sqrt{2}} \, \mathrm{e} ^{i \, \phi \, \mathrm{j}} & -\frac{\sin (\theta)}{\sqrt{2}} 
	& \frac{\cos (\theta)}{\sqrt{2}}\\
	\frac{1}{\sqrt{2}} & - \frac{\sin (\theta)}{\sqrt{2}} \, \mathrm{e} ^{- i \, \phi \, \mathrm{j}} 
	& \frac{\cos (\theta)}{\sqrt{2}} \, \mathrm{e} ^{-i \, \phi \, \mathrm{j}}
\end{array}
\right)
\]
\noindent The angle $\theta$ is determined to be $\tan (2 \, \theta)= \frac{2 \, \sqrt{2} \, b}{c+d-a}$. Therefore it lies in the interval 
$[0, \frac{\pi}{2})$. If the three
eigenvalues are not degenerate, the eigenvectors are determined by them up to phases 
\footnote{Since the eigenvectors should be normalized their length is fixed to one.}. Therefore
the variants of the mixing matrix $U$ are given by permutations of the columns. \noindent With this at hand we can look for possible interesting structures in the mixing matrix which is just the
product of two matrices of this form, as we assume that the up quark
(Dirac neutrino) and the down quark (charged lepton) mass matrix 
 is either of the form $M_{4}$
or $M_{5}$. The mixing matrix is then of the form
$V=W_{1} ^{T} \, W_{2} ^{\star}$ with $W_{i}$ being a variant of the matrix
$U$ above. For $V=V_{CKM}$ we have $W_{1} \equiv U_{u}$ which is the unitary
matrix diagonalizing the up quark mass matrix and $W_{2} \equiv U_{d}$ which
is the corresponding matrix for the down quarks. In case of 
$V=V_{MNS}$, $W_{1}$ is equivalent to $U_{l}$ and $W_{2}$ to $U_{\nu}$ 
\footnote{Throughout the paper we assume that the neutrinos are Dirac particles for
simplicity. Therefore $V_{MNS}$ has the same structure as $V_{CKM}$, i.e.
there are no (additional) Majorana phases present in the lepton sector.}.
The matrix $W_{i}$ contains the group theoretical phase $\phi_{i}$
according to the breaking direction $m_{i}$, the angle $\theta_{i}$ and the phase $\beta_{i}$.
For $W_{1} \equiv U_{u}$ we also use the notation $\phi_{u}$, $m_{u}$, $\theta_{u}$
and $\beta_{u}$. An analogous convention is used for $U_{d}$, $U_{l}$
and $U_{\nu}$. It turns out that one of the elements is
determined by the index $\rm j$ of the representation $\MoreRep{2}{j}$ under which two of
the left-handed fields transform and 
the difference of the group theoretical phases $\phi_{1}$ and 
$\phi_{2}$ only. These 
phases do not depend on the values of the parameters $A,B, ...$ (and therefore also not on $a,b,c,d$
and $\beta$), but only on the index $n$ of the group \Groupname{D}{n} and the indices
$m_{1}$ and $m_{2}$ being the parameters that determine the subgroup to which the Higgs fields 
break \Groupname{D}{n} down. 
Therefore this element is determined by fundamental values of the model only and not by
an arbitrarily tunable number. The actual form of (the absolute value of) the element is 
\begin{equation}
\frac{1}{2}\, \left| 1 + \mathrm{e} ^{i \, (\phi_{1}-\phi_{2}) \, \mathrm{j}}
\right| = \left| \cos ((\phi_{1}-\phi_{2})\, \frac{\mathrm{j}}{2}) \right|
 = \left| \cos (\frac{\pi}{n}\, (m_{1}-m_{2})\, \mathrm{j}) \right|
\end{equation}
Note that this value is only non-trivial, if $m_{1} \neq m_{2}$, i.e.
the (directions of the) subgroups which are preserved in the up quark
(Dirac neutrino) sector and the down quark (charged lepton) sector
are not the same, i.e. only their mismatch leads to non-trivial mixing.
This element can be traced back to the eigenvectors which correspond to the eigenvalue $c-d$ in the up quark (Dirac neutrino) and the down quark (charged lepton)
 sector, i.e. the product of these two eigenvectors gives rise to the element 
$\frac{1}{2} \left( 1 + \mathrm{e} ^{i \, (\phi_{1}-\phi_{2}) \, \mathrm{j}}
\right) $. Therefore
the ordering of the eigenvectors in the up quark (Dirac neutrino) and down quark 
(charged lepton) sector determines in which position of the mixing matrix 
the fixed element appears. 
As $V_{CKM}=U_{u} ^{T} \, U_{d} ^{\star}$, the fact whether
the eigenvalue $c-d$ is associated with up, charm or top quark mass determines the row in
which the element appears while the choice of the eigenvalue $c-d$ to be either $m_{d}$, $m_{s}$ or $m_{b}$
determines the column. Analogously, the choice whether $m_{e}$, $m_{\mu}$
or $m_{\tau}$ is given by $c-d$ determines the row, while the column
in which the element appears is given by the fact which of the (Dirac) neutrino
masses, $m_{1}$, $m_{2}$ or $m_{3}$, is equal to $c-d$.\\
\vspace{0.1in}\\
\noindent In \cite{dntheorypaper} we already mentioned that we can accommodate
the CKM matrix element $|V_{us}|$ by $\cos (\frac{3 \, \pi}{7}) \approx 0.2225$,
i.e. by taking $n=7$ and for example $\mathrm{j}=3$, $m_{u}=1$
and $m_{d}=0$. 
\noindent Here we show first which of the other 
elements of $V_{CKM}$ can also be accommodated well 
by the form 
$\left| \cos (\frac{\pi}{n}\, (m_{u}-m_{d})\, \mathrm{j}) \right|$.\\
\noindent The elements of $V_{CKM}$ are precisely measured \cite{pdg}
\[
|V_{CKM}|= \left( \begin{array}{ccc}
	0.97383 ^{+0.00024} _{-0.00023} & 0.2272 ^{+0.0010} _{-0.0010} & 
	(3.96 ^{+0.09} _{-0.09}) \, \times \, 10^{-3} \\
	0.2271 ^{+0.0010} _{-0.0010} & 0.97296 ^{+0.00024} _{-0.00024} & 
	(42.21 ^{+0.10} _{-0.80}) \, \times \, 10^{-3}\\
	(8.14 ^{+0.32} _{-0.64}) \, \times \, 10^{-3} & (41.61^{+0.12} _{-0.78}) \, \times \, 10^{-3} &
	0.999100 ^{+0.000034} _{-0.000004} 
\end{array}
\right)
\]
together with the Jarlskog invariant \cite{Jarlskog} $J_{CP}= (3.08 ^{+0.16} _{-0.18}) \, \times 10^{-5}$.
 We restrict ourselves to values of $n$ smaller than 30, since the index $n$ of the
dihedral group \Groupname{D}{n} is correlated to its order and larger values of $n$ correspond to larger
groups. Enforcing $n< 30$ leads to a group order smaller than 60 which seems to be reasonable.
Then we see that we can put the elements of the $1-2$ sub-block, i.e. $|V_{ud}|$, $|V_{us}|$, $|V_{cd}|$
and $|V_{cs}|$, into the form $\left| \cos (\frac{\pi}{n}\, (m_{u}-m_{d})\, \mathrm{j}) \right|$. 
As $|V_{cd}| \approx |V_{us}|$ holds to good accuracy,
also $|V_{cd}|$ can be described well by $\cos (\frac{3 \, \pi}{7})$.
 Furthermore $|V_{ud}| \approx
|V_{cs}|$ can be approximated well as $\cos (\frac{\pi}{14}) \approx 0.9749$ 
which points towards the flavor
group \Groupname{D}{14}. Note that the value of $|V_{ud}|$ as well as of $|V_{cs}|$ can be 
accommodated even a bit better with $\cos (\frac{2 \, \pi}{27}) \approx 0.9730$. However,
 we do not use this, as it needs the group \Groupname{D}{27} which is a group of order 54 and
therefore already quite large. Note that, even if $|V_{us}|$ is taken 
to be $\cos (\frac{3 \, \pi}{7})$, there is no unique solution which flavor symmetry
has to be used and to which subgroup it has to be broken, since for
example taking $\mathrm{j}=1$, $m_{u}=3$, $m_{d}=0$ and $n=7$ leads to 
$\left| \cos (\frac{\pi}{n}\, (m_{u}-m_{d})\, \mathrm{j}) \right|=|\cos (\frac{3 \, \pi}{7})|$ 
as well as $\mathrm{j}=3$, $m_{u}=1$, $m_{d}=0$ and $n=7$ and also
$\mathrm{j}=1$, $m_{u}=6$, $m_{d}=0$ and $n=14$. As $|\cos (\frac{4 \, \pi}{7})|$ equals $|\cos (\frac{3 \, \pi}{7})|$, this allows us to deduce further possible values for
$\mathrm{j}$, $m_{u}$, $m_{d}$ and $n$ like $\mathrm{j}=1$, $m_{u}=0$, $m_{d}=4$
and $n=7$.\\ 
\noindent In the next section we study the cases 
$|V_{us}|$ and $|V_{cd}|$ equal to $\cos (\frac{3 \, \pi}{7})$
and $|V_{ud}|$ and $|V_{cs}|$ equal to $\cos (\frac{\pi}{14})$
in greater detail
and thereby check whether we can always adjust the two other mixing angles $\theta_{13}^{q}$ and $\theta_{23}^{q}$
with the free angles $\theta_{u}$ and $\theta_{d}$ and also the Jarlskog invariant $\mathrm{J}_{CP}$
with the difference of the two phases $\beta_{u}$ and $\beta_{d}$.

%%%%%%%%%%%%%%%%%%%%%%%%%%%%%%%%%%%%%%%%%%
\mathversion{bold}
\section{Analysis of $V_{CKM}$ only}
\mathversion{normal}
%%%%%%%%%%%%%%%%%%%%%%%%%%%%%%%%%%%%%%%%%%
\label{sec:vckm}

%%%%%%%%%%%%%%%%%%%%%%%%%%%%%%%%%%%%%%%%%%
\subsection{Remarks}
%%%%%%%%%%%%%%%%%%%%%%%%%%%%%%%%%%%%%%%%%%
\label{subsec:analyticvckm}

\noindent There are six possible forms for $U$ which correspond to different identifications of
the eigenvalues. However, the fact that $m_{u} \ll m_{c} \ll m_{t}$ and $m_{d} \ll m_{s} \ll m_{b}$
allows only three of them, as the eigenvalue $\frac{1}{2} \, (a+c+d -
\sqrt{(a-c-d)^2 + 8 \, b^2})$ is smaller than $\frac{1}{2} \, (a+c+d +
\sqrt{(a-c-d)^2 + 8 \, b^2})$. Therefore, we will only vary the position of the eigenvector belonging 
to the eigenvalue $c-d$, while
 keeping the ordering of the two others fixed.
 The three different forms of the mixing matrix $U$ are then:
\begin{eqnarray}\nonumber
U &=& \left( \begin{array}{ccc}
	0 & \cos (\theta) \, \mathrm{e} ^{i \, \beta} 
	& \sin (\theta) \, \mathrm{e} ^{i \, \beta}\\
	-\frac{1}{\sqrt{2}} \, \mathrm{e} ^{i \, \phi \, \mathrm{j}} & -\frac{\sin (\theta)}{\sqrt{2}} 
	& \frac{\cos (\theta)}{\sqrt{2}}\\
	\frac{1}{\sqrt{2}} & - \frac{\sin (\theta)}{\sqrt{2}} \, \mathrm{e} ^{- i \, \phi \, \mathrm{j}} 
	& \frac{\cos (\theta)}{\sqrt{2}} \, \mathrm{e} ^{-i \, \phi \, \mathrm{j}}
\end{array}
\right)
\\ \nonumber
U^{\prime} &=& \left( \begin{array}{ccc}
	\cos (\theta) \, \mathrm{e} ^{i \, \beta} & 0 & \sin (\theta) \, \mathrm{e} ^{i \, \beta}\\
	-\frac{\sin (\theta)}{\sqrt{2}} & -\frac{1}{\sqrt{2}} \, \mathrm{e} ^{i \, \phi \, \mathrm{j}}
	& \frac{\cos (\theta)}{\sqrt{2}}\\
	-\frac{\sin (\theta)}{\sqrt{2}} \, \mathrm{e} ^{-i \, \phi \, \mathrm{j}} & \frac{1}{\sqrt{2}}
	& \frac{\cos (\theta)}{\sqrt{2}} \, \mathrm{e} ^{-i \, \phi \, \mathrm{j}}
\end{array}
\right)
\\ \nonumber
U^{\prime \, \prime} &=& \left( \begin{array}{ccc}
	\cos (\theta) \, \mathrm{e} ^{i \, \beta} & \sin (\theta) \, \mathrm{e} ^{i \, \beta} & 0\\
	-\frac{\sin (\theta)}{\sqrt{2}} & \frac{\cos (\theta)}{\sqrt{2}}
	& -\frac{1}{\sqrt{2}} \, \mathrm{e} ^{i \, \phi \, \mathrm{j}}\\
	-\frac{\sin (\theta)}{\sqrt{2}} \, \mathrm{e} ^{-i \, \phi \, \mathrm{j}} & 
	\frac{\cos (\theta)}{\sqrt{2}} \, \mathrm{e} ^{-i \, \phi \, \mathrm{j}} 
	& \frac{1}{\sqrt{2}}\\
\end{array}
\right)
\end{eqnarray}
\noindent Combining them leads to nine distinct possibilities for the CKM
matrix whose forms are displayed in \Appref{app:vmix}. 
Since we already mentioned that we want to concentrate on the $1-2$ sub-block
we only need to consider the four possible combinations which involve the matrices $U$
and $U^{\prime}$ as mixing matrices.\\

%%%%%%%%%%%%%%%%%%%%%%%%%%%%%%%%%%%%%%
\subsection{Numerical Study}
%%%%%%%%%%%%%%%%%%%%%%%%%%%%%%%%%%%%%%
\label{subsec:numericalvckm}

\noindent In this section we discuss the results of our fits to the CKM
matrix where we assume that one of the matrix elements in the $1-2$
sub-block is determined by group theory,  
as explained in the preceding section.
There are three free parameters in the mixing matrix: $\theta _{u,d}$ and $\alpha=\beta_{u}-\beta_{d}$. 
We use these to fit the other two mixing angles $\theta_{13} ^{q}$ and $\theta_{23} ^{q}$ as well as 
the CP violation $J_{CP}$.\\
\noindent The forms of $V_{mix}$ presented in \Appref{app:vmix} show that two of the 
elements $|V_{ub}|$,  $|V_{cb}|$, $|V_{td}|$ and $|V_{ts}|$ are determined
 by $\cos (\theta _{u,d})$ in each of the four different cases.
\noindent As these elements are small, the free angles $\theta_{u}$ and $\theta_{d}$ are restricted to
be $\theta _{d,u} \approx \frac{\pi}{2}$. Therefore 
 $\theta _{d,u}$ is expanded around $\frac{\pi}{2}$, $\theta _{d,u}= \frac{\pi}{2} 
- \epsilon _{d,u}$, $\epsilon_{d,u}>0$. The resulting four CKM matrices are (up to the first order
in $\epsilon _{u,d}$) 

\small
\begin{eqnarray}
\label{eq:vckm11}
|V_{CKM} ^{11}| &\approx& \left( \begin{array}{ccc}
	 \cos (\frac{\pi}{14}) & \cos (\frac{3 \, \pi}{7}) & \cos (\frac{3 \, \pi}{7}) \, \epsilon _{d}\\
	 \cos (\frac{3 \, \pi}{7}) &  \cos (\frac{\pi}{14}) & \frac{1}{2} \, 
|(1 + \mathrm{e} ^{\frac{\pi}{7} \, i}) \, \epsilon _{d} -2 \, \mathrm{e} ^ {i \, \alpha} \, \epsilon _{u}|\\
	\cos (\frac{3 \, \pi}{7})  \, \epsilon _{u} & \frac{1}{2} \, 
|(1 + \mathrm{e} ^{\frac{\pi}{7} \, i}) \, \epsilon _{u} -2 \, \mathrm{e} ^{i \, \alpha} \, \epsilon _{d}| & 1
\end{array}
\right)
\\ 
\label{eq:vckm12}
|V_{CKM} ^{12}| &\approx& \left( \begin{array}{ccc}
 	\cos (\frac{\pi}{14}) & \cos (\frac{3 \, \pi}{7}) & \cos (\frac{\pi}{14}) \, \epsilon _{d}\\
	 \cos (\frac{3 \, \pi}{7}) &  \cos (\frac{\pi}{14}) & \frac{1}{2} \, |(1 + \mathrm{e} ^ {\frac{6 \, \pi}{7} \, i}) \, \epsilon _{d} -2 \, \mathrm{e} ^ {i \, \alpha} \, \epsilon _{u}|\\
	 \frac{1}{2} \, |(1 + \mathrm{e} ^ {\frac{6 \, \pi}{7} \, i}) \, \epsilon _{u} -2 \, \mathrm{e} ^{i \, \alpha} \, \epsilon _{d}| & \cos (\frac{\pi}{14})  \, \epsilon _{u} & 1	
\end{array}
\right)
\\
\label{eq:vckm21}
|V_{CKM} ^{21}| &\approx& \left( \begin{array}{ccc}
 	\cos (\frac{\pi}{14}) & \cos (\frac{3 \, \pi}{7}) &  \frac{1}{2} \, |(1 + \mathrm{e} ^ {\frac{6 \, \pi}{7} \, i}) \, \epsilon _{d} -2 \, \mathrm{e} ^ {i \, \alpha} \, \epsilon _{u}|\\
	 \cos (\frac{3 \, \pi}{7}) &  \cos (\frac{\pi}{14}) &  \cos (\frac{\pi}{14}) \, \epsilon _{d}\\
  \cos (\frac{\pi}{14})  \, \epsilon _{u} & \frac{1}{2} \, |(1 + \mathrm{e} ^ {\frac{6 \, \pi}{7} \, i}) \, \epsilon _{u} -2 \, \mathrm{e} ^{i \, \alpha} \, \epsilon _{d}| &  1
\end{array}
\right)
\\
\label{eq:vckm22}
|V_{CKM} ^{22}| &\approx& \left( \begin{array}{ccc}
	\cos (\frac{\pi}{14}) & \cos (\frac{3 \, \pi}{7}) & \frac{1}{2} \, 
|(1+ \mathrm{e} ^{\frac{\pi}{7} \, i}) \, \epsilon _{d} - 2 \, \mathrm{e} ^{i \,\alpha} \, \epsilon _{u}| \\
	\cos (\frac{3 \, \pi}{7}) & \cos (\frac{\pi}{14}) & \cos (\frac{3 \, \pi}{7}) \, \epsilon _{d} \\
        \frac{1}{2} \, |(1+ \mathrm{e} ^{\frac{\pi}{7} \, i}) \, \epsilon _{u} - 2 \, 
\mathrm{e} ^{i \,\alpha} \, \epsilon _{d}|  & \cos (\frac{3 \, \pi}{7}) \, \epsilon _{u} & 1
\end{array}
\right)
\end{eqnarray}
\normalsize

\noindent Without loss of generality we have set the representation index $\rm j$ to 1,
the group theoretical phase $\phi_{u}$ to zero ($m_{u}=0$) and the phase $\phi_{d}$
to $\frac{2 \, \pi}{14}$ ($m_{d}=1$, $n=14$) for \Eqref{eq:vckm11} and 
\Eqref{eq:vckm22}, while we
take it to be $\frac{6 \, \pi}{7}$ ($m_{d}=3$, $n=7$) 
for \Eqref{eq:vckm12} and \Eqref{eq:vckm21}.\\
\noindent Comparing \Eqref{eq:vckm11} to the best fit values of $|V_{ub}|$
and $|V_{td}|$ given in \cite{pdg} leads to
$\epsilon _{u} \approx 0.0366$ and $\epsilon _{d} \approx 0.0178$.
\noindent The phase $\alpha$ is then mainly determined by the values of $|V_{cb}|$ and
$|V_{ts}|$. A numerical computation leads to a best fit for $\alpha \approx 4.810$
\footnote{We performed a $\chi^{2}$ fit of $J_{CP}$ and all elements of $|V_{CKM}|$ 
excluding the one which is fixed by group theory. Instead of taking the (very small)
experimental errors we simply assumed $10 \%$ errors for all quantities.}.
\noindent Furthermore one can calculate $J_{CP}$ in this case:
\begin{eqnarray}\nonumber
J_{CP} ^{11} &=& \frac{1}{8} \, \sin \left(\frac{\pi}{7} \right) \, \sin \left( \frac{\pi}{14} \right) \, \sin \left( 2 \, \theta _{d} \right) \,
\sin \left( 2 \, \theta _{u} \right) \, \sin \left( \frac{\pi}{14} - \alpha \right)\\ \nonumber
 &\approx& \frac{1}{2} \, \sin \left( \frac{\pi}{7} \right) \, \sin  
\left( \frac{\pi}{14} \right) \, \sin \left( \frac{\pi}{14} - \alpha \right) \, 
\epsilon _{u} \, \epsilon _{d}
\end{eqnarray}
\noindent A similar analysis can be carried out for the three other matrices $V_{CKM} ^{12}$,
$V_{CKM} ^{21}$ and $V_{CKM} ^{22}$ with similar results which we have 
collected in \Tabref{tab:vckmfits}. The value of $J_{CP}$ belonging to
$V_{CKM} ^{22}$, i.e. $J_{CP} ^{22}$, is of the same form as $J_{CP} ^{11}$.
For $V_{CKM} ^{12}$ and $V_{CKM} ^{21}$ one finds
\begin{eqnarray}\nonumber
J_{CP} ^{12} = J_{CP} ^{21} &=& -\frac{1}{8} \, \sin \left( \frac{6 \, \pi}{7} \right) \, \sin  
\left( \frac{3 \, \pi}{7} \right) \,
\sin \left( 2 \, \theta _{d} \right) \, \sin \left( 2 \, \theta _{u} \right) \, \sin \left( \frac{3 \, \pi}{7} - \alpha \right) 
\\ \nonumber
&\approx& -\frac{1}{2} \, \sin \left( \frac{6 \, \pi}{7} \right) \, \sin \left( \frac{3 \, \pi}{7} \right) \, \sin \left( \frac{3 \, \pi}{7} - \alpha \right) 
\, \epsilon _{u} \, \epsilon _{d}
\end{eqnarray}
\begin{table}[h!]
\begin{center}
\begin{tabular}{|c|c|c|c|c|}
\hline
Parameters & $V_{CKM} ^{11}$ & $V_{CKM} ^{12}$ & $V_{CKM} ^{21}$ & $V_{CKM} ^{22}$\\
\hline
$\epsilon _{u}$ & +0.0364 & +0.0427 & +0.00831 & +0.188 \\
$\epsilon _{d}$ & +0.0177 & +0.00405 & +0.0433 & +0.191 \\
$\alpha$ & 4.810 & 2.355 & 1.764 & 0.2056\\
\hline
\end{tabular}
\end{center}
\begin{center}
\begin{minipage}[t]{12cm}
\caption[]{Fit results for $\epsilon _{u,d}$ ($\theta _{u,d}$) and the phase $\alpha$
for $V_{CKM}$ with either $|V_{ud}|$, $|V_{us}|$, $|V_{cd}|$ or $|V_{cs}|$ being group theoretically
determined. \label{tab:vckmfits}}
\end{minipage}
\end{center}
\end{table}
\noindent As one can see in \Tabref{tab:vckmfits}, $\epsilon _{u,d}$ have to be larger in 
case of $V_{CKM} ^{22}$, since they are determined by $|V_{cb}|$ and $|V_{ts}|$. In this way the
expansion of $\theta _{u,d}$ around $\frac{\pi}{2}$ gets worse and 
the second order in $\epsilon _{u,d}$ becomes important. This can be seen
best in $|V_{us}| \approx 0.2225$ and $|V_{cd}| \approx 0.2225 $ which are lowered to $0.2186(5)$ 
such that the discrepancy between the experimentally measured value 
and the result of the fit gets larger. However,
 corrections from higher-dimensional operators and 
explicit breakings of the residual subgroups can lead to further contributions
allowing all data to be fitted successfully.

%%%%%%%%%%%%%%%%%%%%%%%%%%%%%%%%%%%%%%%%%%%%%%%%
\section{Analysis of the Quark Sector}
%%%%%%%%%%%%%%%%%%%%%%%%%%%%%%%%%%%%%%%%%%%%%%%%
\label{sec:quarksector}

\noindent After having shown that one element of $V_{CKM}$ can be explained
in terms of group theoretical indices only and studying this issue numerically
we want to go a step further and construct a viable model at least for the quark
sector which includes this issue. The model is viable, if
we find a numerical solution which accommodates not only the
mixing parameters contained in $V_{CKM}$, but also the quark masses. 
Due to the strong hierarchy among the quarks this is a non-trivial task,
although the number of parameters in the mass matrices $M_{u}$ and $M_{d}$ exceeds the
number of observables. Furthermore
we have to show that a Higgs potential exists allowing us to realize the
desired VEV structure. In the simplest case
we assume that all Higgs fields are $SU(2)_{L}$ doublets as the Higgs field
in the SM.

%%%%%%%%%%%%%%%%%%%%%%%%%%%%%%%%%%%%%%%%%%%%%%%%
\mathversion{bold}
\subsection{\Groupname{D}{7} Assignments for Quarks}
\mathversion{normal}
%%%%%%%%%%%%%%%%%%%%%%%%%%%%%%%%%%%%%%%%%%%%%%%%
\label{subsec:assignquarks}

\noindent Here we present ways to produce the two matrix structures 
$M_{4}$ and $M_{5}$ shown in \Eqref{eq:M4andM5fordowns} 
and \Eqref{eq:M4andM5forups}
with the help of the dihedral 
group \Groupname{D}{7}.
Choosing \Groupname{D}{7} as flavor symmetry leaves us the possibility
of either determining $|V_{us}|$ or $|V_{cd}|$ in terms of group
theoretical quantities as $\cos (\frac{3 \, \pi}{7})$.

%%%%%%%%%%%%%%%%%%%%%%%%%%%%%%%%%%%%%%%%%%%%%%%%
\mathversion{bold}
\subsubsection{Matrix Structure $M_{4}$}
\mathversion{normal}
%%%%%%%%%%%%%%%%%%%%%%%%%%%%%%%%%%%%%%%%%%%%%%%%
\label{subsubsec:assign_M4}

\noindent For $M_{4}$, we assign the quarks to
\begin{equation}
\label{eq:quarksM4}
Q_{1} \sim \MoreRep{1}{1} \; , \;\; \left( \begin{array}{c} Q_{2} \\ Q_{3} \end{array} \right)
\sim \MoreRep{2}{1} \; , \;\; u^{c}_{1}, d^{c}_{1} \sim \MoreRep{1}{2} \; , \;\;
u^{c}_{2,3}, d^{c}_{2,3} \sim \MoreRep{1}{1}
\end{equation}
under \Groupname{D}{7}. In this way we can generate the matrix structure 
found in \Eqref{eq:M4andM5fordowns} and \Eqref{eq:M4andM5forups} for the down
as well as the up quarks. Thereby we assume that the theory contains Higgs doublet fields
transforming as $\MoreRep{1}{1}$ and $\MoreRep{2}{1}$, which we call $H_{s}$ and $H_{1,2}$.
As the relation between the mixing parameters of $V_{CKM}$ and the group theoretical
indices only arises, if the flavor symmetry \Groupname{D}{7} is broken down to a subgroup $Z_{2}= <\mathrm{B} \, \mathrm{A} ^{m_{u}}>$ by
fields which couple to up quarks, while it is broken down to 
$Z_{2}= <\mathrm{B} \, \mathrm{A} ^{m_{d}}>$ with $m_{d} \neq m_{u}$ by fields
coupling to down quarks, 
we need an extra symmetry to perform this separation. In the SM this can be 
achieved by a $Z_{2} ^{(aux)}$ symmetry:
\begin{equation}
\label{eq:Z2auxtrafo}
d^{c} _{i} \;\; \rightarrow \;\; - d^{c}_{i} \;\;\; \mbox{and} \;\;\; H_{s}^{d} \;\; 
\rightarrow \;\; - H_{s} ^{d} \; , \;\; H_{i} ^{d} \;\; \rightarrow \;\; - H_{i}^{d}
\end{equation}
while all other fields $Q_{i}$, $u^{c}_{i}$, $H_{s}^{u}$ and $H_{1,2} ^{u}$ 
are invariant under  $Z_{2} ^{(aux)}$. 
Note that in principle also a Higgs field transforming as $\MoreRep{1}{2}$ under
\Groupname{D}{7} could couple directly to the quarks. However, if this field 
acquires a non-vanishing VEV, its VEV breaks the residual \Groupname{Z}{2}
generated by $<\mathrm{B} \, \mathrm{A}^{m}>$. Therefore Higgs fields $\sim \MoreRep{1}{2}$
either do not get a VEV or do not exist in the model at all. In both cases they are
not relevant in the discussion of the fermion mass matrices.
So, we deal with six
Higgs fields coupling to the fermions, $H^{u}_{s} \sim (\MoreRep{1}{1}, +1)$, $H_{1,2} ^{u} \sim (\MoreRep{2}{1},+1)$
and  $H^{d}_{s} \sim (\MoreRep{1}{1}, -1)$, $H_{1,2} ^{d} \sim (\MoreRep{2}{1},-1)$ under \Groupname{D}{7} $\times$ $Z_{2} ^{(aux)}$. The matrices are of the form:
\[
M_{u}= \left( \begin{array}{ccc}
	0 & y_{1} ^{u} \, \VEV{H_{s}^{u}}^{\star} & y_{2} ^{u} \, \VEV{H_{s}^{u}}^{\star}\\
	y_{3} ^{u} \, \VEV{H_{1} ^{u}}^{\star} & y_{4} ^{u} \, \VEV{H_{1} ^{u}}^{\star} & 
  	y_{5} ^{u} \, \VEV{H_{1}^{u}}^{\star}\\
	-y_{3} ^{u} \, \VEV{H_{2} ^{u}}^{\star} & y_{4} ^{u} \, \VEV{H_{2} ^{u}}^{\star} & 
  	y_{5} ^{u} \, \VEV{H_{2}^{u}}^{\star}\\		
\end{array}
\right) \;\;\; \mbox{and} \;\;\; 
M_{d}= \left( \begin{array}{ccc}
	0 & y_{1} ^{d} \, \VEV{H_{s}^{d}} & y_{2} ^{d} \, \VEV{H_{s}^{d}}\\
	y_{3} ^{d} \, \VEV{H_{2} ^{d}} & y_{4} ^{d} \, \VEV{H_{2} ^{d}} & 
  	y_{5} ^{d} \, \VEV{H_{2}^{d}}\\
	-y_{3} ^{d} \, \VEV{H_{1} ^{d}} & y_{4} ^{d} \, \VEV{H_{1} ^{d}} & 
  	y_{5} ^{d} \, \VEV{H_{1}^{d}}\\	
\end{array}
\right) 
\]
where $y^{u,d} _{i}$ denote Yukawa couplings. The VEV structure is taken to be:
\[
\VEV{H_{s}^{d,u}} > 0 \; , \;\; \VEV{H_{1} ^{d}}= \VEV{H_{2} ^{d}} = v_{d} \; , \;\;
\VEV{H_{1} ^{u}} = v_{u} \, \mathrm{e}^{-\frac{3 \, \pi \, i}{7}} \;\;\; \mbox{and} \;\;\; 
\VEV{H_{2} ^{u}} = v_{u} \, \mathrm{e}^{\frac{3 \, \pi \, i}{7}} 
\]
with $v_{d} > 0$ and $v_{u} > 0$. The VEVs are required to be real apart
from the phase $\pm \frac{3 \, \pi}{7}$ which is necessary for the correct 
breaking to the desired
subgroup of \Groupname{D}{7}.

\noindent The parameters $A, B, ...$ shown in \Eqref{eq:M4andM5fordowns} 
and \Eqref{eq:M4andM5forups} can be written in terms
of Yukawa couplings and VEVs:
\begin{eqnarray}\nonumber
&& A_{u}= y^{u}_{1} \, \VEV{H_{s} ^{u}} \; , \;\; B_{u}= y^{u}_{2} \, \VEV{H_{s}^{u}} \; , \;\;
C_{u}= y^{u}_{3} \, v_{u} \, \mathrm{e} ^{-\frac{3 \,\pi \, i}{7}} \; , \;\;
D_{u}= y^{u}_{4} \, v_{u} \, \mathrm{e} ^{-\frac{3 \,\pi \, i}{7}} \; , \;\;
E_{u}= y^{u}_{5} \, v_{u} \, \mathrm{e} ^{-\frac{3 \,\pi \, i}{7}} \; , \\ \nonumber
&& A_{d}= y^{d}_{1} \, \VEV{H_{s} ^{d}} \; , \;\; B_{d}= y^{d}_{2} \, \VEV{H_{s}^{d}} \; , \;\;
C_{d}= y^{d}_{3} \, v_{d} \; , \;\;
D_{d}= y^{d}_{4} \, v_{d} \; , \;\;
E_{d}= y^{d}_{5} \, v_{d}  
\end{eqnarray}
together with $\phi_{u}=\frac{6 \, \pi}{7}$ ($m_{u}=3$), $\phi_{d}=0$ ($m_{d}=0$)
and $\rm j=1$, as the left-handed quark doublets of the second and third generation
transform as $\MoreRep{2}{1}$. The preserved \Groupname{Z}{2} subgroups
are generated by $\mathrm{B} \, \mathrm{A}^{3}$ and $\mathrm{B}$
in the up and the down quark sector, respectively. As we have not fixed the ordering 
of the mass eigenvalues, the question which of the elements
of $V_{CKM}$ is determined by group theoretical quantities to be 
$\cos (\frac{3 \,\pi}{7})$ cannot be answered at this point.

%%%%%%%%%%%%%%%%%%%%%%%%%%%%%%%%%%%%%%%%%%%%%%%%
\mathversion{bold}
\subsubsection{Matrix Structure $M_{5}$}
\mathversion{normal}
%%%%%%%%%%%%%%%%%%%%%%%%%%%%%%%%%%%%%%%%%%%%%%%%
\label{subsubsec:assign_M5}

\noindent For the case of $M_{5}$, see \Eqref{eq:M4andM5fordowns}
and \Eqref{eq:M4andM5forups}, we can assign the quarks to:
\begin{equation}
\label{eq:quarksM5}
Q_{1}, u^{c}_{1}, d^{c}_{1} \sim \MoreRep{1}{1} \; , \;\; \left( 
\begin{array}{c}
Q_{2}\\
Q_{3}
\end{array}
\right), \left( 
\begin{array}{c}
u^{c}_{2}\\
u^{c}_{3}
\end{array}
\right), \left(
\begin{array}{c}
d^{c}_{2}\\
d^{c}_{3}
\end{array}
\right) \sim \MoreRep{2}{1}
\end{equation}
under \Groupname{D}{7}.
We then need five Higgs fields for each sector, i.e. for the up and the down quarks.
These transform as
\begin{eqnarray}\nonumber
&& H_{s} ^{u} \sim (\MoreRep{1}{1},+1) \; , \;\; \left( 
\begin{array}{c} 
H_{1} ^{u}\\
H_{2} ^{u}
\end{array}
\right) \sim (\MoreRep{2}{1},+1) \; , \;\; \left( 
\begin{array}{c} 
h_{1} ^{u}\\
h_{2} ^{u}
\end{array}
\right) \sim (\MoreRep{2}{2},+1) \\ \nonumber
&& H_{s} ^{d} \sim (\MoreRep{1}{1},-1) \; , \;\; \left( 
\begin{array}{c} 
H_{1} ^{d}\\
H_{2} ^{d}
\end{array}
\right) \sim (\MoreRep{2}{1},-1) \; , \;\; \left( 
\begin{array}{c} 
h_{1} ^{d}\\
h_{2} ^{d}
\end{array}
\right) \sim (\MoreRep{2}{2},-1)
\end{eqnarray}
where we again assumed the existence of an extra $Z_{2} ^{(aux)}$ symmetry. 
Under this $Z_{2} ^{(aux)}$ the quarks transform in the same way as in the
example above, i.e. only the down quarks $d^{c}_{i}$ acquire a sign.
The mass matrices 
are then in terms of Yukawa couplings and VEVs:
\[
M_{u}= \left( \begin{array}{ccc}
	y_{1} ^{u} \, \VEV{H_{s}^{u}}^{\star} & y_{2} ^{u} \, \VEV{H_{1} ^{u}} ^{\star} 
	& y_{2} ^{u} \, \VEV{H_{2} ^{u}} ^{\star} \\
	y_{3} ^{u} \, \VEV{H_{1} ^{u}}^{\star} & y_{5} ^{u} \, \VEV{h_{1} ^{u}}^{\star} & 
  	y_{4} ^{u} \,\VEV{H_{s} ^{u}} ^{\star}\\
	y_{3} ^{u} \, \VEV{H_{2} ^{u}}^{\star} & y_{4} ^{u} \, \VEV{H_{s} ^{u}}^{\star} & 
  	y_{5} ^{u} \, \VEV{h_{2} ^{u}}^{\star} \\		
\end{array}
\right) \;\;\; \mbox{and} \;\;\; 
M_{d}= \left( \begin{array}{ccc}
	y_{1} ^{d} \, \VEV{H_{s}^{d}} & y_{2}^{d} \, \VEV{H_{2} ^{d}} & y_{2} ^{d} \, \VEV{H_{1} ^{d}}\\
	y_{3} ^{d} \, \VEV{H_{2} ^{d}} & y_{5} ^{d} \,  \VEV{h_{2} ^{d}} & 
  	y_{4} ^{d} \, \VEV{H_{s} ^{d}} \\
	y_{3} ^{d} \, \VEV{H_{1} ^{d}} & y_{4} ^{d} \,  \VEV{H_{s}^{d}} & 
  	y_{5} ^{d} \, \VEV{h_{1} ^{d}} \\	
\end{array}
\right) 
\]
where $y^{u,d} _{i}$ denote Yukawa couplings. The VEV structure is assumed to be:
\begin{eqnarray}\nonumber
&& \VEV{H_{s}^{d,u}} > 0 \; , \;\; \VEV{H_{1} ^{d}}= \VEV{H_{2} ^{d}} = v_{d} \; , \;\; 
\VEV{h_{1} ^{d}}= \VEV{h_{2} ^{d}} = w_{d} \; , \;\; 
\\ \nonumber
&& \VEV{H_{1} ^{u}} = v_{u} \, \mathrm{e}^{-\frac{3 \, \pi \, i}{7}} \; , \;\;
\VEV{H_{2} ^{u}} = v_{u} \, \mathrm{e}^{\frac{3 \, \pi \, i}{7}}  \; , \;\;
 \VEV{h_{1} ^{u}} = w_{u} \, \mathrm{e}^{-\frac{6 \, \pi \, i}{7}} \;\;\; \mbox{and} \;\;\; 
\VEV{h_{2} ^{u}} = w_{u} \, \mathrm{e}^{\frac{6 \, \pi \, i}{7}} 
\end{eqnarray}
with $v_{d,u} > 0$ and $w_{d,u} > 0$. As above we only consider real values for the VEVs apart from
the phases which are required by the desire to break down to a certain subgroup of \Groupname{D}{7}. \\
\noindent Compared to the form of $M_{5}$ given in 
\Eqref{eq:M4andM5fordowns} and \Eqref{eq:M4andM5forups} we see that the parameters
$A,B, ...$ are given by:
\begin{eqnarray}\nonumber
&& A_{u}= y^{u}_{1} \, \VEV{H^{u}_{s}} \; , \;\; 
B_{u}= y^{u}_{3} \, v_{u} \, \mathrm{e} ^{-\frac{3 \, \pi \, i}{7}} \; , \;\;
C_{u}= y^{u}_{2} \, v_{u} \, \mathrm{e} ^{-\frac{3 \, \pi \, i}{7}} \; , \;\;
D_{u}= y^{u}_{5} \, w_{u} \, \mathrm{e} ^{-\frac{6 \, \pi \, i}{7}} \; , \;\;
E_{u}= y^{u}_{4} \, \VEV{H_{s}^{u}} \; , \\ \nonumber
&& A_{d}= y^{d}_{1} \, \VEV{H_{s}^{d}} \; , \;\;
B_{d}= y^{d}_{3} \, v_{d} \; , \;\;
C_{d}= y^{d}_{2} \, v_{d} \; , \;\;
D_{d}= y^{d}_{5} \, w_{d} \; , \;\;
E_{d}= y^{d}_{4} \, \VEV{H_{s} ^{d}}
\end{eqnarray}
together with $\phi_{u}=\frac{6 \, \pi}{7}$ ($m_{u}=3$), $\phi_{d}=0$ ($m_{d}=0$)
and $\rm j=k=1$ for up as well as down quarks, since all generations transform
as $\MoreRep{1}{1} + \MoreRep{2}{1}$ in this setup. Therefore the preserved
subgroups in the up and down quark sector are again 
$Z_{2}=<\mathrm{B} \, \mathrm{A}^{3}>$ and $Z_{2}=<\mathrm{B}>$.

\vspace{0.1in}

\noindent Note that the shown assignments are not unique, since 
it is also possible to use another two-dimensional representation
 instead of $\MoreRep{2}{1}$ for the fermions. Obviously,
then also the transformation properties of the Higgs fields have to be changed accordingly.\\
\noindent From the viewpoint of unification the second assignment in which the left-handed
as well as the left-handed conjugate fields transform as $\Rep{1} + \Rep{2}$ is more desirable.
However in this case we need at least five Higgs fields transforming as
$\MoreRep{1}{1}$, $\MoreRep{2}{i}$, $\MoreRep{2}{j}$ with $\rm i \neq j$ in order
to arrive at the matrix structure $M_{5}$. As we have to separate the up quark
from the down quark sector, i.e. have to have Higgs fields which either
couple to up quarks or down quarks, we need at least ten such fields. 
Since we want to show the minimal model, we constrain ourselves to the first case,
i.e. matrix structure $M_{4}$,
in the following numerical study and the study of the corresponding
Higgs potential and only give a numerical solution for the second matrix structure
$M_{5}$.

%%%%%%%%%%%%%%%%%%%%%%%%%%%%%%%%%%%%%%%%%%%%%%%%
\subsection{Numerical Analysis of Quark Masses and Mixing Angles}
%%%%%%%%%%%%%%%%%%%%%%%%%%%%%%%%%%%%%%%%%%%%%%%%
\label{subsec:quarkmassesandmixings}

%%%%%%%%%%%%%%%%%%%%%%%%%%%%%%%%%%%%%%%%%%%%%%%%
\mathversion{bold}
\subsubsection{Matrix Structure $M_{4}$}
\mathversion{normal}
%%%%%%%%%%%%%%%%%%%%%%%%%%%%%%%%%%%%%%%%%%%%%%%%
\label{subsubsec:quarkmassesandmixings_M4}

\noindent Coming to our numerical results we take all VEVs to have the same absolute value 61.5 $\GeV$ which equals the electroweak scale $174 \GeV$ divided by $\sqrt{8}$, because our
complete model includes eight Higgs fields  
\footnote{The additional two Higgs fields which do not couple to the fermions directly, are
necessary in order to break accidental symmetries present in the Higgs potential which we discuss in \Secref{sec:Higgs}. The equality of the VEVs is motivated by our numerical 
study of the Higgs
potential which clearly prefers solutions in which the VEVs are of the
same order, otherwise severe fine-tunings of the parameters in the potential
are necessary. However, this does not exclude in general the possibility
that for example $m_{b} \ll m_{t}$ could be explained by a hierarchy
among the VEVs of the Higgs fields coupling only to up quarks and those coupling
only to down ones.}.
 The Yukawa couplings 
are taken to be
\begin{eqnarray}\nonumber
&& y_{1}^{u}= 1.07967 \cdot \mathrm{e}^{i \, (-2.17704)} \; , \;\;
y_{2}^{u}= 2.55955 \cdot \mathrm{e}^{i \, (1.41609)} \; , \;\;
y_{3}^{u}= 1.9546 \cdot 10^{-5} \cdot \mathrm{e}^{i \, (2.43366)} \; , \\ \nonumber
&& y_{4}^{u}= 3.89557 \cdot 10^{-2}\cdot \mathrm{e}^{i \, (-2.28452)} \; , \;\;
y_{5}^{u}= 7.47229 \cdot 10^{-2} \cdot \mathrm{e}^{i \, (1.2469)}  \; , \;\; 
\\ \nonumber
&& y_{1}^{d}=  2.52251 \cdot 10^{-2} \cdot \mathrm{e}^{i \, (3.00267)} \; , \;\;
y_{2}^{d}=  3.92611 \cdot 10^{-2} \cdot \mathrm{e}^{i \, (-2.29202)} \; , \;\;
y_{3}^{d}=  6.20874 \cdot 10^{-4} \cdot \mathrm{e}^{i \, (-0.54014)} \; , \\ \nonumber
&& y_{4}^{d}=  8.95471 \cdot 10^{-5} \cdot \mathrm{e}^{i \, (-2.13972)} \; , \;\;
y_{5}^{d}=  1.04917 \cdot 10^{-4} \cdot \mathrm{e}^{i \, (-1.59912)}
\end{eqnarray}
The values of the quark masses are then
\begin{eqnarray}\nonumber
&m_{u}= 0.0017 \, \GeV \; , \;\; m_{c} = 0.62 \, \GeV \; , \;\; m_{t} = 171 \, \GeV \; ,&\\ \nonumber
&m_{d}= 0.003 \, \GeV \; , \;\; m_{s} = 0.054 \, \GeV \; , \;\; m_{b} = 2.87 \, \GeV &
\end{eqnarray}
which correspond to the values given at $M_{Z}$ \cite{quarkmasses}.
For $V_{CKM}$, we find:
\[
|V_{CKM}|= \left( \begin{array}{ccc}
	0.97492  & 0.2225 & 3.95 \, \times \, 10^{-3} \\
	0.2224 & 0.97404 & 42.23 \, \times \, 10^{-3}\\
	8.11 \, \times \, 10^{-3} & 41.64 \, \times \, 10^{-3} & 0.9991
\end{array}
\right)
\]
and $J_{CP}=3.09 \, \times \, 10^{-5}$. 
All these values are within a $10 \, \%$ error range \cite{pdg}.
Furthermore $|V_{us}|$ is given by
 $\cos (\frac{3 \, \pi}{7})= 0.2225$.
\noindent Due to the ordering of the eigenvalues the mass of the strange as well as the one of the up
quark is determined by $\sqrt{2} \, |C_{d}|$ and $\sqrt{2} \, |C_{u}|$, respectively. They therefore
correspond to the eigenvalue $(c-d)$ in the language of \Secref{sec:basics}.\\
The Yukawa couplings lie in the range $10^{-5} ... 1$ due to the
strong hierarchy of the quark masses. However this can be explained by 
the Froggatt-Nielsen mechanism \cite{FN}. 
The quarks transform in the following way:
\[
q_{FN} (Q_{1})= +1 \; , \;\; q_{FN} (Q_{2,3})= +2 \; , \;\; q_{FN} (d^{c}_{1,2,3}) = 0 \; \;\;
q_{FN} (u^{c}_{1})= +1 \; , \;\; q_{FN} (u^{c} _{2,3})= -1
\]
under the additional $U(1)_{FN}$ symmetry.
As usual we assume a gauge singlet $\vartheta$ with $q_{FN} (\vartheta) = -1$ 
which neither transforms under
\Groupname{D}{7} nor under $Z_{2} ^{(aux)}$ and which 
 acquires a VEV $\VEV{\vartheta}$ at a large energy scale
\footnote{Here we assume that the $U(1)_{FN}$ is
broken explicitly in other parts of the Lagrangian so that no Goldstone boson arises
when $\vartheta$ gets a VEV.}.
According to the choice of the $U(1)_{FN}$ charges
 only the Yukawa couplings $y^{u}_{1,2}$ exist at tree level while all other
couplings require some insertion of the $\vartheta$ field, i.e. become non-renormalizable
involving some power of $\frac{\VEV{\vartheta}}{\Lambda}$ with $\Lambda$ being the cutoff
scale of the theory. One can then re-write the given Yukawa couplings $y_{i} ^{u,d}$
in terms of new couplings $\tilde{y}_{i} ^{u,d}$ and 
$\epsilon \equiv \frac{\VEV{\vartheta}}{\Lambda} = 3 \cdot 10^{-2}$ as $y_{i} ^{u,d} = \tilde{y}_{i} ^{u,d} \, \epsilon ^{x}$ where $x$ is determined by the charges of the quark fields under the 
$U(1)_{FN}$. The values of $\tilde{y}_{i} ^{u,d}$ are then all of natural size:
\begin{eqnarray}\nonumber
&& \tilde{y}_{1}^{u}= 1.07967 \cdot \mathrm{e}^{i \, (-2.17704)} \; , \;\;
\tilde{y}_{2}^{u}= 2.55955 \cdot \mathrm{e}^{i \, (1.41609)} \; , \;\;
\tilde{y}_{3}^{u}= 0.723926 \cdot \mathrm{e}^{i \, (2.43366)} \; , \\ \nonumber
&& \tilde{y}_{4}^{u}= 1.29852 \cdot \mathrm{e}^{i \, (-2.28452)} \; , \;\;
\tilde{y}_{5}^{u}= 2.49076 \cdot \mathrm{e}^{i \, (1.2469)} \; , \;\;  
\\ \nonumber
&& \tilde{y}_{1}^{d}=   0.840837 \cdot \mathrm{e}^{i \, (3.00267)} \; , \;\;
\tilde{y}_{2}^{d}=   1.3087 \cdot \mathrm{e}^{i \, (-2.29202)} \; , \;\;
\tilde{y}_{3}^{d}=   0.68986 \cdot \mathrm{e}^{i \, (-0.54014)} \; , \\ \nonumber
&& \tilde{y}_{4}^{d}=   0.099497 \cdot\mathrm{e}^{i \, (-2.13972)} \; , \;\;
\tilde{y}_{5}^{d}=   0.116574 \cdot \mathrm{e}^{i \, (-1.59912)}
\end{eqnarray}

%%%%%%%%%%%%%%%%%%%%%%%%%%%%%%%%%%%%%%%%%%%%%%%%
\mathversion{bold}
\subsubsection{Matrix Structure $M_{5}$}
\mathversion{normal}
%%%%%%%%%%%%%%%%%%%%%%%%%%%%%%%%%%%%%%%%%%%%%%%%
\label{subsubsec:quarkmassesandmixings_M5}

\noindent For the second matrix structure $M_{5}$, we also performed
a numerical study with the mass matrix structure given above and found for 
example the following possible values for the parameters $A_{u,d}, B_{u,d}, ...$:
\begin{eqnarray}\nonumber
&& A_{u}= 40.40221 \cdot \mathrm{e} ^{i \, (0.185452)} \; , \;\; 
B_{u}= 0.238084 \cdot \mathrm{e} ^{i \, (-2.99845)} \; , \;\; 
C_{u}= 117.4875 \cdot \mathrm{e} ^{i \, (-0.234118)} \; , \\ \nonumber
&& D_{u}= 0.420584 \cdot \mathrm{e} ^{i \, (-3.13931)}\; , \;\; 
E_{u}= 0.984542 \cdot \mathrm{e} ^{i \, (-0.849532)} \; , \;\;\\ \nonumber
&& A_{d}= 2.233447 \cdot \mathrm{e} ^{i \, (-1.91017)} \; , \;\; 
B_{d}= 0.051223 \cdot \mathrm{e} ^{i \, (-3.05165)} \; , \;\; 
C_{d}= 1.271448 \cdot \mathrm{e} ^{i \, (-0.751605)} \; , \\ \nonumber
&& D_{d}= 0.058343 \cdot \mathrm{e} ^{i \, (-2.41411)} \; , \;\; 
E_{d}= 0.056221 \cdot \mathrm{e} ^{i \, (-2.37708)} \; .
\end{eqnarray}
\noindent All values are given in $\GeV$.
The phases $\phi_{u,d}$ can be chosen to be $\phi_{u}= \frac{6 \, \pi}{7}$ and $\phi_{d}=0$.\\
The quark masses are then:
\begin{eqnarray}\nonumber
&m_{u}= 0.0017 \, \GeV \; , \;\; m_{c} = 0.62 \, \GeV \; , \;\; m_{t} = 171 \, \GeV \; ,&\\ \nonumber
&m_{d}= 0.003 \, \GeV \; , \;\; m_{s} = 0.054 \, \GeV \; , \;\; m_{b} = 2.87 \, \GeV &
\end{eqnarray}
and the absolute values of $V_{CKM}$:
\[
|V_{CKM}|=\left( \begin{array}{ccc}
	0.97489 & 0.2226 & 3.95 \, \times \, 10^{-3}\\
	0.2225 & 0.97401 & 42.23 \, \times \, 10^{-3}\\
	8.11 \, \times \, 10^{-3} & 41.64 \, \times \, 10^{-3} & 0.9991
\end{array} \right)
\]
together with $J_{CP}= 3.09 \, \times \, 10^{-5}$. 
 These values match the experimental results quite well. Note here 
that this time not $|V_{us}|$, but now $|V_{cd}|$ is given in terms of the group theoretical
indices, i.e. $|V_{cd}|= \cos (\frac{3 \, \pi}{7})= 0.2225$. Since 
$|V_{us}|_{exp} \approx |V_{cd}|_{exp}$, $|V_{cd}|$ is now $2 \, \%$ below its experimental value. 
This is due to the fact that the eigenvalue $(c-d)$ introduced in \Secref{sec:basics}
is given by $m_{c}$ in the up quark and by $m_{d}$ in the down quark sector. These masses
can be expressed in a simple way in terms of the parameters $D_{u,d}$ and $E_{u,d}$,
namely $m_{c}= |D_{u}-E_{u} \, \mathrm{e} ^{-i \, \phi_{u} \, \mathrm{k}}|$
and $m_{d}= |D_{d} - E_{d} \, \mathrm{e} ^{i \, \phi_{d} \, \mathrm{k}}|$
with $\phi_{u}=\frac{6 \, \pi}{7}$, $\phi_{d}=0$ and $\mathrm k=1$
\footnote{$\rm k$ is here the same for the up and down quark sector due to the 
choice of the transformation properties of the left-handed conjugate fields $u^{c}_{2,3}$
and $d^{c}_{2,3}$. Note, however, that they could be in principle different.}
 as shown above.
Also here the hierarchy among the
parameters $A_{u,d}, B_{u,d}, ...$ which are products of Yukawa couplings and VEVs may
not be explained by a hierarchy among the VEVs. Assuming that all VEVs have the same
absolute value, i.e. $\frac{174}{\sqrt{10}} \GeV \approx 55 \GeV$ 
\footnote{Here we assume the existence of only the ten Higgs fields which couple to the
fermions in order to produce the mass matrix structure $M_{5}$, as it seems unlikely 
that there are accidental symmetries in the Higgs potential which would enforce the
existence of further Higgs fields.}, 
 an additional $U(1)_{FN}$ is responsible for the fermion mass hierarchy.
The quarks have the $U(1)_{FN}$ charges:
\begin{equation}\nonumber
q_{FN} (Q_{2,3})= + 2 \;\;\; \mbox{and}  \;\;\; q_{FN} (d^{c} _{1}) = +1
\end{equation}
and the other fields have zero charge.
The parameter $\epsilon = \frac{\VEV{\vartheta}}{\Lambda}$ ($\vartheta$: $U(1)_{FN}$ breaking field with $q_{FN} (\vartheta)=-1$,
$\Lambda$: cutoff scale) should be around $8 \cdot 10^{-2}$. 

%%%%%%%%%%%%%%%%%%%%%%%%%%%%%%%%%%%%%%%%%%%%%%
\section{Higgs Sector}
%%%%%%%%%%%%%%%%%%%%%%%%%%%%%%%%%%%%%%%%%%%%%%
\label{sec:Higgs}

\noindent In this section, the Higgs sector belonging to the 
 first numerical example given in \Secref{subsubsec:assign_M4} is discussed. 
As already mentioned above, we 
concentrate on a multi-Higgs doublet potential. We are aware of the fact 
that such multi-Higgs 
doublet models usually suffer from the problem that large FCNCs are induced by the 
additional Higgs fields. However, as a proof of principle that we can produce
our required VEV configuration the consideration of such
a multi-Higgs doublet model seems to be reasonable. 
The minimal number of fields needed in order to produce the fermion mass matrices is 
$2 \times 3$, i.e. the model includes three Higgs $SU(2)_{L}$ doublets, called
$H_{s}^{d}$ and $H_{1,2} ^{d}$, coupling
to the down quarks and three doublets, $H_{s}^{u}$ and $H_{1,2} ^{u}$, coupling to the up 
quarks.
This separation is necessary, since the key point of the study lies in the fact
 that a sizeable mixing angle, like the Cabibbo angle, can only
arise from preserved (non-trivial) subgroups, if we break to different (directions of) 
subgroups in the down and up quark sector.
The subgroups correspond to different VEV configurations 
 of the Higgs doublets $\left\{ H_{s}^{d}, H_{1,2} ^{d} \right\}$ and 
 $\left\{ H_{s}^{u}, H_{1,2} ^{u} \right\}$. 
An additional  $Z_{2} ^{(aux)}$ symmetry is introduced in order to perform the
separation.
According to \Eqref{eq:Z2auxtrafo} in \Secref{subsubsec:assign_M4} the Higgs
fields coupling to the down quarks acquire a sign under $Z_{2} ^{(aux)}$.

\noindent We first
construct the three Higgs doublet potential with Higgs fields $H_{s} \sim \MoreRep{1}{1}$
and $\left( \begin{array}{c} H_{1} \\ H_{2} \end{array} \right) \sim \MoreRep{2}{1}$.\\
\noindent The potential has the form: \footnote{Note that $\sigma_{2}$ is complex, but it can be made real by appropriate
redefinition of the field $H_{s}$, for example.}
\small
\begin{eqnarray}
\label{eq:potentialV3}
V_{3} (H_{s}, H_{i}) &=& -\mu_{s} ^{2} \, H_{s}
^{\dagger } \, H_{s} -\mu _{D} ^{2} \, \sum \limits _{i=1} ^{2}
H_{i} ^{\dagger}  \, H _{i} + \lambda _{s} \, \left( H_{s}
  ^{\dagger} \, H_{s} \right)^2 + \lambda _{1} \, \left( \sum \limits
  _{i=1} ^{2} H _{i} ^{\dagger} \, H_{i} \right) ^{2} \\
\nonumber
 &+&  \lambda _{2} \, \left( H _{1} ^{\dagger} \, H _{1}-H
   ^{\dagger} _{2} \, H _{2} \right) ^{2} + \lambda _{3} \, |
   H _{1} ^{\dagger} \, H _{2} |^{2} \\ \nonumber
 &+& \sigma _{1} \, \left( H_{s} ^{\dagger} \, H_{s} \right) \,
 \left(\sum \limits _{i=1} ^{2} H ^{\dagger} _{i} \, H _{i}
 \right) + \left\{ \sigma _{2} \, \left( H_{s} ^{\dagger} \, H
     _{1}\right) \, \left( H_{s} ^{\dagger} \, H _{2} \right) +
   \mathrm{h.c.} \right\} + \sigma _{3} \, \sum \limits _{i=1} ^{2} |H_{s} ^{\dagger} \, H _{i}|^{2}  
\end{eqnarray}
\normalsize
\noindent As already shown in \cite{d5paper} and also mentioned in \cite{dnpotentialpaper},
this potential has an additional $U(1)$ symmetry, i.e. there exists a 
further $U(1)$ symmetry in the potential apart from the $U(1)_{Y}$
symmetry. This further symmetry is necessarily broken by our desired
VEV structure such that a massless Goldstone boson
appears in the Higgs spectrum which is not eaten by a gauge boson. This
problem cannot be solved by taking into account the whole potential
for all six Higgs fields, since even if the terms, coupling the fields
$H^{u}_{s}$, $H^{u}_{1,2}$ and $H^{d}_{s}$, $H_{1,2} ^{d}$ together, are included, 
we find an accidental $U(1)$ symmetry in the potential.
Therefore we have to enlarge the Higgs sector by further
Higgs fields in order to create new \Groupname{D}{7} invariant couplings
which break this accidental symmetry explicitly. We find that this can be
done in the simplest way by adding two Higgs fields transforming as $\MoreRep{2}{2}$
under \Groupname{D}{7}. Due to their transformation properties they do not 
directly couple to the fermions (see \Secref{subsubsec:assign_M4}). 
We decided to add two such fields to 
the three Higgs fields which couple to the down quarks. Therefore
the model contains eight Higgs doublet fields in total: three of them couple to up
and three of them to down quarks, while the other two ones are needed for
a viable Higgs sector:
\begin{eqnarray}
&& H^{u}_{s} \sim (\MoreRep{1}{1}, +1) \; , \;\; 
\left( \begin{array}{c}
H^{u}_{1} \\ H^{u}_{2}
\end{array}
\right) \sim (\MoreRep{2}{1},+1) \; , \;\; \\ \nonumber
&& H^{d}_{s} \sim (\MoreRep{1}{1}, -1) \; , \;\; 
\left( \begin{array}{c}
H^{d}_{1} \\ H^{d}_{2}
\end{array}
\right) \sim (\MoreRep{2}{1},-1) \;\;\; \mbox{and} \;\;\;
\left( \begin{array}{c}
\chi^{d}_{1} \\ \chi^{d}_{2}
\end{array}
\right) \sim (\MoreRep{2}{2},-1) \; .
\end{eqnarray}
\noindent under \Groupname{D}{7} $\times$ $Z_{2} ^{(aux)}$. The complete potential
consists of three parts:
\begin{equation}
V = V_{u} + V_{d} + V_{mixed}
\end{equation}
where $V_{u}$ denotes the part of the potential which only contains
Higgs fields coupling to the up quarks, $V_{d}$ contains the
five Higgs fields which have a non-vanishing $Z_{2} ^{(aux)}$ charge
(three of them give masses to the down quarks),
while $V_{mixed}$ consists of all other terms. The explicit form
of the potential is given in \Appref{app:Higgs}.\\
\noindent The VEV structure
of the fields $H_{s}^{d,u}$ and $H_{1,2} ^{d,u}$ is determined by
our desire to break down to two distinct \Groupname{Z}{2} subgroups
in the up and the down quark sector (see \Secref{subsubsec:assign_M4}):
\[
\VEV{H_{s}^{d,u}} > 0 \; , \;\; \VEV{H_{1} ^{d}}= \VEV{H_{2} ^{d}} = v_{d} \; , \;\;
\VEV{H_{1} ^{u}} = v_{u} \, \mathrm{e}^{-\frac{3 \, \pi \, i}{7}} \;\;\; \mbox{and} \;\;\; 
\VEV{H_{2} ^{u}} = v_{u} \, \mathrm{e}^{\frac{3 \, \pi \, i}{7}} 
\]
with $v_{d} > 0$ and $v_{u} > 0$. In contrast to this,
the VEV structure of the fields $\chi_{1,2} ^{d}$ is not fixed in this
way. However, in order to preserve the \Groupname{Z}{2} subgroup generated
by $\rm B$ not only through the VEVs of the fields $H_{s}^{d}$
and $H_{1,2} ^{d}$, but also by the VEVs of the fields $\chi_{1,2} ^{d}$,
$\langle \chi_{1} ^{d} \rangle = \langle \chi_{2} ^{d} \rangle > 0$ will be
assumed in the following (see \Secref{sec:basics}).\\
\noindent We proceed in the following way in order to find a minimum of this 
potential which allows for our choice of VEVs:
first we treat $V_{u}$ and $V_{d}$ separately to find a viable
solution for these two parts of the potential. Note that we can allow all parameters 
in the potential
$V_{d}$ to be real, as the VEVs of the corresponding Higgs fields
are also real.
Since $V_{u}$ suffers from the above mentioned
accidental $U(1)$ symmetry, we find a fourth massless particle
in the Higgs mass spectrum. In a second step we add as many terms
as necessary from $V_{mixed}$ to get a minimum of the whole
potential $V$ which does not have more than the usual three Goldstone
bosons. It turns out that it is sufficient to take into account three
terms in addition to $V_{u}$ and $V_{d}$ to get a viable solution. The
terms are of the form:

\small
\[
\kappa_{2} \, \left( {H_{s}^{u}}^{\dagger} \, H_{s}^{d}  \right)^{2}  + 
\kappa_{5} \,\left( \sum \limits _{i=1} ^{2} {H_{i} ^{u}}^{\dagger} \, H_{i} ^{d} \right)^{2}
+ \kappa_{19} \, \left( {H_{s}^{u}}^{\dagger} \, H_{s}^{d}  \right) \, 
\left( \sum \limits _{i=1} ^{2} {H_{i} ^{u}}^{\dagger} \, H_{i} ^{d} \right) 
+ \mathrm{h.c.}
\subset V_{mixed}
\]
\normalsize

\noindent Note that we take all VEVs to have the same absolute value, since this considerably 
simplifies the search for a numerical solution, as a fine-tuning of the
parameters in the Higgs potential is avoided. However, in principle other solutions
should also be possible, e.g. the fact that the up quarks are much heavier than the
down ones could be explained by assuming that the VEVs of the fields $H^{u}_{s}$,
$H^{u}_{1,2}$ are (much) larger than the ones of the fields $H^{d}_{s}$, $H^{d}_{1,2}$.\\
\noindent Finally, let us mention that the resulting Higgs masses are usually
in between 50 and 500 $\GeV$. These values are either not favored by
the constraints coming from FCNCs or already excluded by direct searches. 
There are two reasons for the too low Higgs masses: on the one hand
 $V_{u}$ contains an accidental symmetry and on the other hand
all mass parameters of the potential are chosen to be of natural order, i.e. to be around 
the electroweak scale. Additionally, all quartic couplings of the potential 
must be perturbative. However, as already mentioned above, this model
is not intended to be fully realistic. Adding \Groupname{D}{7}
breaking soft masses to the potential might allow to push the masses
of the additional Higgs particles above $10 \, \TeV$.\\
\noindent The rest of the discussion of the potential is delegated to \Appref{app:Higgs} 
where we present a numerical solution for the parameters of the 
Higgs potential and the resulting Higgs masses.

%%%%%%%%%%%%%%%%%%%%%%%%%%%%%%%%%%%%%%%%%%%%%%%
\mathversion{bold}
\section{Ways to generate $\theta_{C}$ only}
\mathversion{normal}
%%%%%%%%%%%%%%%%%%%%%%%%%%%%%%%%%%%%%%%%%%%%%%%
\label{sec:alternatives}

\noindent In the preceding sections we confined ourselves to cases
in which all mixing angles can be reproduced at tree level. Therefore
we only discussed the matrix structures $M_{4}$ and $M_{5}$
of \Eqref{eq:M4andM5fordowns} and \Eqref{eq:M4andM5forups}.
 However, $\theta_{13}^{q}$ and $\theta_{23} ^{q}$ are roughly
an order of magnitude smaller than the Cabibbo angle $\theta_{C} \equiv \theta_{12} ^{q}$
which gives reason for also considering matrix structures which 
lead to only $\theta_{C} \neq 0$ at LO.
For this a block matrix structure (with correlated elements), which
we introduced in \Eqref{eq:blockmatrix}, is suitable. Such a
structure can be achieved in at least two different ways:
$a.)$ we can simply omit some of the Higgs fields which are in 
principle allowed a VEV in order to arrive at the zero elements
of the mass matrix; $b.)$ we can demand that the preserved subgroup is not
just a \Groupname{Z}{2} symmetry, but a dihedral group \Groupname{D}{q}
with $q>1$ \footnote{The general results of the mass matrices from the various preserved
subgroups can be found in \cite{dntheorypaper}.}. Note that due to 
the choice of the scalar fields in case $a.)$ the results of such
a model are a bit arbitrary, since the structure
of the mass matrices and therefore the mixing pattern is not fully
determined by the fermions and the flavor symmetry alone.
In the following, we show examples for the two cases.
For case $a.)$ the simplest example is probably the one
in which the quarks transform as
\[
\left( 
\begin{array}{c}
Q_{1}\\
Q_{2}
\end{array}
\right), \left( 
\begin{array}{c}
u^{c}_{1}\\
u^{c}_{2}
\end{array}
\right), \left(
\begin{array}{c}
d^{c}_{1}\\
d^{c}_{2}
\end{array}
\right) \sim \MoreRep{2}{1} \;\; , \;
Q_{3}, u^{c}_{3}, d^{c}_{3} \sim \MoreRep{1}{1} 
\]
under \Groupname{D}{7} and we assume that there exist two sets of three Higgs 
fields transforming as
\[
H_{s} ^{u} \sim (\MoreRep{1}{1},+1) \; , \;\; \left( 
\begin{array}{c} 
h_{1} ^{u}\\
h_{2} ^{u}
\end{array}
\right) \sim (\MoreRep{2}{2},+1) \; , \;\;
H_{s} ^{d} \sim (\MoreRep{1}{1},-1) \; , \;\; \left( 
\begin{array}{c} 
h_{1} ^{d}\\
h_{2} ^{d}
\end{array}
\right) \sim (\MoreRep{2}{2},-1)
\]
under \Groupname{D}{7} $\times$ $Z_{2} ^{(aux)}$, one of them coupling 
to up and the other one coupling to down quarks. 
The additional $Z_{2} ^{(aux)}$ symmetry is the same as used above 
(see \Secref{subsubsec:assign_M4}).
Then the mass matrices are of the form:
\[
M_{u}= \left( \begin{array}{ccc}
	y_{3} ^{u} \, \VEV{h_{1} ^{u}}^{\star} & y_{2} ^{u} \,\VEV{H_{s} ^{u}} ^{\star} & 0 \\	
	y_{2} ^{u} \, \VEV{H_{s} ^{u}}^{\star} & y_{3} ^{u} \,\VEV{h_{2} ^{u}}^{\star} & 0 \\
	0 & 0 & y_{1} ^{u} \, \VEV{H_{s}^{u}}^{\star}		
\end{array}
\right) \;\;\; \mbox{and} \;\;\; 
M_{d}= \left( \begin{array}{ccc}
	y_{3} ^{d} \,  \VEV{h_{2} ^{d}} & y_{2} ^{d} \, \VEV{H_{s} ^{d}} & 0 \\
	y_{2} ^{d} \,  \VEV{H_{s} ^{d}} & y_{3} ^{d} \, \VEV{h_{1} ^{d}} & 0 \\
	0 & 0 & y_{1} ^{d} \, \VEV{H_{s}^{d}}	
\end{array}
\right) 
\]
Assuming the VEV structure:
\[
 \VEV{H_{s}^{d,u}} > 0 \; , \;\; \VEV{h_{1} ^{d}}= \VEV{h_{2} ^{d}} = w_{d} \; , \;\; 
 \VEV{h_{1} ^{u}} = w_{u} \, \mathrm{e}^{-\frac{6 \, \pi \, i}{7}} \;\;\; \mbox{and} \;\;\; 
 \VEV{h_{2} ^{u}} = w_{u} \, \mathrm{e}^{\frac{6 \, \pi \, i}{7}} 
\]
with $w_{d,u} > 0$ we arrive at 
\[
M_{u}= \left( \begin{array}{ccc}
	y_{3} ^{u} \, w_{u} \, \mathrm{e}^{\frac{6 \, \pi \, i}{7}} & 
	y_{2} ^{u} \,\VEV{H_{s} ^{u}} & 0 \\	
	y_{2} ^{u} \, \VEV{H_{s} ^{u}} 
	& y_{3} ^{u} \, w_{u} \, \mathrm{e}^{\frac{-6 \, \pi \, i}{7}} & 0 \\
	0 & 0 & y_{1} ^{u} \, \VEV{H_{s}^{u}}		
\end{array}
\right) \;\;\; \mbox{and} \;\;\; 
M_{d}= \left( \begin{array}{ccc}
	y_{3} ^{d} \,  w _{d} & y_{2} ^{d} \, \VEV{H_{s} ^{d}} & 0 \\
	y_{2} ^{d} \,  \VEV{H_{s} ^{d}} & y_{3} ^{d} \, w _{d} & 0 \\
	0 & 0 & y_{1} ^{d} \, \VEV{H_{s}^{d}}	
\end{array}
\right) 
\]
and
\begin{equation}
\label{eq:VCKMthetaConly}
|V_{CKM}|= \left( \begin{array}{ccc}
	|\cos (\frac{\pi}{14})| & |\cos (\frac{3 \,\pi}{7})| & 0 \\
	|\cos (\frac{3 \,\pi}{7})| & |\cos (\frac{\pi}{14})| & 0 \\
	0 & 0 & 1
\end{array}
\right) \approx  
\left( \begin{array}{ccc}
	0.97493 & 0.2225 & 0 \\
	0.2225 & 0.97493 & 0 \\
	0 & 0 & 1
\end{array}
\right)
\end{equation}
\noindent The preserved subgroup is a \Groupname{Z}{2} in each sector which is
generated by $\mathrm{B} \, \mathrm{A}^{3}$ and $\rm B$ in the up 
quark and the down quark sector, respectively.
The masses of the quarks are $
(m_{u}^{2}, m_{c}^{2}, m_{t}^{2})=
(|y_{2}^{u} \, \langle H_{s}^{u} \rangle + y_{3} ^{u} \, w_{u}|^{2},
|y_{2}^{u} \, \langle H_{s}^{u} \rangle - y_{3} ^{u} \, w_{u}|^{2}, 
|y_{1} ^{u} \, \langle H_{s} ^{u} \rangle|^{2})
$
and 
$(m_{d}^{2}, m_{s}^{2}, m_{b}^{2})=
(|y_{2}^{d} \, \langle H_{s}^{d} \rangle - y_{3}^{d} \, w_{d}|^{2},
|y_{2}^{d} \, \langle H_{s} ^{d} \rangle + y_{3} ^{d} \, w_{d}|^{2},$
$|y_{1}^{d} \, \langle H_{s} ^{d} \rangle|^{2})
$, i.e. the mass of the third generation
is solely determined by $y_{1} ^{u,d} \VEV{H_{s} ^{u,d}}$. Note that if the
 VEVs of $H_{s} ^{u,d}$ are taken to be large in order to explain the large mass of the 
third generation, the Yukawa couplings $y_{2} ^{u,d}$ have to be suppressed. This might
be viewed as fine-tuning. A possible solution is the assumption of an additional
$U(1)_{FN}$ as already used above or to consider the case $b.)$ instead. The possibility to 
choose the two-dimensional representations for the left-handed and left-handed conjugate
fields to be distinct from each other, such that the trivial representation $\MoreRep{1}{1}$
cannot be coupled to the first and second generation, does not exist, since in this
case also two of the four zeros disappear. The reason can be found by looking at the 
Kronecker products shown in \Appref{app:grouptheory}. 
Actually, this setup is very similar to the one shown in \Secref{subsubsec:assign_M5}.
The main difference is the fact that now there are no Higgs fields
transforming as $\MoreRep{2}{1}$ under \Groupname{D}{7}. The other difference is
that the first and second generation of the left-handed and left-handed
conjugate fields are unified into the doublet under the flavor group instead of the 
second and third one as done above. However, this only leads to a change in 
the appearance of the
mass matrix itself, but does not have any phenomenological consequences, since
this permutation of fields is cancelled in the mixing matrix. The existence
of six instead of ten Higgs fields coupling to the fermions may be advantageous
with regard to the problem of FCNCs mediated by these fields. The corresponding
Higgs potential ought to be of the same form as the one discussed in \Secref{sec:Higgs}.\\
\noindent The second case $b.)$ cannot be maintained with the flavor group
\Groupname{D}{7} which we used throughout this work, since  
it only contains \Groupname{Z}{q} groups
as subgroups, but no dihedral ones \Groupname{D}{q}, $q>1$. 
Therefore we have to consider the
group \Groupname{D}{14} instead. In the study of the $V_{CKM}$ elements
in \Secref{sec:basics} and \Secref{sec:vckm} \Groupname{D}{14} turned out to 
be the smallest group which is appropriate to 
describe the elements $|V_{ud}|$ and $|V_{cs}|$ in terms of group theoretical
indices. As argued in \Secref{sec:basics} and \Secref{sec:vckm} it can also be used in 
order to reproduce the \Groupname{D}{7} results, i.e. either $|V_{us}|=|\cos (\frac{3 \,\pi}{7})|$
or $|V_{cd}|=|\cos (\frac{3 \, \pi}{7})|$. Here we just show a possible example
in which \Groupname{D}{14} is broken to its subgroup \Groupname{D}{2} 
$=<\mathrm{A}^{7}, \mathrm{B} \, \mathrm{A}^{m}>$ ($m=0,1,...,6$) in order 
to reproduce a matrix of block structure. We assign the quarks to
\[
\left( 
\begin{array}{c}
Q_{1}\\
Q_{2}
\end{array}
\right), \left( 
\begin{array}{c}
u^{c}_{1}\\
u^{c}_{2}
\end{array}
\right), \left(
\begin{array}{c}
d^{c}_{1}\\
d^{c}_{2}
\end{array}
\right) \sim \MoreRep{2}{1} \;\; , \;
Q_{3}, u^{c}_{3}, d^{c}_{3} \sim \MoreRep{1}{1} 
\]
\noindent under \Groupname{D}{14}. According to the Kronecker products 
\[
\MoreRep{1}{1} \times \MoreRep{2}{1} = \MoreRep{2}{1} \;\;\; \mbox{and} \;\;\;
\MoreRep{2}{1} \times \MoreRep{2}{1} = \MoreRep{1}{1} + \MoreRep{1}{2} + \MoreRep{2}{2}
\]
the Higgs fields which can in principle couple to form \Groupname{D}{14}-invariants
have to transform as $\MoreRep{1}{1}$, $\MoreRep{1}{2}$, $\MoreRep{2}{1}$ and 
$\MoreRep{2}{2}$. However, $\MoreRep{1}{2}$ is not allowed a VEV and
 the representation index $\rm j$
of $\MoreRep{2}{j}$ has to be even \footnote{In general this index must be 
divisible by the group index of the dihedral subgroup which should be 
preserved.}. Therefore we
take
\[
H_{s} ^{u} \sim \MoreRep{1}{1}  \; , \;\; \left(
\begin{array}{c}
H_{1} ^{u}\\
H_{2} ^{u}
\end{array}
\right) \sim \MoreRep{2}{2} \; , \;\; 
H_{s} ^{d} \sim \MoreRep{1}{1}  \;\;\; \mbox{and} \;\;\; \left(
\begin{array}{c}
H_{1} ^{d}\\
H_{2} ^{d}
\end{array}
\right) \sim \MoreRep{2}{2}
\]
(with implicit $Z_{2} ^{(aux)}$ assignment as above)
and arrive at the matrix forms which are exactly the same as given above for case $a.)$
\footnote{The Clebsch Gordan coefficients
necessary for the calculation of the mass matrices in \Groupname{D}{14} can be found in a general
form in \cite{dntheorypaper}. However, in this special case they
 coincide with those given for the
group \Groupname{D}{7}.}, if we assume the VEVs to be
\[
\langle H_{s} ^{u,d} \rangle > 0 \; , \;\; \langle H_{1} ^{u} \rangle 
= v_{u} \, \mathrm{e} ^{-\frac{6 \, \pi \, i}{7}} 
\; , \;\; \langle H_{2} ^{u} \rangle = v_{u} \, \mathrm{e} ^{\frac{6 \, \pi \, i}{7}} \; , \;\;
\langle H_{1} ^{d} \rangle = \langle H_{2} ^{d} \rangle = v_{d} 
\]
\noindent The subgroups which are preserved by the VEVs in
the up and down quark sector are then of the form \Groupname{D}{2} $=<\mathrm{A}^{7}, 
\mathrm{B} \, \mathrm{A} ^{m}>$ with $m_u=6$ for the up quarks and $m_d=0$ for down quarks.
Also here the Higgs fields $H_{s} ^{d,u}$ couple to all three generations. In order to avoid this
one can assign the quarks to different \Groupname{D}{14} representations, e.g.
\[
\left( 
\begin{array}{c}
Q_{1}\\
Q_{2}
\end{array}
\right)  
 \sim \MoreRep{2}{1} \; , \;\;
\left( 
\begin{array}{c}
u^{c}_{1}\\
u^{c}_{2}
\end{array}
\right), \left(
\begin{array}{c}
d^{c}_{1}\\
d^{c}_{2}
\end{array}
\right) \sim \MoreRep{2}{3} \;\; , \;
Q_{3}, u^{c}_{3}, d^{c}_{3} \sim \MoreRep{1}{1} 
\]
Since $\MoreRep{2}{1} \times \MoreRep{2}{3}$ decomposes into $\MoreRep{2}{2}$
and $\MoreRep{2}{4}$ in \Groupname{D}{14}, the $1-2$ sub-block of the mass matrices
is produced by the VEVs of Higgs fields belonging to \Groupname{D}{14} doublets
instead of the singlet $H_{s}^{u,d}$. As the indices of the representations
$\MoreRep{2}{2}$ and $\MoreRep{2}{4}$ are even, they are allowed a VEV by the
requirement to preserve a \Groupname{D}{2} subgroup of \Groupname{D}{14}. We need five
Higgs fields transforming as $\MoreRep{1}{1} + \MoreRep{2}{2} + \MoreRep{2}{4}$
for the down as well as the up quarks. The general form of the mass matrices reads 
\[
M_{u}= \left( \begin{array}{ccc}
	y_{3} ^{u} \, \VEV{h_{1} ^{u}}^{\star} & y_{2} ^{u} \,\VEV{H_{2} ^{u}} ^{\star} & 0 \\	
	y_{2} ^{u} \, \VEV{H_{1} ^{u}}^{\star} & y_{3} ^{u} \,\VEV{h_{2} ^{u}}^{\star} & 0 \\
	0 & 0 & y_{1} ^{u} \, \VEV{H_{s}^{u}}^{\star}		
\end{array}
\right) \;\;\; \mbox{and} \;\;\; 
M_{d}= \left( \begin{array}{ccc}
	y_{3} ^{d} \,  \VEV{h_{2} ^{d}} & y_{2} ^{d} \, \VEV{H_{1} ^{d}} & 0 \\
	y_{2} ^{d} \,  \VEV{H_{2} ^{d}} & y_{3} ^{d} \, \VEV{h_{1} ^{d}} & 0 \\
	0 & 0 & y_{1} ^{d} \, \VEV{H_{s}^{d}}	
\end{array}
\right) 
\]
With the VEVs 
\begin{eqnarray}\nonumber
&&\VEV{H_{s}^{d,u}} > 0 \; , \;\; \VEV{H_{1} ^{d}}= \VEV{H_{2} ^{d}} = v_{d} \; , \;\; 
 \VEV{h_{1} ^{d}}= \VEV{h_{2} ^{d}} = w_{d} \; , \\ \nonumber
&& \VEV{H_{1} ^{u}} = v_{u} \, \mathrm{e}^{-\frac{6 \, \pi \, i}{7}} \; , \;\;
 \VEV{H_{2} ^{u}} = v_{u} \, \mathrm{e}^{\frac{6 \, \pi \, i}{7}} \; , \;\;
 \VEV{h_{1} ^{u}} = w_{u} \, \mathrm{e}^{-\frac{12 \, \pi \, i}{7}}
\;\;\; \mbox{and} \;\;\; 
 \VEV{h_{2} ^{u}} = w_{u} \, \mathrm{e}^{\frac{12 \, \pi \, i}{7}}
\end{eqnarray}
for $H_{s} ^{u,d} \sim \MoreRep{1}{1}$, $H_{1,2} ^{u,d} \sim \MoreRep{2}{2}$
and $h_{1,2} ^{u,d} \sim \MoreRep{2}{4}$, we can achieve 
\[
M_{u}= \left( \begin{array}{ccc}
	y_{3} ^{u} \, w _{u} \, \mathrm{e}^{\frac{12 \, \pi \, i}{7}} 
        & y_{2} ^{u} \, v_{u} \, \mathrm{e}^{\frac{-6 \, \pi \, i}{7}} & 0 \\	
	y_{2} ^{u} \, v_{u} \, \mathrm{e}^{\frac{6 \, \pi \, i}{7}} 
        & y_{3} ^{u} \, w_{u} \, \mathrm{e}^{-\frac{12 \, \pi \, i}{7}} & 0 \\
	0 & 0 & y_{1} ^{u} \, \VEV{H_{s}^{u}}		
\end{array}
\right) \;\;\; \mbox{and} \;\;\; 
M_{d}= \left( \begin{array}{ccc}
	y_{3} ^{d} \,  w_{d} & y_{2} ^{d} \, v_{d} & 0 \\
	y_{2} ^{d} \,  v_{d} & y_{3} ^{d} \, w_{d} & 0 \\
	0 & 0 & y_{1} ^{d} \, \VEV{H_{s}^{d}}	
\end{array}
\right) 
\]
For $(m_{u}^{2}, m_{c} ^{2}, m_{t} ^{2})=(|y_{2}^{u} \, v_{u} + y_{3} ^{u} \, w_{u}|^{2},
|y_{2}^{u} \, v_{u} - y_{3}^{u} \, w_{u}|^{2}, |y_{1} ^{u} \, \langle H_{s} ^{u} \rangle|^2)$ 
and $(m_{d}^{2}, m_{s}^{2}, m_{b}^{2})=(|y_{2} ^{d} \, v_{d} - y_{3} ^{d} \, w_{d}|^{2},
|y_{2}^{d} \, v_{d} + y_{3} ^{d} \, w_{d}|^{2}, |y_{1} ^{d} \, \langle H_{s} ^{d} \rangle|^{2})$
the CKM matrix is of the form as given in \Eqref{eq:VCKMthetaConly}. Although the Higgs fields
$H_{s} ^{u,d}$ couple in this setup only to the third generation and therefore can have
a large VEV without spoiling the masses of the lighter quarks, there still exists
a source of fine-tuning, since the uncorrelated parameters $y_{2,3}^{d,u}$, $v_{d,u}$ and
$w_{d,u}$ have to be arranged such that $|y_{2} ^{d} \, v_{d} - y_{3} ^{d} \, w_{d}| \ll 
|y_{2} ^{d} \, v_{d} + y_{3} ^{d} \, w_{d}|$ for $m_{d} \ll m_{s}$ and 
$|y_{2}^{u} \, v_{u} + y_{3} ^{u} \, w_{u}| \ll |y_{2}^{u} \, v_{u} - y_{3} ^{u} \, w_{u}|$
for $m_{u} \ll m_{c}$. The preserved subgroups in the up and the down quark sector are 
again \Groupname{D}{2} $=<\mathrm{A}^{7}, \mathrm{B} \, \mathrm{A}^{6}>$ 
and \Groupname{D}{2} $=<\mathrm{A}^{7}, \mathrm{B}>$, respectively.

%%%%%%%%%%%%%%%%%%%%%%%%%%%%%%%%%%%%%%%%%%%%%%%
\mathversion{bold}
\section{Numerical Analysis of $V_{MNS}$}
\mathversion{normal}
%%%%%%%%%%%%%%%%%%%%%%%%%%%%%%%%%%%%%%%%%%%%%%%
\label{sec:vmns}

\noindent A similar analysis as done in the case of $V_{CKM}$ can also be carried out 
for the lepton mixing matrix $V_{MNS}$. We assume that
 the neutrinos are Dirac
particles as all the other fermions and that they have the same ordering 
as the other fermions, i.e. the neutrino mass spectrum is normally ordered. 
This allows us to use the matrix structures found in 
\Appref{app:vmix} also for $V_{MNS}$. 
Since the entries of $V_{MNS}$ are not strongly restricted by 
experiments
\cite{umnsranges} (at $3 \, \sigma$): 
\begin{equation}
\label{eq:umnsrange}
|V_{MNS} ^{\mbox{(range)}}|= \left( \begin{array}{ccc}
    0.79-0.88&0.47-0.61& <0.20\\
    0.19-0.52&0.42-0.73&0.58-0.82\\
    0.20-0.53&0.44-0.74&0.56-0.81
    \end{array} 
   \right)
\end{equation}
there are several more possibilities to accommodate the various matrix elements 
regarding the choice of the group index $n$,
and the values $m_{l}$, $m_{\nu}$ and $\mathrm j$. However, as we intend to build a model which
includes quarks as well as leptons, we stick to the selected values of $n$, $n=7$, $n=14$,
which fit the CKM matrix elements of the $1-2$ sub-block best, if we restrict ourselves
to small $n$. 
We check element by element of $V_{MNS}$ whether we can put it into the
form $|\cos (\frac{l \, \pi}{7})|$ where $l=0,1,2,...,6$ or 
$|\cos(\frac{l \, \pi}{14})|$ with $l=0,1,2,...,13$.
 According to \Eqref{eq:umnsrange} all elements of the second and third row can be approximated
by a cosine of the form $|\cos (\frac{l \, \pi}{7})|$ ($l=0,1,2,...,6$) or 
$|\cos(\frac{l \, \pi}{14})|$  ($l=0,1,2,...,13$) \footnote{We omit the trivial possibility
that the $(13)$ element can be approximated by $0$.}. 
\begin{table}[t]
\begin{center}
\begin{tabular}{|c|c|}
\hline
Element $(ij)$ & Possible cosines\\
\hline
(21)	& $\cos (\frac{3 \, \pi}{7}) \, (\approx 0.2225)$, 
	$\cos (\frac{5 \, \pi}{14}) \, (\approx 0.4339)$\\
(22)	& $\cos (\frac{5 \, \pi}{14}) \, (\approx 0.4339)$, 
	$\cos (\frac{2 \, \pi}{7}) \, (\approx 0.6235)$\\
(23)	& $\cos (\frac{2 \, \pi}{7}) \, (\approx 0.6235)$, 
	$\cos (\frac{3 \, \pi}{14}) \, (\approx 0.7818)$\\
(31)	& $\cos (\frac{3 \, \pi}{7}) \, (\approx 0.2225)$, 
	$\cos (\frac{5 \, \pi}{14}) \, (\approx 0.4339)$\\
(32)	& $\cos (\frac{2 \, \pi}{7}) \, (\approx 0.6235)$\\
(33) 	&  $\cos (\frac{2 \, \pi}{7}) \, (\approx 0.6235)$, 
	$\cos (\frac{3 \, \pi}{14}) \, (\approx 0.7818)$\\
\hline
\end{tabular}
\end{center}
\begin{center}
\begin{minipage}[t]{12cm}
\caption[]{Possibilities for the group theoretically determined element in $V_{MNS}$. Note that,
e.g. $\cos (\frac{3 \, \pi}{7})$ equals $\cos (\frac{6 \, \pi}{14})$, i.e.
it could also be reproduced in the group \Groupname{D}{14} with $\rm j=1$ and $m_{l}-m_{\nu}=6$
and not only in \Groupname{D}{7} with $\rm j=1$ and $m_{l}-m_{\nu}=3$. Furthermore, for example, 
$\cos (\frac{4 \, \pi}{7})$ is also included implicitly in the list, as 
$|\cos (\frac{4 \, \pi}{7})|=|\cos (\frac{3 \, \pi}{7})|$.
\label{tab:vmnspossibilities}}
\end{minipage}
\end{center}
\end{table}
We take into account all possibilities shown in \Tabref{tab:vmnspossibilities} and perform a numerical
fit of the mixing angles $\theta_{12}$, $\theta_{13}$ and $\theta_{23}$. 
In the fit procedure we compute the sines of the three mixing angles
and compare these to the best fit values, which are $\sin^{2} (\theta_{23} ^{bf})=0.5$, $\sin^{2} 
(\theta_{12} ^{bf})=0.3$ and $\sin ^{2} (\theta_{13} ^{bf})=0$ \cite{schwetz} 
\footnote{Note that these best fit values are not presented in the same
global analysis as the above mentioned allowed $3 \, \sigma$ ranges
for the elements of $V_{MNS}$. Nevertheless the deviations are very small
such that we do not consider this to lead to a major difference in our numerical
analysis.}. Again, we replace 
 the experimentally allowed $2 \, \sigma$ or $3 \,\sigma$ ranges by $10 \%$
ranges (around the best fit value). For $\sin ^{2} (\theta_{13})$ we consider two possible
upper bounds: $\sin ^{2} (\theta_{13}) \leq 0.025$ which corresponds to the $2 \, \sigma$ bound \cite{schwetz} and a much more loose bound $\sin ^{2} (\theta_{13}) \leq 0.1$ being
even larger than the $4 \, \sigma$ bound \cite{schwetz}. This is done, since
the numerical study showed that loosening the bound on $\sin ^{2} (\theta_{13})$ leads to several
more solutions. Our results for $\sin^{2} (\theta_{13}) \leq 0.1$
are summarized in \Tabref{tab:vmnsresults} where we also display the
numerical values for $\theta_{l}$, $\theta_{\nu}$ and $\alpha= \beta_{l} - \beta_{\nu}$
together with the resulting mixing angles and the (Dirac) CP phase $\delta$.\\ 
\noindent One can observe the
following: there are some cosines listed in \Tabref{tab:vmnspossibilities} for which 
no fit with $\chi^{2} < 1$ has been found. In all these cases the value
of the fixed $V_{MNS}$ element lies almost outside the ranges shown in 
\Eqref{eq:umnsrange}, e.g. for the $(23)$ element 
the possible cosines are
$\cos (\frac{2 \, \pi}{7}) \approx 0.6235$ and $\cos (\frac{3 \, \pi}{14}) \approx 0.7818$
with the first being quite close to the lower bound $(0.58)$ and the second one
 close to the upper one $(0.82)$ of the allowed range. Furthermore, by having a closer look at the form of $|V_{MNS}^{23}|$ given in \Appref{app:vmix}, one realizes that $\tan (\theta_{23})$ is simply
determined by the expression:
\begin{eqnarray}
\label{eq:VMNS23tan23}
\tan (\theta_{23}) &=& \left| \frac{\cot \left( (\phi_{l}-\phi_{\nu}) \, \frac{\mathrm{j}}{2} \right)}
{\cos (\theta_{l})} \right| =
\left| \frac{\cot \left( \frac{\pi \, (m_{l}-m_{\nu}) \, \mathrm{j}}{n} \right)}
{\cos (\theta_{l})} \right| 
\end{eqnarray}
Taking the argument of the cotangent to be either $\frac{2 \, \pi}{7}$ or $\frac{3 \, \pi}{14}$
leads to the numerical values
\begin{eqnarray}\nonumber
\tan (\theta_{23}) &\approx& 0.7975 \, \left| \frac{1}{\cos (\theta_{l})}\right| \;\;\; \mbox{or} \;\;\;
\tan (\theta_{23}) \approx 1.254 \, \left| \frac{1}{\cos (\theta_{l})}\right| \\ \nonumber
\end{eqnarray}
At the same time the sine of $\theta_{l}$ is determined by the $(13)$ element of $V_{MNS}$, i.e.
by the value of $\sin (\theta_{13})$: $|(V_{MNS} ^{23})_{13}|=
\left| \sin \left( (\phi_{l}-\phi_{\nu}) \, \frac{\mathrm{j}}{2} \right) \, \sin (\theta_{l}) \right|
= \left| \sin  \left( \frac{\pi \, (m_{l}-m_{\nu}) \, \mathrm{j}}{n} \right)
 \, \sin (\theta_{l}) \right|$, which gives for $\frac{2 \, \pi}{7}$ and $\frac{3 \, \pi}{14}$ 
$|(V_{MNS} ^{23})_{13}|=
0.7818 \, |\sin (\theta_{l})|$ and $0.6235 \, |\sin (\theta_{l})|$, respectively, i.e.
$|\sin (\theta_{l})|$ has to be as small as possible to fulfill the experimental
bound on $\sin^{2} (\theta_{13})$. Then $|\cos (\theta_{l})| \approx 1$ holds so that
we can deduce the approximate values $0.7975$ and $1.254$ for 
$\tan (\theta_{23})$  from \Eqref{eq:VMNS23tan23}. These correspond to 
$\theta _{23} \approx 38.57 ^{\circ}$ ($\sin^{2} (\theta_{23}) \approx 0.3888$)
and $\theta_{23} \approx 51.43 ^{\circ}$  ($\sin^{2} (\theta_{23}) \approx 0.6113$), i.e. 
they are at the boundaries of the $2 \, \sigma$ range for $\sin ^{2} (\theta_{23})$ 
\cite{schwetz}. 
Similar statements hold in case of $|V_{MNS} ^{33}|$.\\
\begin{table}[t]
\begin{center}
\begin{tabular}{|c|c||c|c|c||c|c|c|c|}
\hline
Element & Cosine & $\theta_{l}$ & $\theta_{\nu}$ & $\alpha$ 
& \rule[0.15in]{0cm}{0cm} $\sin ^{2} (\theta_{12})$ & $\sin ^{2} (\theta_{23})$
& $\sin ^{2} (\theta_{13})$ & $\delta$\\
\hline \hline
(21)	& \rule[0.15in]{0cm}{0cm} $\cos (\frac{3 \, \pi}{7})$ & $0.9790$ & $0.7881$ & $4.937$ 
& $0.2957$ & $0.5085$ & $7.037 \, \times \, 10^{-2}$ & $\sim\pi$\\ 
\cline{2-9}
	& \rule[0.15in]{0cm}{0cm} $\cos (\frac{5 \, \pi}{14})$ & $1.1829$ & $0.6725$ & $5.161$
& $0.3001$ & $0.4999$ & $6.173 \, \times \, 10^{-3}$ & $\sim 0$\\
\hline
(22)	& \rule[0.15in]{0cm}{0cm} $\cos (\frac{5 \, \pi}{14})$ & $-$ & $-$ & $-$
& $-$ & $-$ & $-$ & $-$\\
\cline{2-9} 
	& \rule[0.15in]{0cm}{0cm} $\cos (\frac{2 \, \pi}{7})$ & $0.7728$ & $0.4486$ & $5.386$
& $0.2999$ & $0.4996$ & $6.668 \, \times \, 10^{-3}$ & $\sim\pi$\\
\hline
(23)	& \rule[0.15in]{0cm}{0cm} $\cos (\frac{2 \, \pi}{7})$ & $-$ & $-$ &  $-$
& $-$ & $-$ & $-$ & $-$\\
\cline{2-9} 
	& \rule[0.15in]{0cm}{0cm} $\cos (\frac{3 \, \pi}{14})$ & $-$ & $-$ & $-$
& $-$ & $-$ & $-$ & $-$\\
\hline
(31)	& \rule[0.15in]{0cm}{0cm} $\cos (\frac{3 \, \pi}{7})$ & $0.9790$ & $0.7881$ & $4.937$  
& $0.2957$ & $0.4915$ & $7.037 \, \times \, 10^{-2}$ & $\sim 0$\\
\cline{2-9} 
	& \rule[0.15in]{0cm}{0cm} $\cos (\frac{5 \, \pi}{14})$ & $1.1829$ & $0.6725$ & $5.161$
& $0.3001$ & $0.5001$ & $6.173 \, \times \, 10^{-3}$ & $\sim\pi$\\
\hline
(32)	& \rule[0.15in]{0cm}{0cm} $\cos (\frac{2 \, \pi}{7})$ & $0.7728$ & $0.4486$ & $5.386$
& $0.2999$ & $0.5004$ & $6.668 \, \times \, 10^{-3}$ & $\sim 0$\\
\hline
(33) 	& \rule[0.15in]{0cm}{0cm} $\cos (\frac{2 \, \pi}{7})$ & $-$ & $-$ & $-$
& $-$ & $-$ & $-$ & $-$\\
\cline{2-9} 
       & \rule[0.15in]{0cm}{0cm} $\cos (\frac{3 \, \pi}{14})$ & $-$ & $-$ & $-$
& $-$ & $-$ & $-$ & $-$\\
\hline
\end{tabular}
\end{center}
\begin{center}
\begin{minipage}[t]{12cm}
\caption[]{Numerical results for $V_{MNS}$ in case of $\sin^{2} (\theta_{13}) \leq 0.1$ 
and $10 \%$ errors for the other two sine squares. $\delta$ is given with
a precision of $\mathcal{O}(10^{-6})$.
\label{tab:vmnsresults}}
\end{minipage}
\end{center}
\end{table}
\noindent Furthermore, one observes that in all cases the CP phase $\delta$ is trivial, 
i.e. $0$ or $\pi$ with a numerical
precision of $\mathcal{O}(10^{-6})$. Therefore $J_{CP}$ always vanishes.
In order to understand this result, we have a look at the formulae given for 
$V_{mix} ^{21}$, $V_{mix} ^{22}$, $V_{mix} ^{31}$ and $V_{mix} ^{32}$ in 
\Appref{app:vmix}. As a common feature the $(13)$ element of the mixing matrix
is given by 
\begin{equation}
\frac{1}{2} \, [-(1+ \mathrm{e} ^{-i \, (\phi_{l}-\phi_{\nu}) \, \mathrm{j}})
\, \sin (\theta_{l}) \, \cos (\theta_{\nu}) + 2 \, \mathrm{e} ^{i \, \alpha} \, \cos (\theta_{l})
\, \sin (\theta_{\nu})]
\end{equation}
\noindent In all cases, $\theta_{l}$ and $\theta_{\nu}$ are predominantly determined 
 by one element of the first row and the third column of $V_{MNS}$, respectively. Then 
 $\alpha$ can be used in order to minimize the absolute value of the $(13)$ element of $V_{MNS}$. A minimization 
with respect to $\alpha$ shows
\begin{equation}
\label{eq:alphaexplanation}
\alpha = -(\phi_{l}-\phi_{\nu}) \, \frac{\mathrm{j}}{2} + \pi \, y= 
-\frac{\pi}{n} \, (m_{l}-m_{\nu}) \, \mathrm{j} + \pi \, y \;\;\;\;\; \mbox{with} \;\;\; y \in 
\mathbb{Z}_{0}
\end{equation} 
\noindent The minimum value for $|\sin (\theta _{13})|$ is then 
$|\cos ((\phi_{l}-\phi_{\nu}) \, \frac{\mathrm{j}}{2}) \, \sin (\theta_{l}) \, \cos (\theta_{\nu})+ (-1)^{y+1} \cos (\theta_{l}) \, \sin (\theta_{\nu})|$.
However, in all cases the expression is only minimized for $y=0,2,...$,
as the involved sines and cosines are all positive 
(remember that $\theta_{l}$ and $\theta_{\nu}$ are restricted to be smaller
than $\frac{\pi}{2}$ by definition and also $(\phi_{l}-\phi_{\nu}) \, \frac{\mathrm{j}}{2}
= \frac{\pi}{n} \, (m_{l}-m_{\nu}) \, \mathrm{j}$ which is the argument of
the cosine displayed in the tables is always smaller than $\frac{\pi}{2}$). As 
 $J_{CP}$ is proportional to $\sin ((\phi_{l}-\phi_{\nu}) \, \frac{\mathrm{j}}{2} 
+ \alpha)$, it is zero for the calculated value of $\alpha$. 
Therefore $\delta$ must be either $0$ or $\pi$. 
Additionally, we found an explanation for 
the values of $\alpha$ shown in \Tabref{tab:vmnsresults} given
in terms of the group theoretical quantities, i.e. $2 \, \pi -\frac{3 \,\pi}{7} \approx 4.937$,
$2 \, \pi - \frac{5 \,\pi}{14} \approx 5.161$ and $2 \, \pi - \frac{2 \,\pi}{7} \approx 5.386$.
Since $\alpha$ has to lie in $[0, 2 \, \pi)$, $y$ equals two in all cases, 
see \Eqref{eq:alphaexplanation}.\\
\noindent As a last observation we report that there exist similarities among the 
different cases, 
e.g. fixing the $(21)$ element to be 
$\cos (\frac{3 \,\pi}{7})$ is similar
to fixing the $(31)$ element to the same value. In both cases the fit values of $\theta_{l}$, $\theta_{\nu}$
and $\alpha$ are the same. Therefore, the results for $\sin ^{2} (\theta_{12})$ and 
$\sin ^{2} (\theta_{13})$ coincide (up to $\mathcal{O} (10^{-6})$), while
$\sin ^{2} (\theta_{23})$ is shifted from being $0.5 + \epsilon$ to $0.5 - \epsilon$
with $\epsilon \approx 0.0085$ and the CP phase $\delta$ shifts from $\pi$ to $0$.
Looking at the mixing matrices one recognizes that these
similarities are due to the fact that the second and the third row are interchanged.\\
\noindent Using the $2 \,\sigma$ bound $\sin ^{2} (\theta_{13}) \leq 0.025$ no solution with 
$\chi^{2} < 1$ is found in the cases in which the $(21)$ or the $(31)$ element is fixed to the value 
$\cos (\frac{3 \,\pi}{7})$, since the values for $\sin^{2} (\theta_{13})$ shown in 
\Tabref{tab:vmnsresults} are quite large. For the other configurations
we again find viable fits in which the values $\theta_{l}$, $\theta_{\nu}$ and $\alpha$
are very similar to the ones given in \Tabref{tab:vmnsresults}.\\
Until now we only investigated the cases in which the group theoretically
fixed element is given by one of the cosines shown in 
\Tabref{tab:vmnspossibilities}. However,
as already remarked several times we can also look at cases in which
the cosine is for example $\cos (\frac{4 \, \pi}{7})$ instead of
$\cos (\frac{3 \, \pi}{7})$, since $\cos (\frac{4 \, \pi}{7})$ is just 
the negative of $\cos (\frac{3 \, \pi}{7})$. In terms of group theoretical
quantities this corresponds to sending $m_{l}$ to $n-m_{l}$ and
therefore $\phi_{l}$ to $2 \, \pi - \phi_{l}$ \footnote{Thereby, we have set $\phi_{\nu}$ to zero and $\rm j$ to 1 without loss of
generality.}. The general forms of the mixing matrices
given in \Appref{app:vmix} show that such a transformation does not
change the absolute values of the matrix elements, if we replace 
 the phase $\alpha$ by $-\alpha$ ($2 \, \pi - \alpha$) at the same time. In contrast to this
$J_{CP}$ is not invariant and changes its sign. In the analysis of the
leptonic mixing parameters this is not relevant, since the phase(s) have not
been measured. Moreover, in all cases considered here $J_{CP}$ is almost zero
(up to $\mathcal{O}(10^{-6})$). Therefore, we get the same results 
for these equivalent cases. Note, that in case of the quark mixing
matrix we would have to expect different results, since there 
 $J_{CP}$ is known from experiment and its sign change 
 leads to a distinct solution.\\
\begin{table}
\begin{center}
\begin{tabular}{|c|c|}
\hline
Element $(ij)$ & Possible cosines\\
\hline
(11)	& $\cos (\frac{3 \, \pi}{14}) \, (\approx 0.7818)$\\
(12)	& $\cos (\frac{2 \, \pi}{7}) \, (\approx 0.6235)$\\
(21)	& $\cos (\frac{5 \, \pi}{14}) \, (\approx 0.4339)$\\
(22)	& $\cos (\frac{2 \, \pi}{7}) \, (\approx 0.6235)$\\
(31)	& $\cos (\frac{5 \, \pi}{14}) \, (\approx 0.4339)$\\
(32)	& $\cos (\frac{2 \, \pi}{7}) \, (\approx 0.6235)$\\
\hline
\end{tabular}
\end{center}
\begin{center}
\begin{minipage}[t]{12cm}
\caption[]{Possibilities for the group theoretically determined element in $V_{MNS}$, if
TBM is assumed to be the best fit. For further conventions, see 
\Tabref{tab:vmnspossibilities}.
\label{tab:vmnsTBMpossibilities}}
\end{minipage}
\end{center}
\end{table}
\noindent Apart from studying how well one can accommodate the experimentally allowed ranges, it is
also interesting to see whether one can reproduce some special mixing pattern in the lepton sector.
 In the following we discuss the TBM scenario which has 
initially been discussed in \cite{hps}, since all elements of the lepton mixing matrix can be written in terms
of fractions of square roots $\frac{1}{\sqrt{2}}$, $\frac{1}{\sqrt{3}}$ and $\frac{1}{\sqrt{6}}$:
\begin{equation}
\label{eq:vmnsTBM}
V_{MNS} ^{TBM}=\left(
\begin{array}{ccc}
\frac{2}{\sqrt{6}}& \frac{1}{\sqrt{3}}& 0\\
-\frac{1}{\sqrt{6}}& \frac{1}{\sqrt{3}}& -\frac{1}{\sqrt{2}}\\
-\frac{1}{\sqrt{6}}& \frac{1}{\sqrt{3}}& \frac{1}{\sqrt{2}}
\end{array}
\right)
\end{equation}
corresponding to sines of the mixing angles:
\[
\sin ^{2} (\theta_{23} ^{TBM})= \frac{1}{2} \; , \;\; 
\sin ^{2} (\theta_{12} ^{TBM})= \frac{1}{3} \;\;\; \mbox{and} \;\;\;
\sin ^{2} (\theta_{13} ^{TBM})= 0 \; .
\]
However, it turned out to be not just an assumption of a special form of $V_{MNS}$, but it is a robust 
outcome of certain models based on the discrete non-abelian symmetries \Groupname{A}{4} or \Doub{T}{}
\cite{a4,tprime}.
Therefore we want to analyze whether we can also accommodate the TBM with mixing matrices of the form 
$V_{mix}$ as given in \Appref{app:vmix}. The uncertainty in the mixing matrix elements is taken to be 
$10 \, \%$, i.e. the fixed element given by cosine  $|\cos (\frac{l \, \pi}{7})|$ 
for $l=0,1,2,...,6$ or $|\cos(\frac{l \, \pi}{14})|$ with $l=0,1,2,...,13$ should lie
in one of the ranges:
\begin{equation}
\label{eq:vmnsTBMrange}
V_{MNS} ^{TBM \, \mbox{(range)}}=\left(
\begin{array}{ccc}
 0.73 - 0.90 & 0.52 - 0.64 & <0.20\\
 0.37 - 0.45 & 0.52 - 0.64 & 0.64 - 0.78\\
 0.37 - 0.45 & 0.52 - 0.64 & 0.64 - 0.78 
\end{array}
\right)
\end{equation}
\noindent The bound on the $(13)$ element is taken to be the same as in \Eqref{eq:umnsrange}. 
As shown in \Tabref{tab:vmnsTBMpossibilities}, the elements $(11)$ and $(12)$ can
now be described by a cosine of the announced form, while we find less possibilities 
for the other elements compared to the case of the experimentally allowed range, see
\Tabref{tab:vmnspossibilities}. This analysis is analogous to the one above. Again, we display the results for the fits using the loose bound
for $\sin ^{2} (\theta_{13})$, $\sin ^{2} (\theta_{13}) \leq 0.1$. Similar to
above, there is a case in which we have not found a fit with $\chi^{2}<1$.
For the cases in which either the $(21)$, $(22)$, $(31)$ or $(32)$ element
is determined by group theory all statements made above can also be
applied here, i.e. the CP phase $\delta$ is either $0$ or $\pi$, the
phase $\alpha$ is fixed to a certain value which minimizes 
$|\sin (\theta_{13})|$ and there exists a similarity among the cases 
with a fixed $(21)$ ($(22)$) element and a fixed $(31)$ ($(32)$) element.
\begin{table}
\hspace{-0.45in}\parbox{6in}{\begin{tabular}{|c|c||c|c|c||c|c|c|c|}
\hline
Element & Cosine & $\theta_{l}$ & $\theta_{\nu}$ & $\alpha$ 
& \rule[0.15in]{0cm}{0cm} $\sin ^{2} (\theta_{12})$ & $\sin ^{2} (\theta_{23})$
& $\sin ^{2} (\theta_{13})$ & $\delta$\\
\hline \hline
(11)	& \rule[0.15in]{0cm}{0cm} $\cos (\frac{3 \, \pi}{14})$ & $0.4396 - 1.131$ & $1.139$ & 
$\in [0, 2 \pi)$ & $0.3441$ & $0.5000$ & $6.808 \, \times \, 10^{-2}$ & $\in [\sim 0,\sim 2 \pi)$\\ 
\hline
(12)	& \rule[0.15in]{0cm}{0cm} $\cos (\frac{2 \, \pi}{7})$ & $-$ & $-$ & $-$
& $-$ & $-$ & $-$ & $-$\\
\hline
(21)	& \rule[0.15in]{0cm}{0cm} $\cos (\frac{5 \, \pi}{14})$ & $1.132$ & $0.6697$ & $5.161$
& $0.3334$ & $0.5000$ & $1.968 \, \times \, 10^{-3}$ & $\sim 0$\\
\hline
(22)	& \rule[0.15in]{0cm}{0cm} $\cos (\frac{2 \, \pi}{7})$ & $0.8235$ & $0.4557$ & $5.386$
& $0.3331$ & $0.4991$ & $1.245 \, \times \, 10^{-2}$ & $\sim\pi$\\
\hline
(31)	& \rule[0.15in]{0cm}{0cm} $\cos (\frac{5 \, \pi}{14})$ & $1.132$ & $0.6697$ & $5.161$
& $0.3334$ & $0.5000$ & $1.968 \, \times \, 10^{-3}$ & $\sim\pi$\\
\hline
(32)	& \rule[0.15in]{0cm}{0cm} $\cos (\frac{2 \, \pi}{7})$ & $0.8235$ & $0.4557$ & $5.386$
& $0.3331$ & $0.5009$ & $1.245 \, \times \, 10^{-2}$ & $\sim 0$\\
\hline
\end{tabular}}
\begin{center}
\begin{minipage}[t]{12cm}
\caption[]{Numerical results in the case of TBM. We assume that the bound on 
 $\sin^{2} (\theta_{13})$ is $0.1$ and $10 \%$ errors for the other two sine squares.
  The values of $\delta$ have a numerical precision of 
$\mathcal{O}(10^{-6})$.
Note that in case of the $(11)$ element being $\cos (\frac{3 \,\pi}{14})$ $\delta$
can take arbitrary values. (for details see text).
\label{tab:vmnsTBMresults}}
\end{minipage}
\end{center}
\end{table}
Therefore, we focus on the discussion of a group theoretically
determined $(11)$ element of $V_{MNS}$. This case exhibits some new features
not present in the other ones. First of all, we find that $\theta_{l}$
can take values in a certain range instead of being fixed to a single
value. All of them lead to the same mixing angles. The same is true for
$\alpha$ which varies between $0$ and $2 \, \pi$.
This is related to the fact that we do not fit the CP phase $\delta$ (or 
equivalently the Jarlskog invariant $J_{CP}$). As a result $J_{CP}$ can take
any value in the range $(-5.776...5.776) \, \times \, 10^{-2}$. 
We observe that $\theta_{\nu}$ is fixed by the fit
of $\sin ^{2} (\theta_{12})$ and $\sin^{2} (\theta_{13})$. Fitting them
at the same time leads, unfortunately, to a too large value for 
$\sin^{2} (\theta_{13})$ (see \Tabref{tab:vmnsTBMresults}). The allowed
range for $\theta_{l}$ can then be found analytically under the
assumption that $\sin ^{2} (\theta_{23})=\frac{1}{2}$, since in this
case the $(23)$ and $(33)$ element of $V_{MNS}$ have to be equal. Equating the
expressions $|(V_{mix}^{11})_{23}|^{2}$ and  $|(V_{mix}^{11})_{33}|^{2}$ found
in \Appref{app:vmix} leads to
\begin{equation}
\label{eq:thetalequation}
\tan (2 \, \theta_{l})= \frac{\sin ^{2} (\theta_{\nu})-\cos ^{2} 
((\phi_{l}-\phi_{\nu}) \, \frac{\mathrm{j}}{2}) \, \cos ^{2} (\theta_{\nu})}
{\cos ((\phi_{l}-\phi_{\nu}) \, \frac{\mathrm{j}}{2} + \alpha) \, 
\cos ((\phi_{l}-\phi_{\nu}) \, \frac{\mathrm{j}}{2}) \, 
\sin (2 \, \theta_{\nu})}
\end{equation}
with $\theta_{\nu}$ determined by $\sin ^{2} (\theta_{12,13})$.
Allowing $\alpha \in [0, 2 \,\pi)$ one finds the maximal
range of $\theta_{l}$ to be $z \leq \theta_{l} \leq \frac{\pi}{2} -z$
with  $z \approx 0.4396 \;\;\; \mbox{for} \;\;\; \theta_{\nu} \approx 1.139 \;\;\; \mbox{and}
\;\;\; (\phi_{l}-\phi_{\nu}) \, \frac{\rm j}{2} = \frac{3 \,\pi}{14}$
which corresponds to the numerical values given in 
\Tabref{tab:vmnsTBMresults}. Furthermore, \Eqref{eq:thetalequation} shows that $\theta_{l}$
is a function of $\alpha$.\\
\noindent Demanding $\sin ^{2}(\theta_{13}) \leq 0.025$ removes the 
possibility that
the $(11)$ element of $V_{MNS}$ is determined by group theory, while it
leads to expected slight changes in the results of the fits for the
rest of the cases. As expected, a comparison of
 these results to the ones for the
experimental best fit values shows that there are only small changes in
the precise values of $\theta_{l}$, $\theta_{\nu}$ and $\alpha$. 

\vspace{0.1in}

\noindent In this section we have shown that it is possible to fit the lepton mixing
angles \cite{schwetz} in a framework in which one of the elements of $V_{MNS}$ is completely
determined by group theoretical quantities of a dihedral flavor symmetry.
We restricted ourselves to the dihedral groups \Groupname{D}{7} and
\Groupname{D}{14}, since they allow us to explain the Cabibbo angle via group
theory. A main result of the analysis is that $J_{CP}$ vanishes in all cases.
Using the formulae given in \Appref{app:vmix} one can show that the 
vanishing of $J_{CP}$ is correlated with the minimization of $|\sin (\theta_{13})|$.
As the bound on $|\sin (\theta_{13})|$ is the strongest constraint on the solution,
we showed our results for a very loose bound. Furthermore, we analyzed how well 
one can mimic the TBM scenario. We found several possible solutions. One of these is
of special interest, since it also allows for non-trivial CP violation. However,
the corresponding value of $\sin (\theta_{13})$ is very large and therefore this solution is
disfavored. These results demonstrate that it is possible to treat the lepton mixings
in the same way as the ones of the quarks. Small corrections are expected in a complete model,
e.g. due to explicit breakings of the preserved subgroups.

%%%%%%%%%%%%%%%%%%%%%%%%%%%%%%%%%%%%%%%%%%%%%%%
\section{Summary and Conclusions}
%%%%%%%%%%%%%%%%%%%%%%%%%%%%%%%%%%%%%%%%%%%%%%%
\label{sec:summary}

\noindent In \cite{dntheorypaper} we studied the dihedral
groups as possible flavor symmetries. The key feature there
is the fact that the flavor symmetry is not broken in an arbitrary 
way, but one requires that a subgroup has to be preserved
in all cases. It turned out that the number of possible mass matrix structures
which arise from such a setup is very limited, if we assume that the mass matrix has a non-vanishing
determinant. As a first application
we discussed in \cite{dntheorypaper} the possibility to describe
one element of the CKM mixing matrix only in terms of group theoretical
quantities, i.e. the index $n$ of the dihedral group \Groupname{D}{n},
the index $\rm j$ of the representation $\MoreRep{2}{j}$ under which the fermions transform
and the indices $m_{u}$ and $m_{d}$ of the residual subgroups
\Groupname{Z}{2} $=<\mathrm{B} \, \mathrm{A}^{m_{u}}>$ and 
\Groupname{Z}{2} $=<\mathrm{B} \, \mathrm{A}^{m_{d}}>$:
\begin{equation}
\label{eq:grouptheoryfixedelement}
\frac{1}{2}\, \left| 1 + \mathrm{e} ^{i \, (\phi_{u}-\phi_{d}) \, \mathrm{j}}
\right| = \left| \cos ((\phi_{u}-\phi_{d})\, \frac{\mathrm{j}}{2}) \right|
 = \left| \cos (\frac{\pi}{n}\, (m_{u}-m_{d})\, \mathrm{j}) \right|
\end{equation}
\noindent where $\phi_{u}=\frac{2 \,\pi}{n} \, m_{u}$ and
$\phi_{d}=\frac{2 \,\pi}{n} \, m_{d}$. \Eqref{eq:grouptheoryfixedelement} shows 
that a non-trivial mixing angle demands $m_{u} \neq m_{d}$, i.e.
the two \Groupname{Z}{2} subgroups have to be distinct. 
It has been pointed out that $|V_{us}|$ can be fitted well with 
$\cos (\frac{3 \,\pi}{7}) \approx 0.2225$. In this work, we first studied which of the
other elements of $V_{CKM}$ can also be approximated well by \Eqref{eq:grouptheoryfixedelement}
for certain values of the group index $n$. For the smallest two appropriate values of $n$,
$n=7$ and $n=14$, each element of the $1-2$ sub-block of $V_{CKM}$ can be
put into this form, i.e. $|V_{ud}| \approx |V_{cs}| \approx \cos (\frac{\pi}{14})$
and $|V_{us}| \approx |V_{cd}| \approx \cos (\frac{3 \, \pi}{7})$. A numerical analysis
showed that the other two mixing angles, $\theta_{13} ^{q}$ and $\theta_{23} ^{q}$,
and the CP phase $\delta$ can be fitted well with the free angles $\theta_{u,d}$ and
the phase $\alpha=\beta_{u}-\beta_{d}$. Since the fixed element 
cannot be fitted, the results for $V_{CKM}$
are very close to the experimental values, but not within the (very small)
experimental errors \cite{pdg}.
However, several sources of corrections exist in a complete model, e.g. possible
higher-dimensional operators as well as small, but explicit, breakings of the preserved
subgroups. In a next step, we presented a
low energy model for the quark sector which incorporates the described idea. The flavor
symmetry is taken to be \Groupname{D}{7}. It is broken only spontaneously at the 
electroweak scale by Higgs fields transforming as doublets under $SU(2)_{L}$.
With a numerical fit we showed that all quark masses and mixing parameters 
can be fitted well
at the same time. As the VEV configuration determines the subgroup to which the flavor
symmetry is broken, it is necessary to investigate whether this can be achieved by
the Higgs potential. We studied this issue for the minimal model in which the three Higgs fields
$H_{s}^{u}$ and $H_{1,2}^{u}$ couple to up quarks, while the fields $H_{s}^{d}$
and $H_{1,2}^{d}$ couple to down quarks only. Unfortunately, the Higgs potential
containing only $H_{s}^{d,u}$ and $H_{1,2} ^{d,u}$ has an accidental symmetry
which is necessarily broken by the desired VEV configuration. For this reason we had
to add two further Higgs fields to the model. These do not directly couple to the fermions due to their
\Groupname{D}{7} transformation properties. The couplings of $H_{s}^{d,u}$
and $H_{1,2}^{d,u}$ to these Higgs fields
break all accidental symmetries of the potential. A numerical study showed that
the needed VEV configuration can be achieved with this potential. However, there are two obstacles: 
first of all  if the quartic couplings of the Higgs potential are in the perturbative range and 
the mass parameters are taken to be around the electroweak scale, the Higgs masses  
 turn out to be too small, i.e. some of them are even below the
LEP bound  \cite{lepbound}. This could be cured by adding mass terms which break the flavor symmetry
softly in the Higgs potential and allow for larger Higgs masses. However, even then 
this model might
suffer from the problem that FCNCs induced by the additional Higgs fields are too large
to pass the experimental bounds. The second obstacle is the fact that we are only
able to accommodate the VEV configuration as one possible solution of the Higgs potential,
but not as a favored solution. Moreover, there is in general no way to stabilize
such a configuration against further corrections in a multi-Higgs doublet
potential. Therefore this model is meant as
a proof of principle rather than a realistic model.
A way to circumvent these problems is to disentangle the
scales of the electroweak and the flavor symmetry breaking by using flavored gauge singlets
instead of Higgs doublets and thereby break the dihedral symmetry at higher energies \cite{AF}.\\
\noindent Accounting for the fact that the Cabibbo angle $\theta_C$ is roughly an order of 
magnitude larger than the two other mixing angles $\theta_{13}^{q}$ and $\theta_{23}^{q}$
one can look for models in which $\theta_C$ is given in terms of group theoretical quantities
and $\theta_{13}^{q}$ and $\theta_{23}^{q}$ vanish at LO. As shown in 
\Secref{sec:alternatives} this can be implemented successfully in at least two different
ways: $a.)$ one can simply reduce the number of Higgs fields in the model by
omitting some fields which are allowed to have a non-trivial VEV in principle; $b.)$
one can break the dihedral symmetry down to one of its dihedral subgroups, \Groupname{D}{q},
$q>1$, instead of \Groupname{Z}{2}. Case $a.)$ has the slight disadvantage
that the resulting mass matrices now also depend on the choice of the scalar fields
and are not only determined by the representations under which the fermions transform
and the group theory of the dihedral symmetry. Case $b.)$ on the other hand does not suffer
from this sort of arbitrariness. However, it cannot be realized with all dihedral symmetries,
since not all of them have dihedral subgroups \Groupname{D}{q} with $q>1$. The group
\Groupname{D}{7} which has been used in this paper only has \Groupname{Z}{2} and 
\Groupname{Z}{7} as subgroups, since its group index is prime. Therefore in the shown 
examples (for case $b.)$) the flavor symmetry is taken to
be \Groupname{D}{14} instead. The preserved subgroups are of the form \Groupname{D}{2}
$=<\mathrm{A}^{7}, \mathrm{B} \, \mathrm{A}^{m}>$. Also here it is necessary to
break down to two different \Groupname{D}{2} groups in the up quark and down quark sector in order
to generate a non-vanishing Cabibbo angle. One possible choice is $m_{u}=6$ and $m_{d}=0$.\\ 
\noindent Finally, we also studied the lepton mixing matrix $V_{MNS}$ numerically. 
In order to apply the results of the mixing matrices found in \Secref{sec:basics}
we restricted ourselves to the discussion of Dirac neutrinos with a normally ordered
spectrum, i.e. the neutrinos have the same properties as the other fermions. 
Since the elements of $V_{MNS}$ are much less constrained than the ones of 
$V_{CKM}$ much more combinations of the group theoretical
quantities $n$, $\mathrm{j}$, $m_{l}$ and $m_{\nu}$ can be used in order to describe an
element of $V_{MNS}$. However, since we expect that the leptons transform under
the same flavor symmetry as the quarks, we only considered cases in which the group
index $n$ is fixed to $n=7$ or $n=14$. A numerical analysis shows that the experimental
fit values of the mixing angles can be accommodated well in most of the cases. The strongest
constraint seems to arise from the upper bound on the reactor mixing angle $\theta_{13}$.
Therefore we performed fits with two different bounds on $\sin ^{2}(\theta_{13})$. 
The results which are shown in \Secref{sec:vmns} correspond to a very loose bound, 
$\sin ^{2} (\theta_{13}) \leq 0.1$ (which exceeds the $4 \, \sigma$ bound \cite{schwetz}),
while in the other fit the $2 \, \sigma$ bound, $\sin ^{2} (\theta_{13}) \leq 0.025$, has
been used. A common feature of all fits is the fact that $J_{CP}$ vanishes. As shown in
\Secref{sec:vmns} the condition which minimizes the $(13)$ element of $V_{MNS}$ whose absolute
value is $\sin (\theta_{13})$ also implies $J_{CP}=0$. Furthermore, it turns out that 
the fit parameters $\theta_{l}$, $\theta_{\nu}$ and $\alpha$ are fixed to a
unique value in all cases. In addition to this, we also studied how well one could mimic the TBM scenario
with a mixing matrix resulting from a dihedral flavor symmetry with preserved subgroups.
Again, we only considered the cases $n=7$ and $n=14$. Our results are similar to the ones
found in case of the fit to the experimental best fit values, i.e. a successful fit is possible
in several cases. Similar to above, the main restriction seems to come from the requirement
to pass the bound on $\sin^{2} (\theta_{13})$. For this reason again two different bounds on
$\sin^{2} (\theta_{13})$  have been used. Apart from the cases which lead to similar
results as above, we observe one additional case, namely if $(V_{MNS})_{11}$ is fixed
to be $\cos (\frac{3 \, \pi}{14}) \approx 0.7818$. Unlike in the other cases we can observe
CP violation here, i.e. $J_{CP}$ can take any value between $-5.776 \, \times 10^{-2}$
and $5.776 \, \times 10^{-2}$. In contrast to this, the results of the fit of the mixing
angles do not vary. Unfortunately, $\sin ^{2} (\theta_{13})$ is very large and therefore
a model incorporating this solution is disfavored, if contributions from, for example,
higher-dimensional operators are not able to lower $\sin ^{2} (\theta_{13})$.\\
\noindent In the whole discussion we focussed on the case of Dirac neutrinos, since
then all formulae found in case of the quarks are applicable also to the lepton sector.
However, in general neutrinos can be Majorana particles. If we assume that they
acquire masses from Higgs triplets only, i.e. there are no right-handed neutrinos,
the possible matrix structures are $M_{5}$ (\Eqref{eq:M4andM5forups}) 
and a block structure (\Eqref{eq:blockmatrix}), see
\Secref{sec:basics}. Compared to the case of Dirac neutrinos the mixing matrix
$U_{\nu}$ is now determined by $U^{\dagger}_{\nu} \, M_{\nu} \, U_{\nu} ^{\star}= \diag (m_{1},m_{2},m_{3})$
and therefore in general contains Majorana phases. This, however, does not matter
for the analysis done in \Secref{sec:vmns}, since there only the absolute 
values of the matrix elements are relevant. Things can change, if we consider
the type 1 seesaw instead. As we then deal with the Dirac neutrino mass matrix
and the right-handed Majorana mass matrix, these mass matrices can preserve
different subgroups of the flavor symmetry. This is, for example, the case in 
the models \cite{Grimus1,Grimus2} by Grimus and Lavoura \footnote{In their models
the Dirac neutrino mass matrix is actually invariant under the whole dihedral symmetry, 
i.e. stems solely from VEVs of fields which transform trivially under the flavor group.}. 
The situation can be even more complicated, if the model also includes Higgs 
triplets. Then all these three matrices, i.e. the Majorana mass matrix of the left-handed 
neutrinos, the one of the right-handed ones and the Dirac neutrino mass matrix, 
can conserve different subgroups of the original
symmetry and in general no definite statements can be made about the resulting
mixing matrix. 
Furthermore we assumed  throughout our analysis that the neutrinos have the same mass 
ordering as all the other fermions. However, due to the unknown
sign of the atmospheric mass squared difference it is also possible
that the neutrino mass hierarchy is inverted ($m_{3} < m_{1} < m_{2}$).\\
\noindent Our study is by no means a complete study of all possible mixing structures
which can in principle arise from a dihedral flavor symmetry with preserved subgroups.
For example, in all cases we presented here the subgroups which are preserved in the up and down quark sector
 have the same group structure, i.e. they are either both \Groupname{Z}{2}
or \Groupname{D}{2} groups. In general, however, these group structures could be
different. Successful examples which employ subgroups of different structures are 
the \Groupname{A}{4} (\Doub{T}{}) models as well as the models by Grimus and Lavoura. 
As already mentioned in the Introduction, in the \Groupname{A}{4} (\Doub{T}{}) model the conserved 
subgroups are \Groupname{Z}{2} (\Groupname{Z}{4}) and \Groupname{Z}{3} in order to predict
TBM in the lepton sector. In the first model \cite{Grimus1} by Grimus and Lavoura 
the flavor symmetry \Groupname{D}{4} $\times$ $Z_{2}^{(aux)}$ is
broken either to \Groupname{D}{2}, \Groupname{Z}{2} or is left intact (see
\cite{dntheorypaper}). Similarly,
in their second model \cite{Grimus2} with \Groupname{D}{3} $\times$ $Z_{2} ^{(aux)}$
as flavor symmetry \Groupname{Z}{3}, \Groupname{Z}{2} and \Groupname{D}{3} are the
preserved subgroups (see also \cite{dntheorypaper}). Both models lead to vanishing
$\theta_{13}$ and maximal atmospheric mixing (for the leptons). 
This shows that the usage of subgroups of different group
structure leaves much more possibilities than the ones shown here. As the complete 
 study of mass matrix structures (with $\det (M)\neq 0$) which can arise from a dihedral 
symmetry, if a subgroup
remains preserved, already exists \cite{dntheorypaper}, it is only the question how to combine
these results in order to get further interesting predictions for the mixing patterns of 
quarks and leptons. One interesting example, namely the explanation of the
Cabibbo angle, has been studied in detail in this work.\\
\noindent Finally, let us remark that a common feature of the model(s) shown here and the 
successful \Groupname{A}{4}
(\Doub{T}{}) models is the need for an additional $\mbox{\Groupname{Z}{n}}^{(aux)}$ symmetry 
which can separate the different sectors according to the different subgroups of the
flavor symmetry which should be preserved. Due to such an additional symmetry an 
embedding of these models into an $SO(10)$ GUT is in general not straightforward. However,
assigning the quarks to
\begin{eqnarray}
Q_{1}, u^{c}_{1} \sim (\MoreRep{1}{1},+1) \; , \;\; \left( 
\begin{array}{c}
Q_{2}\\
Q_{3}
\end{array}
\right), \left( 
\begin{array}{c}
u^{c}_{2}\\
u^{c}_{3}
\end{array}
\right) \sim (\MoreRep{2}{1},+1) \; , \;\;
d^{c}_{1} \sim (\MoreRep{1}{1},-1) \; , \;\; \left(
\begin{array}{c}
d^{c}_{2}\\
d^{c}_{3}
\end{array}
\right) \sim (\MoreRep{2}{1},-1) \;\;
\end{eqnarray}
under \Groupname{D}{7} $\times$ $Z_{2} ^{(aux)}$ as done in  \Secref{subsubsec:assign_M5}
still allows an embedding into
$SU(5)$ multiplets.

\vspace{0.1in}
\noindent \textit{Note added:} At the final stages of this work the 
paper \cite{Lam} by C. S. Lam appeared. He also deals with the fact that 
non-trivial subgroups of some 
discrete flavor symmetry can help to explain a certain mixing pattern and also very briefly mentions 
that the Cabibbo angle might be the result of some dihedral group.

%%%%%%%%%%%%%%%%%%%%%%%%%%%%%%%%%%%%%%%%%%%%%%%%
\subsection*{Acknowledgements}
%%%%%%%%%%%%%%%%%%%%%%%%%%%%%%%%%%%%%%%%%%%%%%%%
\noindent A.B. acknowledges support from the Studienstiftung des Deutschen Volkes.
C.H. was supported by the ``Sonderforschungsbereich'' TR27.

\newpage
\appendix

%%%%%%%%%%%%%%%%%%%%%%%%%%%%%%%%%%%%%%%%%%%%%%%%
\mathversion{bold}
\section{Possible Forms of $V_{mix}$}
\mathversion{normal}
%%%%%%%%%%%%%%%%%%%%%%%%%%%%%%%%%%%%%%%%%%%%%%%%
\label{app:vmix}

\noindent According to the three possible identifications of the eigenvalue $c-d$ there exist three possible diagonalization matrices in each sector (up and down sector, charged lepton and neutrino sector) $U$, $U^{\prime}$ and $U^{\prime \, \prime}$ which are shown in 
\Secref{subsec:analyticvckm}. Out of these one can 
form nine possible mixing matrices $V_{mix} ^{ab}=W ^{T} _{1} \, W^{\star} _{2}$ with $a,b=1,2,3$
and $W_{i} \in \left\{ U, U^{\prime}, U^{\prime \, \prime} \right\}$ where $W_{i}$ depends on the
group theoretical phase $\phi_{i}$
(the index $m_{i}$) and contains the parameters $\theta_{i}$ and $\beta_{i}$. They all
have the property that one of their matrix elements, namely the element $(ab)$, 
is completely determined by 
group theory, i.e. by the index $n$ of the dihedral group, by the index $\mathrm{j}$ of 
the two-dimensional representation $\MoreRep{2}{j}$ under which two of three generations of $SU(2)_{L}$ 
doublets transform and by the breaking directions described by 
$m_{1}$ and $m_{2}$ in the two different sectors. In the following we abbreviate $\beta_{1}-\beta_{2}$
with $\alpha$, $\sin (\theta_{i})$ with $s_{i}$
and $\cos (\theta_{i})$ with $c_{i}$.

\footnotesize
\begin{eqnarray}\nonumber
V_{mix} ^{11} &=& \frac{1}{2} \, \left( \begin{array}{ccc}
	1 + \mathrm{e} ^{i \, (\phi_{1}-\phi_{2}) \, \mathrm{j}}
	& (\mathrm{e} ^{i \, \phi_{1} \, \mathrm{j}} - \mathrm{e} ^{i \, \phi_{2} \, \mathrm{j}}) \, s_{2} 
	&  -(\mathrm{e} ^{i \, \phi_{1} \, \mathrm{j}} - \mathrm{e} ^{i \, \phi_{2} \, \mathrm{j}}) \, c_{2}\\
	 -(\mathrm{e} ^{-i \, \phi_{1} \, \mathrm{j}} - \mathrm{e} ^{- i \, \phi_{2} \, \mathrm{j}}) \, s_{1} 
	& (1 + \mathrm{e} ^{- i \, (\phi_{1}-\phi_{2}) \, \mathrm{j}}) \, s_{1} \, s_{2} + 2 \, \mathrm{e} ^{i \, 
\alpha} \, c_{1} \, c_{2}
	& -(1 + \mathrm{e} ^{- i \, (\phi_{1}-\phi_{2}) \, \mathrm{j}}) \, s_{1} \, c_{2} + 2 \, \mathrm{e} ^{i \, 
\alpha} \, c_{1} \, s_{2}\\
	 (\mathrm{e} ^{-i \, \phi_{1} \, \mathrm{j}} - \mathrm{e} ^{- i \, \phi_{2} \, \mathrm{j}}) \, c_{1} 
	& -(1 + \mathrm{e} ^{- i \, (\phi_{1}-\phi_{2}) \, \mathrm{j}}) \, c_{1} \, s_{2} + 2 \, \mathrm{e} ^{i \, 
\alpha} \, s_{1} \, c_{2} 
	& (1 + \mathrm{e} ^{- i \, (\phi_{1}-\phi_{2}) \, \mathrm{j}}) \, c_{1} \, c_{2} 
+ 2 \, \mathrm{e} ^{i \, \alpha} \, s_{1} \, s_{2}\\
\end{array}
\right)\\ \nonumber
V_{mix} ^{12} &=& \frac{1}{2} \, \left( \begin{array}{ccc}
	(\mathrm{e} ^{i \, \phi_{1} \, \mathrm{j}} - \mathrm{e} ^{i \, \phi_{2} \, \mathrm{j}}) \, s_{2}  
	& 1 + \mathrm{e} ^{i \, (\phi_{1}-\phi_{2}) \, \mathrm{j}} 
	&  -(\mathrm{e} ^{i \, \phi_{1} \, \mathrm{j}} - \mathrm{e} ^{i \, \phi_{2} \, \mathrm{j}}) \, c_{2}\\
	(1 + \mathrm{e} ^{- i \, (\phi_{1}-\phi_{2}) \, \mathrm{j}}) \, s_{1} \, s_{2} + 2 \, \mathrm{e} ^{i \, 
\alpha} \, c_{1} \, c_{2}
	&  -(\mathrm{e} ^{-i \, \phi_{1} \, \mathrm{j}} - \mathrm{e} ^{- i \, \phi_{2} \, \mathrm{j}}) \, s_{1}
	&  -(1 + \mathrm{e} ^{- i \, (\phi_{1}-\phi_{2}) \, \mathrm{j}}) \, s_{1} \, c_{2} + 2 \, \mathrm{e} ^{i \, 
\alpha} \, c_{1} \, s_{2}\\
	 -(1 + \mathrm{e} ^{- i \, (\phi_{1}-\phi_{2}) \, \mathrm{j}}) \, c_{1} \, s_{2} + 2 \, \mathrm{e} ^{i \, 
\alpha} \, s_{1} \, c_{2} 
	&  (\mathrm{e} ^{-i \, \phi_{1} \, \mathrm{j}} - \mathrm{e} ^{- i \, \phi_{2} \, \mathrm{j}}) \, c_{1}
	& (1 + \mathrm{e} ^{- i \, (\phi_{1}-\phi_{2}) \, \mathrm{j}}) \, c_{1} \, c_{2} 
+ 2 \, \mathrm{e} ^{i \, \alpha} \, s_{1} \, s_{2}\\
\end{array} 
\right)\\ \nonumber
V_{mix} ^{13} &=&  \frac{1}{2} \,\left( \begin{array}{ccc}
	(\mathrm{e} ^{i \, \phi_{1} \, \mathrm{j}} - \mathrm{e} ^{i \, \phi_{2} \, \mathrm{j}}) \, s_{2}
	& -(\mathrm{e} ^{i \, \phi_{1} \, \mathrm{j}} - \mathrm{e} ^{i \, \phi_{2} \, \mathrm{j}}) \, c_{2}
	& 1 + \mathrm{e} ^{i \, (\phi_{1}-\phi_{2}) \, \mathrm{j}}\\
	(1 + \mathrm{e} ^{- i \, (\phi_{1}-\phi_{2}) \, \mathrm{j}}) \, s_{1} \, s_{2} + 2 \, \mathrm{e} ^{i \, 
\alpha} \, c_{1} \, c_{2}
	& -(1 + \mathrm{e} ^{- i \, (\phi_{1}-\phi_{2}) \, \mathrm{j}}) \, s_{1} \, c_{2} + 2 \, \mathrm{e} ^{i \, 
\alpha} \, c_{1} \, s_{2}
	& -(\mathrm{e} ^{-i \, \phi_{1} \, \mathrm{j}} - \mathrm{e} ^{- i \, \phi_{2} \, \mathrm{j}}) \, s_{1}\\
	 -(1 + \mathrm{e} ^{- i \, (\phi_{1}-\phi_{2}) \, \mathrm{j}}) \, c_{1} \, s_{2} + 2 \, \mathrm{e} ^{i \, 
\alpha} \, s_{1} \, c_{2}
	& (1 + \mathrm{e} ^{- i \, (\phi_{1}-\phi_{2}) \, \mathrm{j}}) \, c_{1} \, c_{2} 
+ 2 \, \mathrm{e} ^{i \, \alpha} \, s_{1} \, s_{2}
	& (\mathrm{e} ^{-i \, \phi_{1} \, \mathrm{j}} - \mathrm{e} ^{- i \, \phi_{2} \, \mathrm{j}}) \, c_{1}\\
\end{array}
\right)\\ \nonumber
V_{mix} ^{21} &=&  \frac{1}{2} \,\left( \begin{array}{ccc}
	-(\mathrm{e} ^{-i \, \phi_{1} \, \mathrm{j}} - \mathrm{e} ^{- i \, \phi_{2} \, \mathrm{j}}) \, s_{1}
	& (1 + \mathrm{e} ^{- i \, (\phi_{1}-\phi_{2}) \, \mathrm{j}}) \, s_{1} \, s_{2} + 2 \, \mathrm{e} ^{i \, 
\alpha} \, c_{1} \, c_{2}
	&  -(1 + \mathrm{e} ^{- i \, (\phi_{1}-\phi_{2}) \, \mathrm{j}}) \, s_{1} \, c_{2} + 2 \, \mathrm{e} ^{i \, 
\alpha} \, c_{1} \, s_{2}\\
	1 + \mathrm{e} ^{i \, (\phi_{1}-\phi_{2}) \, \mathrm{j}}
	& (\mathrm{e} ^{i \, \phi_{1} \, \mathrm{j}} - \mathrm{e} ^{i \, \phi_{2} \, \mathrm{j}}) \, s_{2}
	&  -(\mathrm{e} ^{i \, \phi_{1} \, \mathrm{j}} - \mathrm{e} ^{i \, \phi_{2} \, \mathrm{j}}) \, c_{2}\\
	(\mathrm{e} ^{-i \, \phi_{1} \, \mathrm{j}} - \mathrm{e} ^{- i \, \phi_{2} \, \mathrm{j}}) \, c_{1}
	& -(1 + \mathrm{e} ^{- i \, (\phi_{1}-\phi_{2}) \, \mathrm{j}}) \, c_{1} \, s_{2} + 2 \, \mathrm{e} ^{i \, 
\alpha} \, s_{1} \, c_{2} 
	& (1 + \mathrm{e} ^{- i \, (\phi_{1}-\phi_{2}) \, \mathrm{j}}) \, c_{1} \, c_{2} 
+ 2 \, \mathrm{e} ^{i \, \alpha} \, s_{1} \, s_{2}\\
\end{array}
\right)\\ \nonumber
V_{mix} ^{22} &=&  \frac{1}{2} \,\left( \begin{array}{ccc}
	(1 + \mathrm{e} ^{- i \, (\phi_{1}-\phi_{2}) \, \mathrm{j}}) \, s_{1} \, s_{2} + 2 \, \mathrm{e} ^{i \, 
\alpha} \, c_{1} \, c_{2}
	& -(\mathrm{e} ^{-i \, \phi_{1} \, \mathrm{j}} - \mathrm{e} ^{- i \, \phi_{2} \, \mathrm{j}}) \, s_{1}
	& -(1 + \mathrm{e} ^{- i \, (\phi_{1}-\phi_{2}) \, \mathrm{j}}) \, s_{1} \, c_{2} + 2 \, \mathrm{e} ^{i \, 
\alpha} \, c_{1} \, s_{2}\\
	(\mathrm{e} ^{i \, \phi_{1} \, \mathrm{j}} - \mathrm{e} ^{i \, \phi_{2} \, \mathrm{j}}) \, s_{2}
	& 1 + \mathrm{e} ^{i \, (\phi_{1}-\phi_{2}) \, \mathrm{j}}
	&  -(\mathrm{e} ^{i \, \phi_{1} \, \mathrm{j}} - \mathrm{e} ^{i \, \phi_{2} \, \mathrm{j}}) \, c_{2}\\
	-(1 + \mathrm{e} ^{- i \, (\phi_{1}-\phi_{2}) \, \mathrm{j}}) \, c_{1} \, s_{2} + 2 \, \mathrm{e} ^{i \, 
\alpha} \, s_{1} \, c_{2} 
	& (\mathrm{e} ^{-i \, \phi_{1} \, \mathrm{j}} - \mathrm{e} ^{- i \, \phi_{2} \, \mathrm{j}}) \, c_{1}
	& (1 + \mathrm{e} ^{- i \, (\phi_{1}-\phi_{2}) \, \mathrm{j}}) \, c_{1} \, c_{2} 
+ 2 \, \mathrm{e} ^{i \, \alpha} \, s_{1} \, s_{2}\\
\end{array}
\right)\\ \nonumber
V_{mix} ^{23} &=&  \frac{1}{2} \,\left( \begin{array}{ccc}
	(1 + \mathrm{e} ^{- i \, (\phi_{1}-\phi_{2}) \, \mathrm{j}}) \, s_{1} \, s_{2} + 2 \, \mathrm{e} ^{i \, 
\alpha} \, c_{1} \, c_{2}
	&  -(1 + \mathrm{e} ^{- i \, (\phi_{1}-\phi_{2}) \, \mathrm{j}}) \, s_{1} \, c_{2} + 2 \, \mathrm{e} ^{i \, 
\alpha} \, c_{1} \, s_{2}
	&  -(\mathrm{e} ^{-i \, \phi_{1} \, \mathrm{j}} - \mathrm{e} ^{- i \, \phi_{2} \, \mathrm{j}}) \, s_{1}\\
	(\mathrm{e} ^{i \, \phi_{1} \, \mathrm{j}} - \mathrm{e} ^{i \, \phi_{2} \, \mathrm{j}}) \, s_{2}
	&  -(\mathrm{e} ^{i \, \phi_{1} \, \mathrm{j}} - \mathrm{e} ^{i \, \phi_{2} \, \mathrm{j}}) \, c_{2}
	&  1 + \mathrm{e} ^{i \, (\phi_{1}-\phi_{2}) \, \mathrm{j}}\\
	-(1 + \mathrm{e} ^{- i \, (\phi_{1}-\phi_{2}) \, \mathrm{j}}) \, c_{1} \, s_{2} + 2 \, \mathrm{e} ^{i \, 
\alpha} \, s_{1} \, c_{2} 
	&  (1 + \mathrm{e} ^{- i \, (\phi_{1}-\phi_{2}) \, \mathrm{j}}) \, c_{1} \, c_{2} 
+ 2 \, \mathrm{e} ^{i \, \alpha} \, s_{1} \, s_{2}
	& (\mathrm{e} ^{-i \, \phi_{1} \, \mathrm{j}} - \mathrm{e} ^{- i \, \phi_{2} \, \mathrm{j}}) \, c_{1}\\
\end{array}
\right)\\ \nonumber
V_{mix} ^{31} &=&  \frac{1}{2} \,\left( \begin{array}{ccc}
	-(\mathrm{e} ^{-i \, \phi_{1} \, \mathrm{j}} - \mathrm{e} ^{- i \, \phi_{2} \, \mathrm{j}}) \, s_{1}
	& (1 + \mathrm{e} ^{- i \, (\phi_{1}-\phi_{2}) \, \mathrm{j}}) \, s_{1} \, s_{2} 
+ 2 \, \mathrm{e} ^{i \, \alpha} \, c_{1} \, c_{2}
	& -(1 + \mathrm{e} ^{- i \, (\phi_{1}-\phi_{2}) \, \mathrm{j}}) \, s_{1} \, c_{2} + 2 \, \mathrm{e} ^{i \, 
\alpha} \, c_{1} \, s_{2}\\
	(\mathrm{e} ^{-i \, \phi_{1} \, \mathrm{j}} - \mathrm{e} ^{- i \, \phi_{2} \, \mathrm{j}}) \, c_{1}
	& -(1 + \mathrm{e} ^{- i \, (\phi_{1}-\phi_{2}) \, \mathrm{j}}) \, c_{1} \, s_{2} + 2 \, \mathrm{e} ^{i \, 
\alpha} \, s_{1} \, c_{2} 
	& (1 + \mathrm{e} ^{- i \, (\phi_{1}-\phi_{2}) \, \mathrm{j}}) \, c_{1} \, c_{2} 
+ 2 \, \mathrm{e} ^{i \, \alpha} \, s_{1} \, s_{2}\\
	1 + \mathrm{e} ^{i \, (\phi_{1}-\phi_{2}) \, \mathrm{j}}
	& (\mathrm{e} ^{i \, \phi_{1} \, \mathrm{j}} - \mathrm{e} ^{i \, \phi_{2} \, \mathrm{j}}) \, s_{2}
	&  -(\mathrm{e} ^{i \, \phi_{1} \, \mathrm{j}} - \mathrm{e} ^{i \, \phi_{2} \, \mathrm{j}}) \, c_{2}\\
\end{array}
\right)\\ \nonumber
V_{mix} ^{32} &=&  \frac{1}{2} \,\left( \begin{array}{ccc}
	(1 + \mathrm{e} ^{- i \, (\phi_{1}-\phi_{2}) \, \mathrm{j}}) \, s_{1} \, s_{2} 
+ 2 \, \mathrm{e} ^{i \, \alpha} \, c_{1} \, c_{2}
	& -(\mathrm{e} ^{-i \, \phi_{1} \, \mathrm{j}} - \mathrm{e} ^{- i \, \phi_{2} \, \mathrm{j}}) \, s_{1}
	&  -(1 + \mathrm{e} ^{- i \, (\phi_{1}-\phi_{2}) \, \mathrm{j}}) \, s_{1} \, c_{2} + 2 \, \mathrm{e} ^{i \, 
\alpha} \, c_{1} \, s_{2}\\
	-(1 + \mathrm{e} ^{- i \, (\phi_{1}-\phi_{2}) \, \mathrm{j}}) \, c_{1} \, s_{2} + 2 \, \mathrm{e} ^{i \, 
\alpha} \, s_{1} \, c_{2} 
	& (\mathrm{e} ^{-i \, \phi_{1} \, \mathrm{j}} - \mathrm{e} ^{- i \, \phi_{2} \, \mathrm{j}}) \, c_{1}
	& (1 + \mathrm{e} ^{- i \, (\phi_{1}-\phi_{2}) \, \mathrm{j}}) \, c_{1} \, c_{2} 
+ 2 \, \mathrm{e} ^{i \, \alpha} \, s_{1} \, s_{2}\\
	(\mathrm{e} ^{i \, \phi_{1} \, \mathrm{j}} - \mathrm{e} ^{i \, \phi_{2} \, \mathrm{j}}) \, s_{2}
	& 1 + \mathrm{e} ^{i \, (\phi_{1}-\phi_{2}) \, \mathrm{j}}
	&  -(\mathrm{e} ^{i \, \phi_{1} \, \mathrm{j}} - \mathrm{e} ^{i \, \phi_{2} \, \mathrm{j}}) \, c_{2}\\
\end{array}
\right)\\ \nonumber
V_{mix} ^{33} &=&  \frac{1}{2} \,\left( \begin{array}{ccc}
	(1 + \mathrm{e} ^{- i \, (\phi_{1}-\phi_{2}) \, \mathrm{j}}) \, s_{1} \, s_{2} 
+ 2 \, \mathrm{e} ^{i \, \alpha} \, c_{1} \, c_{2}
	&  -(1 + \mathrm{e} ^{- i \, (\phi_{1}-\phi_{2}) \, \mathrm{j}}) \, s_{1} \, c_{2} + 2 \, \mathrm{e} ^{i \, 
\alpha} \, c_{1} \, s_{2}
	&  -(\mathrm{e} ^{-i \, \phi_{1} \, \mathrm{j}} - \mathrm{e} ^{- i \, \phi_{2} \, \mathrm{j}}) \, s_{1}\\
	-(1 + \mathrm{e} ^{- i \, (\phi_{1}-\phi_{2}) \, \mathrm{j}}) \, c_{1} \, s_{2} + 2 \, \mathrm{e} ^{i \, 
\alpha} \, s_{1} \, c_{2}
	& (1 + \mathrm{e} ^{- i \, (\phi_{1}-\phi_{2}) \, \mathrm{j}}) \, c_{1} \, c_{2} 
+ 2 \, \mathrm{e} ^{i \, \alpha} \, s_{1} \, s_{2}
	& (\mathrm{e} ^{-i \, \phi_{1} \, \mathrm{j}} - \mathrm{e} ^{- i \, \phi_{2} \, \mathrm{j}}) \, c_{1}\\
	(\mathrm{e} ^{i \, \phi_{1} \, \mathrm{j}} - \mathrm{e} ^{i \, \phi_{2} \, \mathrm{j}}) \, s_{2}
	&  -(\mathrm{e} ^{i \, \phi_{1} \, \mathrm{j}} - \mathrm{e} ^{i \, \phi_{2} \, \mathrm{j}}) \, c_{2}
	&  1 + \mathrm{e} ^{i \, (\phi_{1}-\phi_{2}) \, \mathrm{j}}\\
\end{array}
\right)
\end{eqnarray}

\small
\noindent The measure of CP-violation $J_{CP}^{ab}$ is given for the matrices $V_{mix} ^{ab}$
as
\begin{eqnarray}
&& J_{CP} ^{11} = J_{CP} (\mathrm{j},\phi_{1},\phi_{2};\theta_{1},\theta_{2},\alpha) \; , \;\;
J_{CP} ^{12} = - J_{CP} (\mathrm{j},\phi_{1},\phi_{2};\theta_{1},\theta_{2},\alpha) \; , \;\;
J_{CP} ^{13} = J_{CP} (\mathrm{j},\phi_{1},\phi_{2};\theta_{1},\theta_{2},\alpha)\\
&& J_{CP} ^{21} = - J_{CP} (\mathrm{j},\phi_{1},\phi_{2};\theta_{1},\theta_{2},\alpha) \; , \;\;
J_{CP} ^{22} = J_{CP} (\mathrm{j},\phi_{1},\phi_{2};\theta_{1},\theta_{2},\alpha) \; , \;\;
J_{CP} ^{23} = - J_{CP} (\mathrm{j},\phi_{1},\phi_{2};\theta_{1},\theta_{2},\alpha)\\
&& J_{CP} ^{31} = J_{CP} (\mathrm{j},\phi_{1},\phi_{2};\theta_{1},\theta_{2},\alpha) \; , \;\;
J_{CP} ^{32} = -J_{CP} (\mathrm{j},\phi_{1},\phi_{2};\theta_{1},\theta_{2},\alpha) \; , \;\;
J_{CP} ^{33} = J_{CP} (\mathrm{j},\phi_{1},\phi_{2};\theta_{1},\theta_{2},\alpha) \\ \nonumber
&& \mbox{with} \;\;\; J_{CP} (\mathrm{j},\phi_{1},\phi_{2};\theta_{1},\theta_{2},\alpha) = -\frac{1}{8} \, \sin ((\phi_{1}-\phi_{2}) \, \mathrm{j}) \, \sin (\frac{1}{2} \, (\phi_{1} -\phi_{2}) \, \mathrm{j})
	\, \sin (2 \, \theta_{1}) \, \sin (2 \, \theta_{2}) \, 
\sin (\frac{1}{2}\, (\phi_{1}-\phi_{2}) \, \mathrm{j} + \alpha)\\
\end{eqnarray}

\normalsize
\newpage

%%%%%%%%%%%%%%%%%%%%%%%%%%%%%%%%%%%%%%%%%%%%%%%
\mathversion{bold}
\section{Group Theory of \Groupname{D}{7}}
\mathversion{normal}
%%%%%%%%%%%%%%%%%%%%%%%%%%%%%%%%%%%%%%%%%%%%%%%
\label{app:grouptheory}

\noindent The group \Groupname{D}{7} has two one- and three two-dimensional irreducible
representations which we denote as $\MoreRep{1}{1}$, $\MoreRep{1}{2}$, $\MoreRep{2}{1}$,
$\MoreRep{2}{2}$ and $\MoreRep{2}{3}$. $\MoreRep{1}{1}$ is the trivial representation of
the group. All two-dimensional representations are faithful. The order of the group is 14.
\noindent The generator relations for the two generators $\rm A$ and $\rm B$ are:
\[
\mathrm{A}^{7}=\mathrm{1} \; , \;\;\; \mathrm{B}^{2}=\mathrm{1} \; , 
\;\;\; \mathrm{A} \mathrm{B} \mathrm{A} =\mathrm{B} \; .
\]
$\rm A$ and $\rm B$ can be chosen to be 
\begin{eqnarray}\nonumber
\mathrm{A} = \left( \begin{array}{cc}
	\mathrm{e} ^{\frac{2 \, \pi \, i}{7}} & 0\\
	0 & \mathrm{e} ^{-\frac{2 \, \pi \, i}{7}}
\end{array} \right) \; , \;\; 
\mathrm{B} = \left( \begin{array}{cc}
	0 & 1\\
	1 & 0
\end{array} \right) \;\;\; \mbox{for} \;\;\; \MoreRep{2}{1}\\ \nonumber
\mathrm{A} = \left( \begin{array}{cc}
	\mathrm{e} ^{\frac{4 \, \pi \, i}{7}} & 0\\
	0 & \mathrm{e} ^{-\frac{4 \, \pi \, i}{7}}
\end{array} \right) \; , \;\; 
\mathrm{B} = \left( \begin{array}{cc}
	0 & 1\\
	1 & 0
\end{array} \right) \;\;\; \mbox{for} \;\;\; \MoreRep{2}{2}\\ \nonumber
\mathrm{A} = \left( \begin{array}{cc}
	\mathrm{e} ^{\frac{6 \, \pi \, i}{7}} & 0\\
	0 & \mathrm{e} ^{-\frac{6 \, \pi \, i}{7}}
\end{array} \right) \; , \;\; 
\mathrm{B} = \left( \begin{array}{cc}
	0 & 1\\
	1 & 0
\end{array} \right) \;\;\; \mbox{for} \;\;\; \MoreRep{2}{3}
\end{eqnarray}
For the one-dimensional representations $\MoreRep{1}{1}$ and $\MoreRep{1}{2}$ $\rm A$ and $\rm B$
can be found in the character table \Tabref{tab:charactertable}.
\begin{table}
\begin{center}
\begin{tabular}{|l|ccccc|}
\hline
 &\multicolumn{5}{|c|}{classes}                                                 \\ \cline{2-6}
 &$\Cl{1}$&$\Cl{2}$&$\Cl{3}$&$\Cl{4}$&$\Cl{5}$\\
\cline{1-6}
\rule[0.1cm]{0cm}{0cm} $\rm G$                 &$\rm \mathbb{1}$&$\rm
A$&$\rm A^{2}$               & $\mathrm{A}^{3}$ & $\rm B$     \\
\cline{1-6}
$\OrdCl{i}$          &1      &2      &2         &2      &7                \\
\cline{1-6}
$\Ord{h}{\Cl{i}}$                   &1      &7      &7
&  7 &  2  \\
\hline
\rule[0in]{0.3cm}{0cm}$\MoreRep{1}{1}$    &1 & 1 & 1 & 1 & 1\\                             
\rule[0in]{0.3cm}{0cm}$\MoreRep{1}{2}$    &1 & 1 & 1 & 1 & -1\\                             
\rule[0in]{0.3cm}{0cm}$\MoreRep{2}{1}$    &2 & $2 \, \cos (\varphi)$ & $2
\, \cos (2 \, \varphi)$ & $2 \, \cos (3 \, \varphi)$ & 0\\  
\rule[0in]{0.3cm}{0cm}$\MoreRep{2}{2}$    &2 & $2 \, \cos (2 \, \varphi)$
& $2 \, \cos (4 \, \varphi)$ & $2 \, \cos (6 \, \varphi)$ & 0\\
\rule[0in]{0.3cm}{0cm}$\MoreRep{2}{3}$      &2 & $2 \,
\cos (3 \, \varphi)$ & $2 \, \cos (6 \,
\varphi)$ & $2 \, \cos (9 \, \varphi)$ & 0\\[0.1cm]
\hline
\end{tabular}
\end{center}
\begin{center}
\begin{minipage}[t]{12cm}
\caption[]{Character table of the group
  \Groupname{D}{7}. $\varphi$ is $\frac{2 \pi}{7}$. $\Cl{i}$ are the classes of the
group, $\OrdCl{i}$ is the order of the $i ^{\mathrm{th}}$ class, i.e. the number of distinct elements contained in this class, $\Ord{h}{\Cl{i}}$
is the order of the elements $S$ in the class $\Cl{i}$, i.e. the smallest
integer ($>0$) for which the equation $S ^{\Ord{h}{\Cl{i}}}= \mathbb{1}$
holds. Furthermore the table contains one representative for each
class $\Cl{i}$ given as product of the generators $\rm
A$ and $\rm B$ of the group. \label{tab:charactertable}}
\end{minipage}
\end{center}
\end{table}

\noindent The Kronecker products are:
\begin{eqnarray}\nonumber
&&\MoreRep{1}{1} \times \mu = \mu \; , \;\; \MoreRep{1}{2} \times \MoreRep{1}{2} = \MoreRep{1}{1} \; , \;\;
\MoreRep{1}{2} \times \MoreRep{2}{i}= \MoreRep{2}{i}\\ \nonumber
&&\left[ \MoreRep{2}{1} \times \MoreRep{2}{1} \right] =  \MoreRep{1}{1} + \MoreRep{2}{2}\; , \;\;
\left\{ \MoreRep{2}{1} \times \MoreRep{2}{1} \right\} =  \MoreRep{1}{2}\\ \nonumber
&&\left[ \MoreRep{2}{2} \times \MoreRep{2}{2} \right] = \MoreRep{1}{1} + \MoreRep{2}{3} \; , \;\;
\left\{ \MoreRep{2}{2} \times \MoreRep{2}{2} \right\} =  \MoreRep{1}{2}\\ \nonumber
&&\left[ \MoreRep{2}{3} \times \MoreRep{2}{3} \right] =  \MoreRep{1}{1} + \MoreRep{2}{1}\; , \;\;
\left\{ \MoreRep{2}{3} \times \MoreRep{2}{3} \right\} = \MoreRep{1}{2} \\ \nonumber
&&\MoreRep{2}{1} \times \MoreRep{2}{2} = \MoreRep{2}{1} + \MoreRep{2}{3} \; , \;\; 
\MoreRep{2}{1} \times \MoreRep{2}{3} = \MoreRep{2}{2} + \MoreRep{2}{3} \; , \\ \nonumber
&&\MoreRep{2}{2} \times \MoreRep{2}{3} = \MoreRep{2}{1} + \MoreRep{2}{2} \; ,   
\end{eqnarray}
where $\mu$ is any representation of the group and $\left[ \nu \times \nu \right]$ denotes the
symmetric part of the product $\nu \times \nu$, while $\left\{ \nu \times \nu \right\}$ is the anti-
symmetric one.

\noindent The Clebsch Gordan coefficients are trivial for $\MoreRep{1}{1} \times \mu$
and $\MoreRep{1}{2} \times \MoreRep{1}{2}$. For $\MoreRep{1}{2} \times \MoreRep{2}{i}$ a
non-trivial sign appears 
\[
\left( \begin{array}{c} B \, a_{1} \\ -B \, a_{2} \end{array} \right) \sim \MoreRep{2}{i}
\]
for $B \sim \MoreRep{1}{2}$ and $\left( \begin{array}{c} 
						a_{1}\\
						a_{2}
					\end{array} \right) \sim \MoreRep{2}{i}$.
$\MoreRep{1}{1}$ and $\MoreRep{1}{2}$ of $\MoreRep{2}{i} \times \MoreRep{2}{i}$ are
of the form
\[
a_{1} \, a_{2} ^{\prime} + a_{2} \, a_{1} ^{\prime} \sim \MoreRep{1}{1} \; , \;\;
a_{1} \, a_{2} ^{\prime} - a_{2} \, a_{1} ^{\prime} \sim \MoreRep{1}{2} 
\]
for $\left( \begin{array}{c} 
						a_{1}\\
						a_{2}
	\end{array} \right), 
	\left( \begin{array}{c} 
						a_{1} ^{\prime}\\
						a_{2} ^{\prime}			
	\end{array} \right)				
 \sim \MoreRep{2}{i}$. The two-dimensional representations also contained in these products
read:
\begin{eqnarray}\nonumber
&\mbox{for} \;\; \mathrm{i}=1:& \; \left( \begin{array}{c} 
	a_{1} \, a_{1}^{\prime}\\
	a_{2} \, a_{2}^{\prime}
\end{array} \right) \sim \MoreRep{2}{2}\\ \nonumber
&\mbox{for} \;\; \mathrm{i}=2:& \; \left( \begin{array}{c} 
	a_{2} \, a_{2}^{\prime}\\
	a_{1} \, a_{1}^{\prime}
\end{array} \right) \sim \MoreRep{2}{3}\\ \nonumber
&\mbox{for} \;\; \mathrm{i}=3:& \; \left( \begin{array}{c} 
	a_{2} \, a_{2}^{\prime}\\
	a_{1} \, a_{1}^{\prime}
\end{array} \right) \sim \MoreRep{2}{1}
\end{eqnarray}

\noindent For the rest of the products $\MoreRep{2}{i} \times \MoreRep{2}{j}$ we get:
\begin{eqnarray}\nonumber
\left( \begin{array}{c} 
						a_{1}\\
						a_{2}
	\end{array} \right) \sim \MoreRep{2}{1} \; , \;\; 
	\left( \begin{array}{c} 
						b_{1}\\
						b_{2}			
	\end{array} \right) \sim \MoreRep{2}{2} \; : \;\;\; 
	\left( \begin{array}{c}
		a_{2} \, b_{1}\\
		a_{1} \, b_{2}
	\end{array}
	\right) \sim \MoreRep{2}{1} \; , \;\;
	\left( \begin{array}{c}
		a_{1} \, b_{1}\\
		a_{2} \, b_{2}
	\end{array}
	\right) \sim \MoreRep{2}{3}
\\ \nonumber
\left( \begin{array}{c} 
						a_{1}\\
						a_{2}
	\end{array} \right) \sim \MoreRep{2}{1} \; , \;\; 
	\left( \begin{array}{c} 
						b_{1}\\
						b_{2}			
	\end{array} \right) \sim \MoreRep{2}{3} \; : \;\;\; 
	\left( \begin{array}{c}
		a_{2} \, b_{1}\\
		a_{1} \, b_{2}
	\end{array}
	\right) \sim \MoreRep{2}{2} \; , \;\;
	\left( \begin{array}{c}
		a_{2} \, b_{2}\\
		a_{1} \, b_{1}
	\end{array}
	\right) \sim \MoreRep{2}{3}\\ \nonumber
\left( \begin{array}{c} 
						a_{1}\\
						a_{2}
	\end{array} \right) \sim \MoreRep{2}{2} \; , \;\; 
	\left( \begin{array}{c} 
						b_{1}\\
						b_{2}			
	\end{array} \right) \sim \MoreRep{2}{3} \; : \;\;\; 
	\left( \begin{array}{c}
		a_{2} \, b_{1}\\
		a_{1} \, b_{2}
	\end{array}
	\right) \sim \MoreRep{2}{1} \; , \;\;
	\left( \begin{array}{c}
		a_{2} \, b_{2}\\
		a_{1} \, b_{1}
	\end{array}
	\right) \sim \MoreRep{2}{2}
\end{eqnarray}

\noindent All these formulae are just special cases of the more general formulae given in
\cite{dnpotentialpaper, dntheorypaper} which hold for dihedral groups \Groupname{D}{n} with
an arbitrary index $n$. 

\normalsize
\newpage

%%%%%%%%%%%%%%%%%%%%%%%%%%%%%%%%
\section{Higgs Potential}
%%%%%%%%%%%%%%%%%%%%%%%%%%%%%%%%
\label{app:Higgs}

\noindent We begin by writing down the potential for the three Higgs fields $H_s^u$ and $H_{1,2}^u$, 
which couple only to up quarks, i.e. are even under the additional 
$Z_{2} ^{(aux)}$ symmetry. 
 The potential is of the same form as $V_3$ in \Eqref{eq:potentialV3}.
As mentioned above, it has an accidental $U(1)$ symmetry.

\footnotesize
\begin{eqnarray}
V_u & = & -(\mu_s^u)^2 {H_s^u}^\dagger H_s^u 
-(\mu_D^u)^2 \left( \sum_{i=1}^2 {H_i^u}^\dagger H_i^u \right) + \lambda_s^u ({H_s^u}^\dagger H_s^u)^2  
+\lambda_1^u \left( \sum_{i=1}^2 {H_i^u}^\dagger H_i^u \right)^2\\
&+& \lambda_2^u ({H_1^u}^\dagger H_1^u - {H_2^u}^\dagger H_2^u)^2 
+ \lambda_3^u \vert {H_1^u}^\dagger H_2^u \vert^2 \nonumber\\
&+& \sigma^u_{1} ({H_s^u}^\dagger H_s^u) \left( \sum_{i=1}^2 {H_i^u}^\dagger H_i^u \right)
+ \{ \sigma_2^u ({H_s^u}^\dagger H_1^u) ({H_s^u}^\dagger 
H_2^u) + \mbox{h.c.}\} + \sigma_3^u  \left( \sum_{i=1}^2 \vert {H_s^u}^\dagger H_i^u \vert^2 \right) 
\nonumber
\end{eqnarray}
\normalsize

\noindent We have in addition five Higgs fields which are odd under the extra 
$Z_{2} ^{(aux)}$. These are 
$H_s^d$, $H_{1,2}^d$ and $\chi_{1,2}^d$. The most general potential for these five scalar fields is

\footnotesize
\begin{eqnarray}
V_d &=& - (\mu_s^d)^2 {H_s^d}^\dagger H_s^d 
- (\mu_{D}^d)^2 \left( \sum_{i=1}^2 {H_i^d}^\dagger H_i^d \right) 
- (\tilde\mu_{D}^d)^2 \left( \sum_{i=1}^2 {\chi_i^d}^\dagger \chi_i^d \right) \\
&+& \lambda_s^d ({H_s^d}^\dagger H_s^d)^2 
+ \lambda_1^d \left( \sum_{i=1}^2 {H_i^d}^\dagger H_i^d \right)^2 
+ \tilde\lambda_1^d \left( \sum_{i=1}^2 {\chi_i^d}^\dagger \chi_i^d \right)^2 
+ \lambda_2^d ({H_1^d}^\dagger H_1^d - {H_2^d}^\dagger H_2^d)^2 
+ \tilde\lambda_2^d ({\chi_1^d}^\dagger \chi_1^d -{\chi_2^d}^\dagger \chi_2^d)^2 \nonumber\\
&+& \lambda_3^d \vert {H_1^d}^\dagger H_2^d \vert^2 
+ \tilde\lambda_3^d \vert {\chi_1^d}^\dagger \chi_2^d \vert^2 
+ \sigma_1^d ({H_s^d}^\dagger H_s^d) \left( \sum_{i=1}^2 {H_i^d}^\dagger H_i^d \right) 
+ \tilde\sigma_1^d ({H_s^d}^\dagger H_s^d) \left( \sum_{i=1}^2 {\chi_i^d}^\dagger \chi_i^d \right) 
\nonumber\\
&+& \{\sigma_2^d ({H_s^d}^\dagger H_1^d) ({H_s^d}^\dagger H_2^d) + \mbox{h.c.} \} 
+ \{\tilde\sigma_2^d ({H_s^d}^\dagger \chi_1^d) ({H_s^d}^\dagger  \chi_2^d) + \mbox{h.c.}\} 
+ \sigma_3^d \left( \sum_{i=1}^2 \vert {H_s^d}^\dagger H_i^d \vert^2 \right) 
+ \tilde\sigma_3^d \left( \sum_{i=1}^2 \vert {H_s^d}^\dagger 
      \chi_i^d \vert^2 \right) \nonumber\\
&+& \tau_1^d \left(\sum_{i=1}^2 {H_i^d}^\dagger H_i^d \right) 
\left(\sum_{i=1}^2 {\chi_i^d}^\dagger \chi_i^d \right) 
+ \tau_2^d ({H_1^d}^\dagger H_1^d - 
    {H_2^d}^\dagger H_2^d) ({\chi_1^d}^\dagger \chi_1^d - 
    {\chi_2^d}^\dagger \chi_2^d)  \nonumber\\
&+& \{\tau_3^d ({H_1^d}^\dagger \chi_1^d) ({H_2^d}^\dagger \chi_2^d) + \mbox{h.c.}\} 
+ \tau_4^d \left( \sum_{i=1}^2  \vert {H_i^d}^\dagger \chi_i^d \vert^2 \right) 
+ \{\tau_5^d ({H_1^d}^\dagger \chi_2^d) ({H_2^d}^\dagger \chi_1^d) + \mbox{h.c.} \} 
+ \tau_6^d ( \, \vert {H_1^d}^\dagger \chi_2^d \vert^2 + 
    \vert {H_2^d}^\dagger \chi_1^d \vert^2 )  \nonumber\\
&+& \{\tau_7^d \{({H_2^d}^\dagger \chi_1^d) ({\chi_2^d}^\dagger \chi_1^d) + 
    ({H_1^d}^\dagger \chi_2^d) ({\chi_1^d}^\dagger \chi_2^d)\} + \mbox{h.c.} \} \nonumber\\
&+& \{\omega_1^d \{({H_s^d}^\dagger H_1^d) ({H_2^d}^\dagger \chi_2^d) + ({H_s^d}^\dagger H_2^d) 
({H_1^d}^\dagger \chi_1^d)\} + \mbox{h.c.}\} 
+ \{\omega_2^d \{({H_s^d}^\dagger H_1^d) ({\chi_1^d}^\dagger H_1^d) + 
    ({H_s^d}^\dagger H_2^d) ({\chi_2^d}^\dagger H_2^d)\} + \mbox{h.c.}\} \nonumber\\
&+& \{\omega_3^d \{({H_s^d}^\dagger \chi_1^d) ({H_1^d}^\dagger H_2^d) + 
    ({H_s^d}^\dagger \chi_2^d) ({H_2^d}^\dagger H_1^d)\} + \mbox{h.c.}\} \nonumber
\end{eqnarray}
\normalsize

\noindent This five Higgs potential is free from accidental symmetries. However, the combined potential
 $V_u + V_d$ has an accidental $SU(2) \times U(1) \times U(1)$ symmetry. It is broken explicitly by 
mixing terms, which couple the Higgs fields $H^{u}_{s,1,2}$ and $H^{d}_{s,1,2}$/$\chi^{d}_{1,2}$.
\newpage
\noindent
 The following potential $V_{mixed}$ contains all such terms, which are invariant under the symmetry
\Groupname{D}{7} $\times$ $Z_{2} ^{(aux)}$:

\footnotesize
\begin{eqnarray}
V_{mixed} &=&  \kappa_1 ({H_s^u}^\dagger H_s^u)({H_s^d}^\dagger H_s^d) 
+ \{ \kappa_2 ({H_s^u}^\dagger H_s^d)^2 + \mbox{h.c.} \} 
+ \kappa_3 \vert {H_s^u}^\dagger H_s^d \vert^2 \\
&+& \kappa_4 \left( \sum_{i=1}^2 {H_i^u}^\dagger H_i^u \right) 
\left( \sum_{i=1}^2 {H_i^d}^\dagger H_i^d \right) 
+ \tilde\kappa_4 \left( \sum_{i=1}^2 {H_i^u}^\dagger H_i^u \right) 
\left( \sum_{i=1}^2 {\chi_i^d}^\dagger \chi_i^d \right) \nonumber\\
&+& \{ \kappa_5 \left( \sum_{i=1}^2 {H_i^u}^\dagger H_i^d \right)^2 + \mbox{h.c.} \} 
+ \kappa_6 \vert {H_1^u}^\dagger H_1^d + {H_2^u}^\dagger H_2^d \vert^2 \nonumber\\
&+& \kappa_7 ({H_1^u}^\dagger H_1^u - {H_2^u}^\dagger H_2^u) 
({H_1^d}^\dagger H_1^d - {H_2^d}^\dagger H_2^d) 
+ \tilde\kappa_7 ({H_1^u}^\dagger H_1^u - {H_2^u}^\dagger H_2^u) 
({\chi_1^d}^\dagger \chi_1^d - {\chi_2^d}^\dagger \chi_2^d) \nonumber\\
&+& \{ \kappa_8 ({H_1^u}^\dagger H_1^d - {H_2^u}^\dagger H_2^d)^2 + \mbox{h.c.} \} 
+ \{\tilde\kappa_{[5-8]} ({H_1^u}^\dagger \chi_1^d) ({H_2^u}^\dagger \chi_2^d) + \mbox{h.c.}\} \nonumber\\
&+& \kappa_9 \vert {H_1^u}^\dagger H_1^d - {H_2^u}^\dagger H_2^d \vert^2  
+ \tilde\kappa_{[6+9]} ( \, \vert {H_1^u}^\dagger \chi_1^d \vert^2  + 
    \vert {H_2^u}^\dagger \chi_2^d \vert^2 ) 
+ \kappa_{10} \{ ({H_2^u}^\dagger H_1^u)({H_1^d}^\dagger H_2^d) +\mbox{h.c.} \} \nonumber \\ \nonumber 
&+& \{ \kappa_{11}({H_2^u}^\dagger H_1^d)({H_1^u}^\dagger H_2^d) +\mbox{h.c.} \} 
+ \{\tilde\kappa_{11} ({H_1^u}^\dagger \chi_2^d) ({H_2^u}^\dagger \chi_1^d) + \mbox{h.c.} \} 
+  \kappa_{12} ( \, \vert {H_2^u}^\dagger H_1^d \vert^2 + \vert {H_1^u}^\dagger H_2^d \vert^2 ) \\ \nonumber
&+& \tilde\kappa_{12} ( \, \vert {H_1^u}^\dagger \chi_2^d \vert^2 + 
    \vert {H_2^u}^\dagger \chi_1^d \vert^2 )
+ \kappa_{13} ({H_s^u}^\dagger H_s^u) \left( \sum_{i=1}^2 {H_i^d}^\dagger H_i^d \right) 
+ \tilde\kappa_{13} ({H_s^u}^\dagger H_s^u) \left( \sum_{i=1}^2 {\chi_i^d}^\dagger \chi_i^d \right) \\ \nonumber 
&+& \{\kappa_{14} ({H_s^u}^\dagger  H_1^d) ({H_s^u}^\dagger H_2^d) + \mbox{h.c.}\} 
+ \{\tilde\kappa_{14} ({H_s^u}^\dagger \chi_1^d) ({H_s^u}^\dagger \chi_2^d) + \mbox{h.c.}\} 
+ \kappa_{15} ( \, \vert {H_s^u}^\dagger H_1^d \vert^2 + \vert {H_s^u}^\dagger H_2^d \vert^2 ) \\ \nonumber
&+& \tilde\kappa_{15} ( \, \vert {H_s^u}^\dagger \chi_1^d \vert^2 
 + \vert {H_s^u}^\dagger \chi_2^d \vert^2 ) 
+ \kappa_{16} ({H_s^d}^\dagger H_s^d) \left( \sum_{i=1}^2 {H_i^u}^\dagger H_i^u \right) 
+ \{\kappa_{17} ({H_s^d}^\dagger H_1^u) ({H_s^d}^\dagger H_2^u) + \mbox{h.c.}\} 
+ \kappa_{18} \left(\sum_{i=1}^2 \vert {H_s^d}^\dagger H_i^u \vert^2 \right) \nonumber\\
&+& \{\kappa_{19} ({H_s^u}^\dagger H_s^d) 
 \left( \sum_{i=1}^2 {H_i^u}^\dagger H_i^d \right) +\mbox{h.c.} \} 
+ \{\kappa_{20} ({H_s^u}^\dagger H_s^d) 
 \left( \sum_{i=1}^2 {H_i^d}^\dagger H_i^u \right) +\mbox{h.c.} \} \nonumber\\
&+& \{\kappa_{21} \{({H_s^u}^\dagger H_1^u) ({H_s^d}^\dagger H_2^d) +  ({H_s^u}^\dagger H_2^u) 
 ({H_s^d}^\dagger H_1^d) \} + \mbox{h.c.}\} 
+ \{\kappa_{22} \{({H_s^u}^\dagger H_1^d) ({H_s^d}^\dagger H_2^u) + ({H_s^u}^\dagger H_2^d) 
 ({H_s^d}^\dagger  H_1^u)  \} + \mbox{h.c.}\} \nonumber\\
&+& \{ \kappa_{23} \{({H_s^u}^\dagger H_1^u)({H_1^d}^\dagger H_s^d) + ({H_s^u}^\dagger H_2^u)
 ({H_2^d}^\dagger H_s^d) \} +\mbox{h.c.} \} 
+ \{ \kappa_{24} \{({H_s^u}^\dagger H_1^d)({H_1^u}^\dagger H_s^d) + ({H_s^u}^\dagger H_2^d)
 ({H_2^u}^\dagger H_s^d) \} +\mbox{h.c.} \} \nonumber\\
&+& \{\kappa_{25} \{({H_s^d}^\dagger H_1^u) ({H_2^u}^\dagger \chi_2^d) + ({H_s^d}^\dagger H_2^u) 
 ({H_1^u}^\dagger \chi_1^d)\} + \mbox{h.c.}\} 
+ \{\kappa_{26} \{({H_s^d}^\dagger H_1^u) ({\chi_1^d}^\dagger H_1^u) + 
    ({H_s^d}^\dagger H_2^u) ({\chi_2^d}^\dagger H_2^u)\} + \mbox{h.c.}\} \nonumber\\
&+& \{\kappa_{27} \{({H_s^d}^\dagger \chi_1^d) ({H_1^u}^\dagger H_2^u) + 
    ({H_s^d}^\dagger \chi_2^d) ({H_2^u}^\dagger H_1^u)\} + \mbox{h.c.}\} 
+ \{\kappa_{28} \{({H_s^u}^\dagger H_1^d) ({H_2^u}^\dagger \chi_2^d) 
 + ({H_s^u}^\dagger H_2^d) ({H_1^u}^\dagger \chi_1^d)\} + \mbox{h.c.}\} \nonumber\\
&+& \{\kappa_{29} \{({H_s^u}^\dagger H_1^d) ({\chi_1^d}^\dagger H_1^u) + 
    ({H_s^u}^\dagger H_2^d) ({\chi_2^d}^\dagger H_2^u)\} + \mbox{h.c.}\}
+ \{\kappa_{30} \{({H_s^u}^\dagger \chi_1^d) ({H_1^d}^\dagger H_2^u) + 
    ({H_s^u}^\dagger \chi_2^d) ({H_2^d}^\dagger H_1^u)\} + \mbox{h.c.}\} \nonumber\\
&+& \{\kappa_{31} \{({H_s^u}^\dagger \chi_1^d) ({H_1^u}^\dagger H_2^d) + 
    ({H_s^u}^\dagger \chi_2^d) ({H_2^u}^\dagger H_1^d)\} + \mbox{h.c.}\} 
+ \{\kappa_{32} \{({H_s^u}^\dagger H_1^u) ({H_2^d}^\dagger \chi_2^d) 
 + ({H_s^u}^\dagger H_2^u) ({H_1^d}^\dagger \chi_1^d)\} + \mbox{h.c.}\} \nonumber\\
&+& \{\kappa_{33} \{({H_s^u}^\dagger H_1^u) ({\chi_1^d}^\dagger H_1^d) + 
    ({H_s^u}^\dagger H_2^u) ({\chi_2^d}^\dagger H_2^d)\} + \mbox{h.c.}\} \nonumber
\end{eqnarray}
\normalsize

\noindent In our numerical analysis we restricted ourselves to the inclusion of a minimal number of terms 
from $V_{mixed}$ which break all accidental symmetries such that only three Higgs particles remain massless
which are eaten by the $W^{\pm}$ and $Z^{0}$ boson. As explained in the main part of the text, the
three terms  $\kappa_2$, $\kappa_5$ and $\kappa_{19}$ are sufficient.\\
\noindent The numerical example in \Secref{subsubsec:assign_M4} 
and \Secref{subsubsec:quarkmassesandmixings_M4}
needs the following VEV configuration
\begin{eqnarray}\nonumber
&&\VEV{H_{s}^{d,u}} = 61.5 \, \GeV \; , \;\; \VEV{H_{1} ^{d}}= \VEV{H_{2} ^{d}} = 
\VEV{\chi_{1} ^{d}}= \VEV{\chi_{2} ^{d}} = 61.5 \, \GeV \; , \;\;
\VEV{H_{1} ^{u}} = 61.5 \, \mathrm{e}^{-\frac{3 \, \pi \, i}{7}} \, \GeV 
\\ \nonumber &&\mbox{and} \;\;\; 
\VEV{H_{2} ^{u}} = 61.5 \, \mathrm{e}^{\frac{3 \, \pi \, i}{7}}  \, \GeV
\end{eqnarray}
\noindent which allows real parameters in the potential $V_{d}$, as all fields $H^{d}_{s}$, 
$H^{d} _{1,2}$ and $\chi ^{d}_{1,2}$ have real VEVs. Furthermore we can remove the phase of
$\sigma_2^u$ such that we are left with three complex parameters stemming from $V_{mixed}$.\\
\noindent The mass parameters are at the electroweak scale:
\begin{equation}\nonumber
\mu_s^u = 100 \, \mbox{GeV} \; , \;\;  \mu_D^u = 200 \, \mbox{GeV} \;  , \;\;
\mu_s^d = 100 \, \mbox{GeV} \; , \;\; \mu_D^d = 200 \, \mbox{GeV} \;\;\; 
\mbox{and} \;\;\; \tilde\mu_D^d = 150 \, \mbox{GeV} \; .
\end{equation}
\noindent One possible setup of parameters is then:\\
\noindent For $V_{u}$ we take:
\begin{eqnarray}\nonumber
&& \lambda_s^u = 0.959337 \; , \;\; \lambda_1^u = 2.52548 \; , \;\; \lambda_2^u = 0.374967 \; , \;\; 
\lambda_3^u = -0.588842 \; , \;\; \sigma^u_1 = 1.62353 \; , \;\; \\ \nonumber
&& \sigma_2^u = -0.276964 \; , \;\; \sigma_3^u = -0.283914 \; .
\end{eqnarray}
\noindent For $V_d$ we set:
\begin{eqnarray}\nonumber
&& \lambda_s^d = 1.70438 \; , \;\; \lambda_1^d = 3.76598 \; , \;\; \tilde\lambda_1^d =  1.47549 \; , \;\;
\lambda_2^d = -0.344036 \; , \;\; \tilde\lambda_2^d = -0.185157 \; , \;\; \lambda_3^d = -0.304589 \; , 
\nonumber\\
&& \tilde\lambda_3^d = -0.733236 \; , \;\; \sigma_1^d = 0.22429 \; , \;\; \tilde\sigma_1^d = 4.6792 
\; , \;\;
\sigma_2^d = -0.87457 \; , \;\; \tilde\sigma_2^d = -2.0284 \; , \;\; \sigma_3^d = 0.961454 \; , 
\nonumber\\
&& \tilde\sigma_3^d = 0.649984 \; , \;\; \tau_1^d = 2.96557 \; , \;\; \tau_2^d =  1.22903 \; , \;\;
\tau_3^d = -2.02133 \; , \;\; \tau_4^d = -1.22242 \; , \;\; \tau_5^d =  -2.31577 \; , \nonumber\\
&& \tau_6^d = 2.38236 \; , \;\; \tau_7^d = -0.660102 \; , \;\; \omega_1^d = 0.452165 \; , \;\; 
\omega_2^d = -2.112 \; , \;\; \omega_3^d = -1.63452 \; . \nonumber
\end{eqnarray}
\noindent and for the three complex nonzero couplings from $V_{mixed}$:
\begin{equation}\nonumber
\kappa_2 = -0.638073 + i \, 0.0277608 \; , \;\; \kappa_5 = 0.312782+ i \, 0.140162 \; , \;\; 
\kappa_{19} = -0.278402 - i \, 0.124756 \; .
\end{equation}
\noindent Note that all parameters have absolute values smaller than 5 and hence they are still in the
perturbative regime.\\
\noindent With these parameter values we obtain the desired VEV structure.\\ 
\noindent The Higgs masses are then 

\small
\begin{eqnarray}\nonumber
&& 513 \GeV  , \;\; 499 \GeV  , \;\; 426 \GeV  , \;\; 414 \GeV  , \;\; 386 \GeV  , \;\;
365 \GeV  , \;\; 321 \GeV  , \;\; 266 \GeV  , \;\; 246 \GeV  , \;\; 227 \GeV  , \;\; \\ \nonumber
&& 178 \GeV  , \;\; 
 159 \GeV  , \;\;  134 \GeV  , \;\; 81 \GeV \;\;\; \mbox{and} \;\;\; 55 \GeV
\end{eqnarray}
\normalsize

\noindent for the neutral scalars. Due to the explicit CP violation in the potential we can no longer 
distinguish between scalars and pseudo-scalars. For the charged scalar fields we get

\small
\[
367 \GeV  , \;\; 333 \GeV  , \;\; 294 \GeV  , \;\; 261 \GeV  , \;\; 145 \GeV  , \;\; 115 \GeV 
\;\;\; \mbox{and} \;\;\; 55 \GeV \; .
\]
\normalsize

\noindent They are therefore in general too light to pass the constraints coming from direct
searches as well as from bounds on FCNCs. Nevertheless, soft breaking terms of mass dimension
two of the order of $10 \, \TeV$ could lift the masses above these experimental bounds.

\end{document}